\documentclass[final,3p,times,sort&compress]{elsarticle}

\usepackage{multirow,setspace,times,amssymb,amsmath,graphicx,color,rotating,subfigure,url}
\usepackage[bookmarks=true,colorlinks=true,linkcolor=blue,anchorcolor=blue,citecolor=blue,pdfauthor=author,unicode]{hyperref}
\usepackage{natbib}
\usepackage{booktabs}
\usepackage{mathrsfs}

\graphicspath{{Figures/}}

\hypersetup{CJKbookmarks=true}%
\journal{Physica A}

\begin{document}

\begin{frontmatter}

\title{Visibility graph analysis of the grains and oilseeds indices}

\author[SB,SIE]{Hao-Ran Liu}
\author[SB,SIE]{Ming-Xia Li\corref{CorrAuth}}
\ead{mxli@ecust.edu.cn}
\author[SB,RCE,SM]{Wei-Xing Zhou\corref{CorrAuth}}
\ead{wxzhou@ecust.edu.cn}
\cortext[CorrAuth]{Corresponding authors.} 

\address[SB]{School of Business, East China University of Science and Technology, Shanghai 200237, China}
\address[SIE]{School of Information Science and Engineering, East China University of Science and Technology, Shanghai 200237, China}
\address[RCE]{Research Center for Econophysics, East China University of Science and Technology, Shanghai 200237, China}
\address[SM]{School of Mathematics, East China University of Science and Technology, Shanghai 200237, China}

\begin{abstract}
    The Grains and Oilseeds Index (GOI) and its sub-indices of wheat, maize, soyabeans, rice, and barley are daily price indexes reflect the price changes of the global spot markets of staple agro-food crops. In this paper, we carry out a visibility graph (VG) analysis of the GOI and its five sub-indices. Our findings reveal that the degree distributions of the VGs, except for rice, exhibit exponentially truncated power-law tails, while the rice VG conforms to a power-law tail. The average clustering coefficients of the six VGs are quite large ($>0.5$) and exhibit a nice power-law relation with respect to the average degrees of the VGs. For each VG, the clustering coefficients of nodes are inversely proportional to their degrees for large degrees and are correlated to their degrees as a power law for small degrees. All the six VGs exhibit small-world characteristics. The degree-degree correlation coefficients shows that the VGs for maize and soyabeans indices exhibit weak assortative mixing patterns, while the other four VGs are weakly disassortative. The average nearest neighbor degree functions have similar patterns, and each function shows a more complex mixing pattern which decreases for small degrees, increases for mediate degrees, and decreases again for large degrees.
\end{abstract}


\end{frontmatter}

\section{Introduction}
\label{S1:Introduction}

Food prices have a very strong impact on food security, which is the connerstone of state security and social stability. Rising food prices often cause inflation, famine, and social unrest. Therefore, there is a great need to study the changes in food prices. The Grains and Oilseeds Index (GOI) and its five sub-indices of wheat, maize, soyabeans, rice, and barley are daily price indices compiled by the International Grains Council that reflect the price changes in the global spot markets of staple agro-food crops. In this paper, we will investigate the changes in the GOI index from the perspective of complex networks.

The fundamental motivation lies in the transformative potential of mapping time series onto complex networks, thereby harnessing the rich toolkit and theoretical framework of network science to dissect the multifaceted complexity of temporal data. This conversion not only captures the nonlinear associations and long-range dependencies inherent in time series but also unveils the underlying topological structure and dynamic characteristics that elude conventional statistical approaches.
There are a variety of ways to transform time series into complex networks \cite{Zou-Donner-Marwan-Donges-Kurths-2019-PhysRep}.
Li and Wang transformed the Hang Seng Index into discrete symbolic sequences, which can be transformed into a complex network and analyzed \cite{Li-Wang-2006-CSB, Li-Wang-2007-PhysicaA}. Zhang and Small construct complex networks from pseudoperiodic time series and investigate the relationship between time series and network topological structures \cite{Zhang-Small-2006-PhysRevLett}. Xu et al. transform time series into nearest neighbor networks \cite{Xu-Zhang-Small-2008-ProcNatlAcadSciUSA}. Yang et al. proposed segment correlation networks to transform time series into complex networks \cite{Yang-Yang-2008-PhysicaA}. Lucasa et al. proposed an algorithm to construct time series as visibility graphs \cite{Lacasa-Luque-Ballesteros-Luque-Nuno-2008-ProcNatlAcadSciUSA, Luque-Lacasa-Ballesteros-Luque-2009-PhysRevE}. Then, various extensions have been proposed to the visibility graph \cite{Lacasa-Luque-Luque-Nuno-2009-EPL, Xu-Zhang-Deng-2018-ChaosSolitonsFractals, Ahadpour-Sadra-2012-IS, Ahadpour-Sadra-ArastehFard-2014-IS, Bezsudnov-Snarskii-2014-PhysicaA, Bianchi-Livi-Alippi-Jenssen-2017-SR, Zhou-Jin-Gao-Luo-2012-APS, Gao-Cai-Yang-Dang-2017-PhysicaA, Zou-Donner-Marwan-Small-Kurths-2014-NPG, Gao-Cai-Yang-Dang-Zhang-2016-SR, Chen-Hu-Mahadevan-Deng-2014-PhysicaA, Snarskii-Bezsudnov-2016-PhysRevE}.

Consider the case of converting a time series into a visibility graph, where points in the series become nodes connected based on their mutual "visibility" or adjacency, leading to an intuitive visualization of the data's connectivity landscape. Through the lens of complex network analysis, metrics such as degree distribution, clustering coefficient, and community structure can be employed to identify patterns, anomalies, and trends within the time series that might otherwise remain obscured. This methodology is not confined to single-dimensional sequences; it extends seamlessly to multivariate and multidimensional time series, even spatiotemporal data, offering a unifying analytical perspective across disciplines \cite{Qian-Jiang-Zhou-2010-JPhysA, Ni-Jiang-Zhou-2009-PhysLettA, Zhang-Shang-Xiong-Xia-2018-FNL, Xie-Han-Zhou-2019-EPL, Nguyen-Nguyen-Nguyen-2019-PhysicaA, Vamvakaris-Pantelous-Zuev-2018-PhysicaA, Yang-Wang-Yang-Mang-2009-PhysicaA, Wang-Zheng-Wang-2019-PhysicaA, Elsner-Jagger-Fogarty-2009-GRL, Liu-Zhou-Yuan-2010-PhysicaA, Zhang-Zou-Zhou-Gao-Guan-2017-CNSNS, Xie-Han-Jiang-Wei-Zhou-2017-EPL, Wang-Li-Wang-2012-PhysicaA, Xie-Zhou-2011-PhysicaA, Ravetti-Carpi-Goncalves-Frery-Rosso-2014-PLoS1}. Dai et al. investigated the statistical properties of the visibility graphs of the economic policy uncertainty indices \cite{Dai-Xiong-Zhou-2019-PhysicaA}. Yu studied the characteristics of the visibility graph of the gold price time series \cite{Yu-2013-PhysicaA}. Partida et al. investigated the hierarchical structure of the visibility graphs and horizontal visibility graphs of the Bitcoin and Ethereum price time series \cite{Partida-Gerassis-Criado-Romance-Giraldez-Taboada-2022-ChaosSolitonsFractals}. Xie et al. investigated the network motif of the horizontal visibility graphs of heartbeat rate time series \cite{Xie-Han-Zhou-2019-CommunNonlinearSciNumerSimul}. Fan et al. investigated the dynamic behavior of carbon prices from the perspective of visibility graph \cite{Fan-Li-Yin-Tian-Liang-2019-AEn}. Sun et al. investigated the visibility graph of the natural gas price time series \cite{Sun-Wang-Gao-2016-PhysicaA}.

However, the GOI indices have not been studied from the perspective of visibility graphs. In this paper, we carry out a visibility graph analysis of the GOI and its five subindices. The remainder of this work is organized as follows. In Section~\ref{S1:DataDescription}, we describe the sources and specifications of the data used in this paper and the algorithm to construct GOI visibility graphs from GOI indices and draw six GOI visibility graphs constructed with this algorithm. In Section~\ref{S1:EmpiricalAnalysis}, we investigate six GOI indices and a few structural properties of their visibility graphs, including the Hurst exponent, macro properties, degree distribution, clustering coefficient, small-world property
, and mixing pattern. In Section~\ref{S1:Summary}, we summarize and conclude.

\section{Data Description}
\label{S1:DataDescription}

\subsection{Data source}

The IGC Grains and Oilseeds Index data sets used in this paper were obtained from the International Grains Council (available at \url{https://www.igc.int}, accessed on April 12, 2023). The GOI indices are six daily basis indices with the base day of January, 2000 and the base point of 100. The GOI indices are comprised of the IGC GOI Index and five sub-indices, which are the Wheat Index, the Maize Index, the Soybeans Index, the Rice Index, and the Barley Index. In the dataset used in this paper, the start date of all time series is January 3, 2000, and the end date is December 30, 2022, with a total of 5990 data points. 

\subsection{Visibility graph construction}

In order to investigate the attributes of GOI indices, we construct the visibility graph $\mathscr{G}(\mathscr{V},\mathscr{E})$ for each GOI Index, where $\mathscr{V}$ is the set of nodes and $\mathscr{E}$ is the set of edges. For the GOI Index time series $I^{Index}(t)$ of length $T$, the set of nodes $\mathscr{V}$ is
\begin{equation}
    \mathscr{V} = \{0,1,2,\cdots,T\},
\end{equation}
where $Index \in \{ \mathrm{IGC~GOI, Wheat, Maize, Soyabeans, Rice, Barley}\}$. For any two nodes $t_i$ and $t_j$, the edge $(t_i,t_j)$ will exists if visibility exists between two data points $(t_i,I(t_i))$ and $(t_j,I(t_j))$, which means for all $t_i < t_k < t_j$ meets the condition \cite{Lacasa-Luque-Ballesteros-Luque-Nuno-2008-ProcNatlAcadSciUSA}
\begin{equation}
    I(t_k) < I(t_j) + \left[ I(t_i)-I(t_j) \right] \frac{t_j-t_k}{t_j-t_i}.
\end{equation}

\begin{figure}[!ht]
    \centering
    \includegraphics[width=0.325\linewidth]{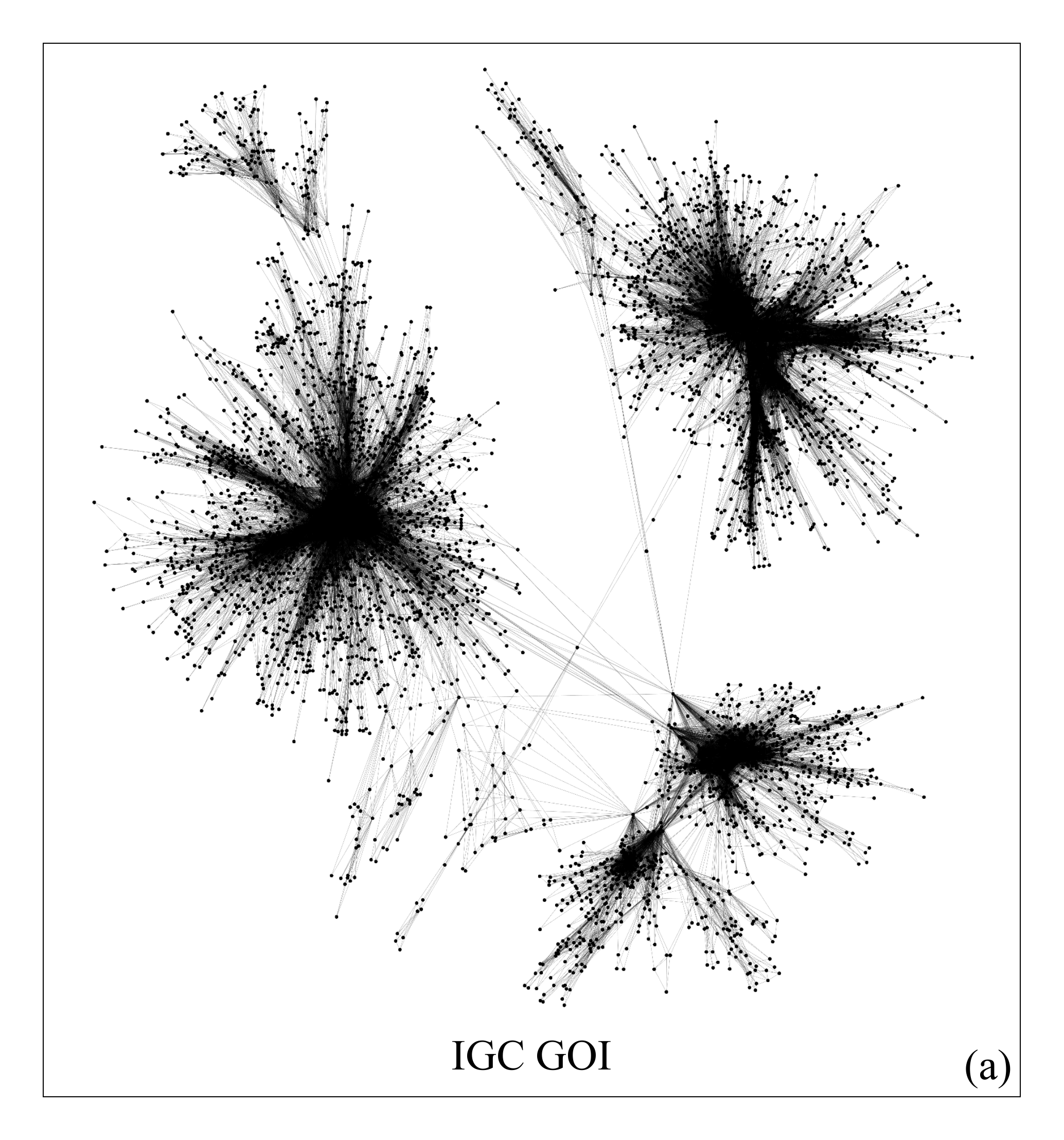}
    \includegraphics[width=0.325\linewidth]{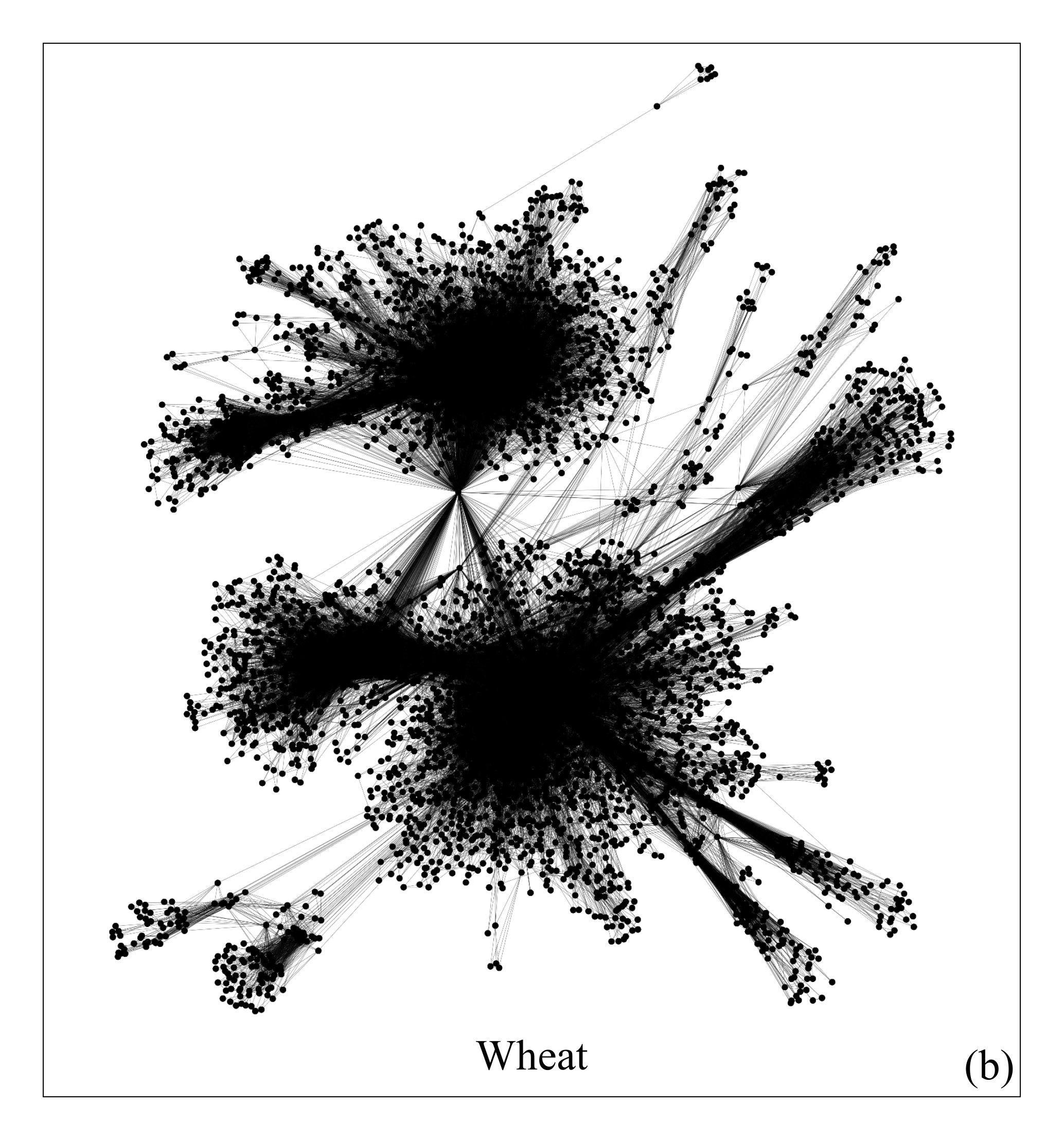}
    \includegraphics[width=0.325\linewidth]{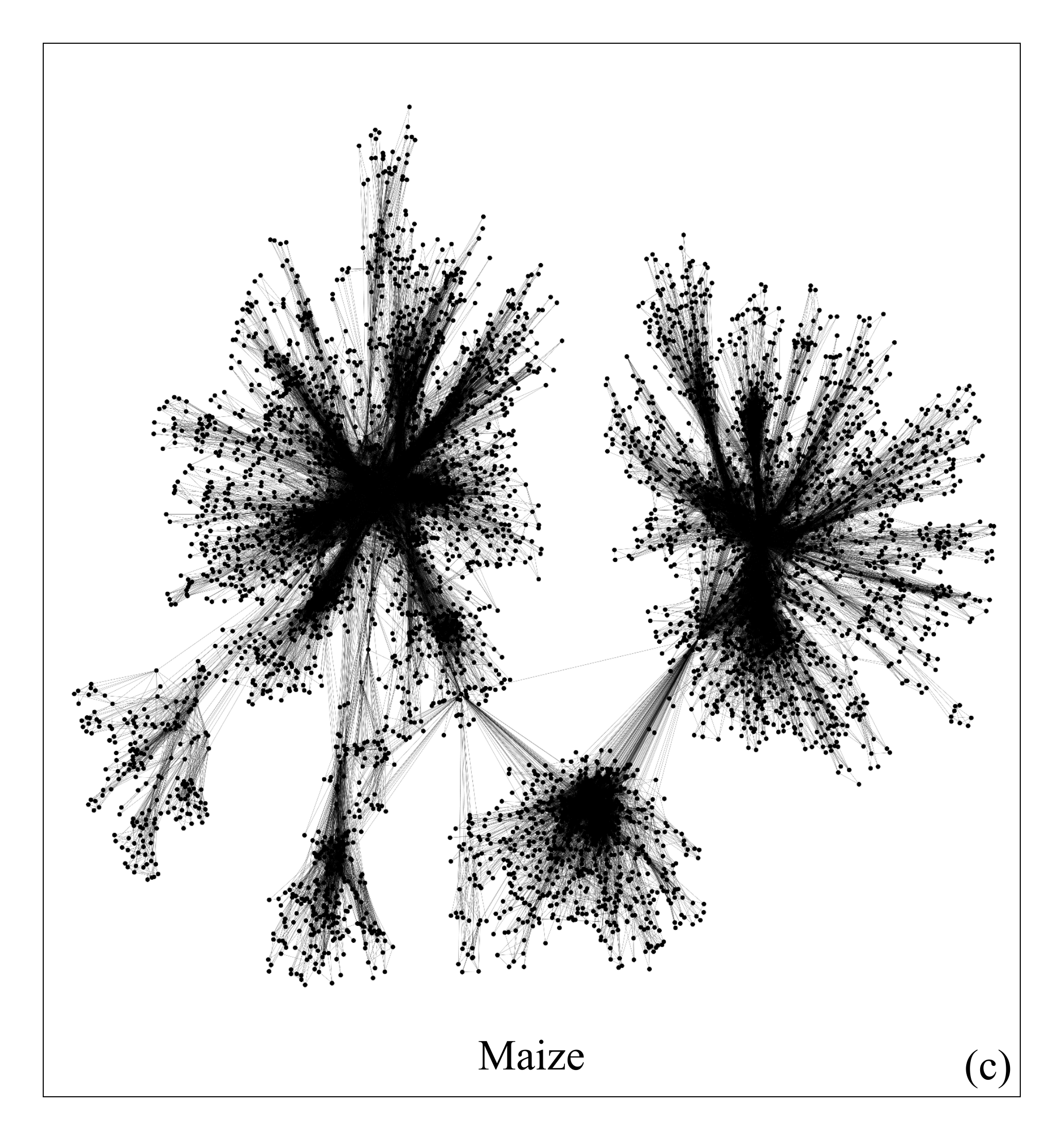}\\
    \includegraphics[width=0.325\linewidth]{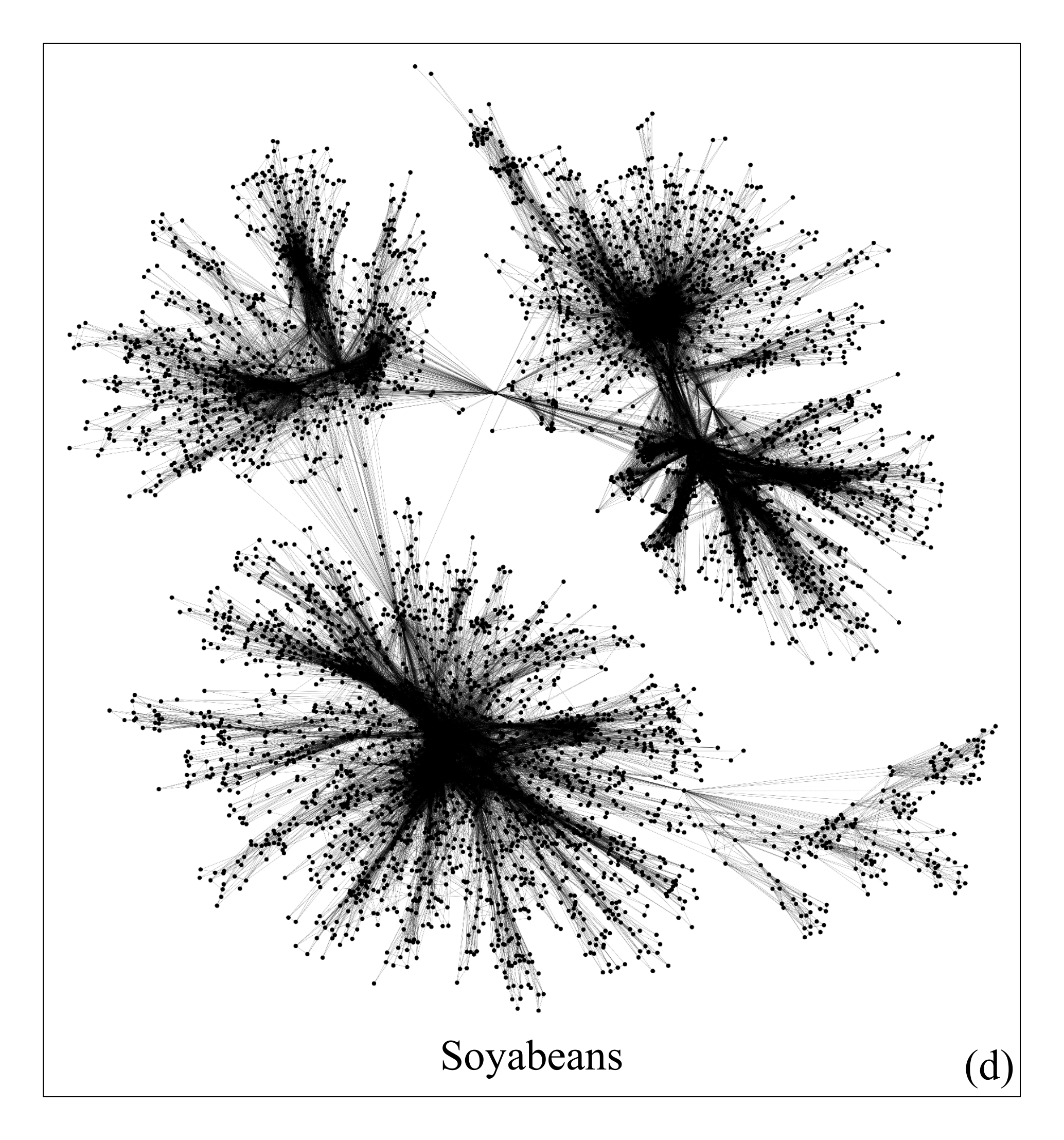}
    \includegraphics[width=0.325\linewidth]{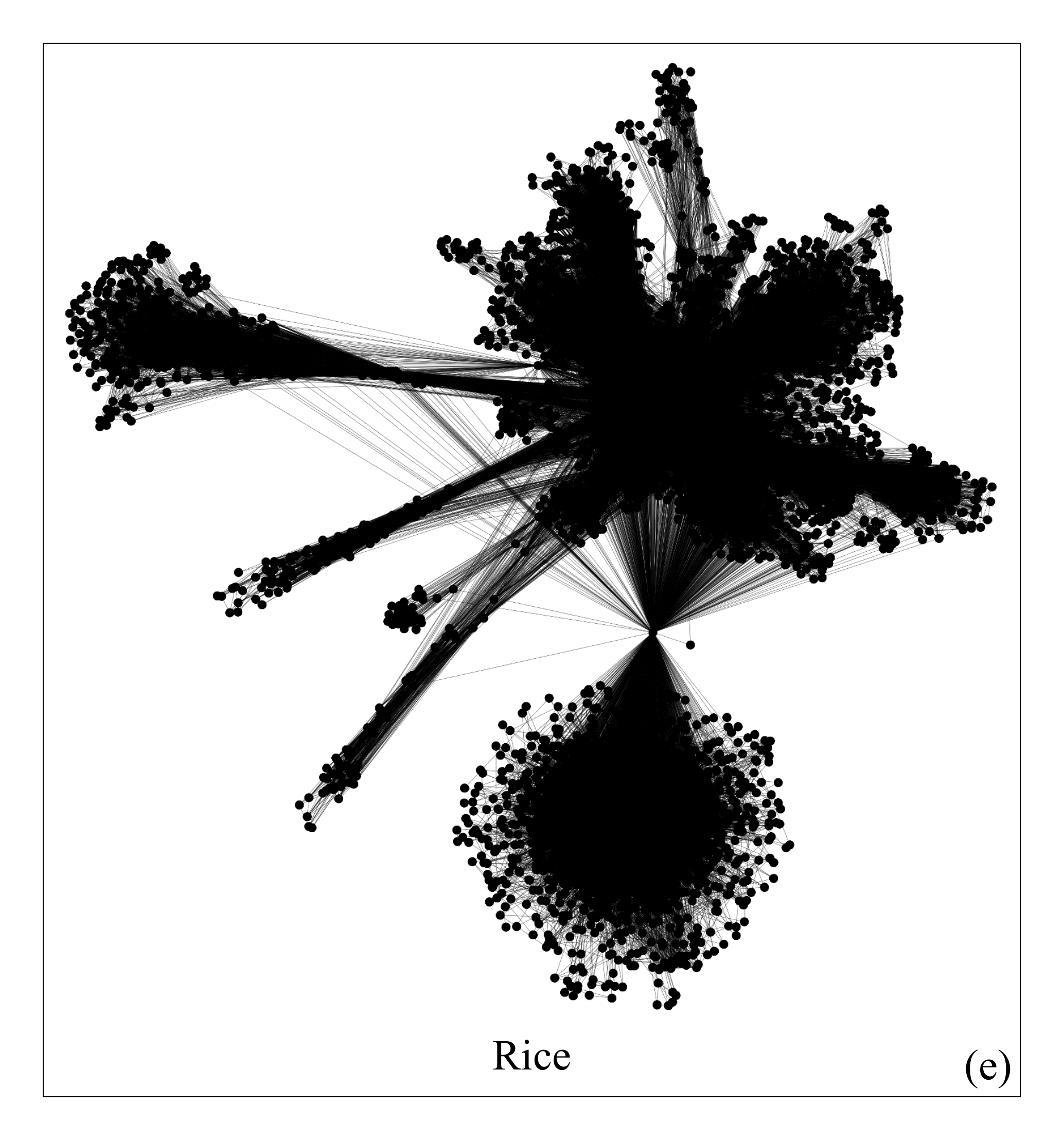}
    \includegraphics[width=0.325\linewidth]{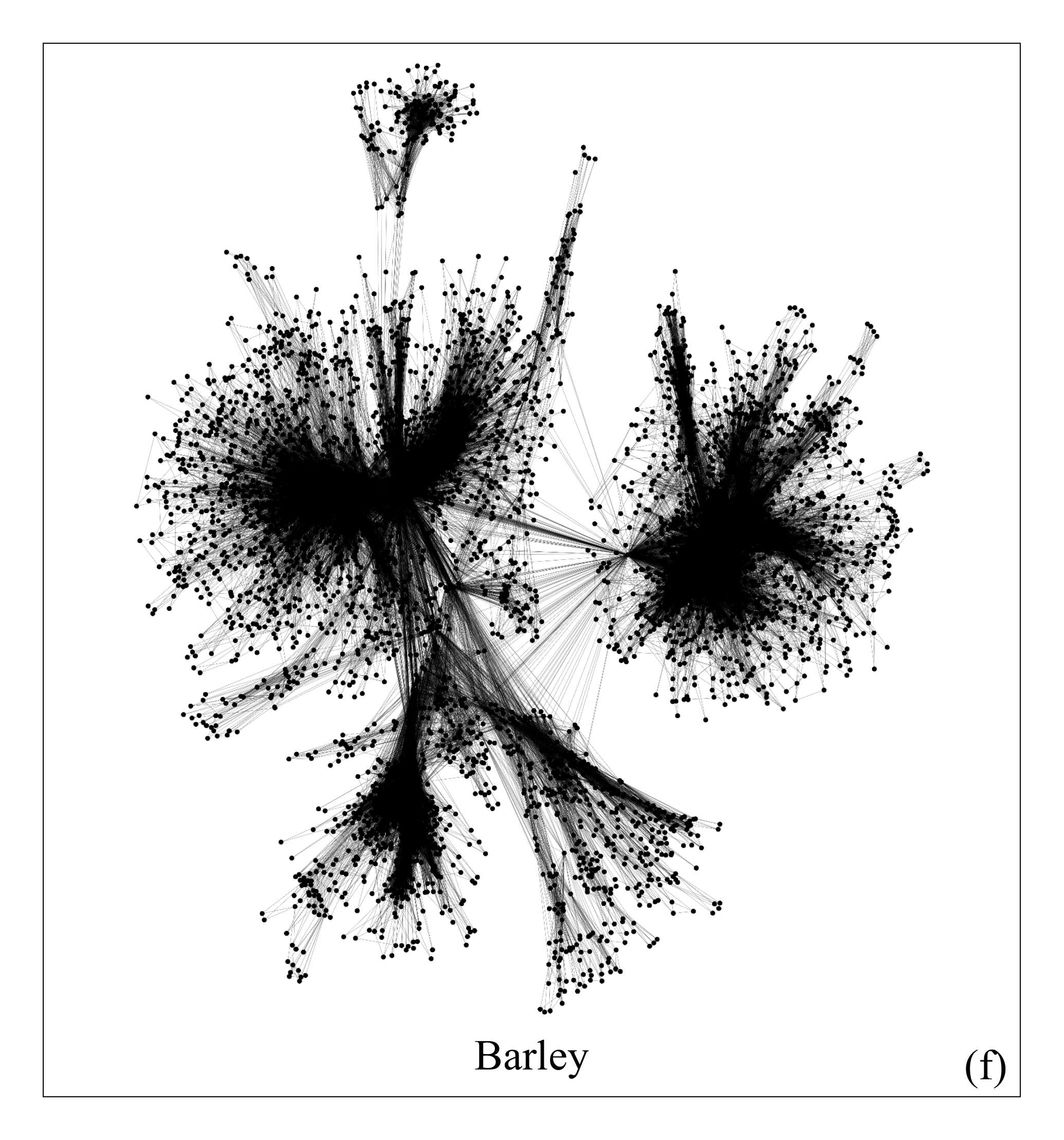}
    \caption{The VGs constructed by six GOI indices. (a) IGC GOI VG. (b) Wheat VG. (c) Maize VG. (d) Soyabeans VG. (e) Rice VG. (f) Barley VG.}
    \label{Fig:GOIVG:VisibilityGraph}
\end{figure}

In this manner, we convert a time series into its visibility graph (VG), which is an undirected graph with $T$ nodes and a maximum of $\frac{1}{2}T(T-1)$ edges. The six VGs constructed by six GOI indices are illustrated in Fig.~\ref{Fig:GOIVG:VisibilityGraph}.

\section{Empirical analysis}
\label{S1:EmpiricalAnalysis}

\subsection{Hurst exponent of six GOI indices}
The Hurst exponent is used as a measure of long-term memory for time series. Since the rescaled range method, the traditional way of measuring Hurst exponent, usually overestimates Hurst exponent, we choose detrended fluctuation analysis to measure Hurst exponent \cite{Kantelhardt-Zschiegner-KoscielnyBunde-Havlin-Bunde-Stanley-2002-PhysicaA}. The steps of detrended fluctuation analysis are as follows:

First, we calculate the profile of a time series $x(i)$.
\begin{equation}
    y(u)=\sum_{i=1}^u{\left[x(i)-\langle x \rangle\right]}, u=1,\dots,N,
\end{equation}
where $\langle x \rangle$ is the average value of $x(i)$. 

Next, the profile $y$ is divided into $n$ non-overlapping segments $Y_k(u)$ of equal length $s$.
\begin{equation}
    Y_k(u)=\left\{y(u)|(k-1)s+1\leq u\leq ks\right\}, k=1,\dot,n.
\end{equation}

Within each segment, a least squares fit is performed, and the fitted trend is subtracted from the data points within that segment. The detrended profile function on scale $s$ is determined by
\begin{equation}
    \tilde{Y}_k(u)=Y_k(u)-g(u),
\end{equation}
where $g(u)$ is a linear function. This detrending step is crucial to isolate the intrinsic fluctuations from any remaining local trends. 

Then, we calculate the variance for each detrended segment.
\begin{equation}
    F(s)=\left\{\frac{1}{n}\sum_{k=1}^{n}\left[\tilde{Y}_k(u)\right]^2\right\}^{1/2}
\end{equation}
This quantifies the magnitude of the fluctuations within each segment after removing the local trend.

\begin{figure}[!ht]
    \centering
    \includegraphics[width=0.5\linewidth]{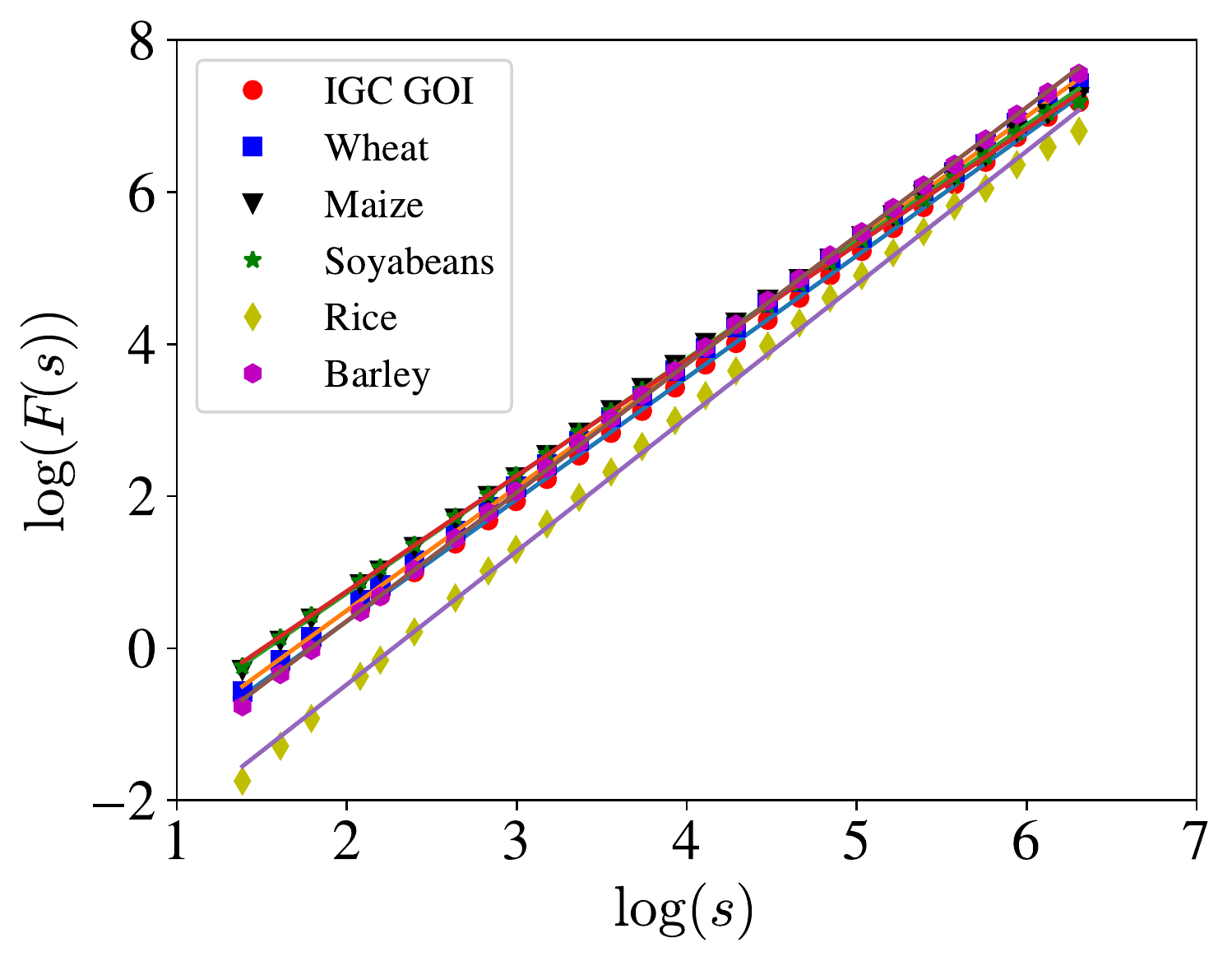}
    \caption{The fluctuation measure $F(s)$ as a function of the scale $s$ of six GOI Index times series.}
    \label{Fig:GOIVG:DFA}
\end{figure}

Finally, we plot fluctuation measure $F(s)$ versus the scale $s$ (Fig.~\ref{Fig:GOIVG:DFA}), which often reveals a power-law relationship, 
\begin{equation}
    F(s)\sim s^H,
\end{equation}
where $H$ is the Hurst exponent. If $0<H<0.5$, the GOI Index is anti-persistent or negative autocorrelated. If $H=0.5$, the GOI Index is completely uncorrelated. If $0.5<H<1$, the GOI Index is persistent or positive autocorrelated.

\begin{table}[!ht]
    \centering
    \caption{The hurst exponent $H$ of six GOI Index time series.}
    \smallskip
    \setlength{\tabcolsep}{5.2mm}
    \begin{tabular}{ccccccc}
    \toprule
      & IGC GOI & Wheat  & Maize  & Soyabeans & Rice& Barley \\ \midrule
    Hurst exponent & 0.6024  & 0.6240 & 0.5405 & 0.5200 & 0.7540 & 0.6894  \\ \bottomrule
    \end{tabular}
    \label{Table:GOIVG:HurstExponent}
\end{table}

The Hurst exponents of the six VGs are illustrated in Table~\ref{Table:GOIVG:HurstExponent}. The GOI Index with the largest Hurst exponent is the Rice GOI Index, with a Hurst exponent of 0.7540, which indicates a strong persistent of the Rice GOI Index. On the contrary, the Hurst exponent of the GOI indices of maize and soyabeans are 0.5405 and 0.5200, which are very close to 0.5, implying that the GOI indices of maize and soyabeans are close to Brownian motion. These results agree with the Hurst exponents obtained by using the detrending moving average analysis \cite{Gao-Shao-Yang-Zhou-2022-ChaosSolitonsFractals}.

\subsection{Global statistical quantities}

We first calculate the main global statistical quantities of the VGs. 
The node number $N_\mathscr{V}$ is the number of nodes in a VG
\begin{equation}
    N_\mathscr{V}=\sharp(\mathscr{V}),
\end{equation}
where $\mathscr{V}$ is the set of nodes and $\sharp(\mathbf{X})$ is the cardinal number of set $\mathbf{X}$. 
The edge number $N_\mathscr{E}$ is the number of edges in a VG
\begin{equation}
    N_\mathscr{E}=\sharp(\mathscr{E}),
\end{equation}
where $\mathscr{E}$ is the set of edges. 
As an undirected network, the density $\rho$ of a VG is defined as
\begin{equation}
    \rho = \frac{2N_\mathscr{E}}{N_\mathscr{V}(N_\mathscr{V}-1)}.
\end{equation}
The node degree $k_i$ of node $i$ is defined as the number of edges connected to node $i$
\begin{equation}
    k_i = \sharp(\{(i,j)~|~ (i,j) \in \mathscr{E}\}),
\end{equation}
and the average degree $\left\langle k \right\rangle$ of a VG is defined as average of node degrees
\begin{equation}
    \left\langle k \right\rangle = \frac{1}{N_\mathscr{V}} \sum_{i \in \mathscr{V}} k_i = \frac{1}{N_\mathscr{V}} \sum_{i \in \mathscr{V}} \sharp(\{(i,j)| (i,j) \in \mathscr{E}\}) = \frac{2 N_\mathscr{E}}{N_\mathscr{V}}.
\end{equation}
The maximum degree $k_{i,\max}$ is defined as the maximum of all node degrees
\begin{equation}
    k_{i,\max} = \max_{i \in \mathscr{V}} \{k_i\}.
\end{equation}

To investigate the macro characteristics of the VGs constructed by GOI indices, we calculated node numbers, edge numbers, network density, average degree and maximum degree of six VGs and illustrated them in Table~\ref{Table:GOIVG:MacroProperties}. As shown in Table~\ref{Table:GOIVG:MacroProperties}, the maize VG has the lowest number of edges with 69,708 edges, where the rice VG has the highest number of edges with 224,657 edges. Moreover, the density of six VGs is quite small, which means every VG is quite sparse.

\begin{table}[!ht]
    \centering
    \caption{Global statistical quantities of the six VGs conveted from the six GOI indices.}
    \smallskip
    \begin{tabular}{ccccccc}
    \toprule
    Series & Node number & Edge number & Density & Average degree & Maximum degree  \\ \midrule
    IGC GOI  & 5990  & 97682 & 0.0054  & 32.62 & 638              \\
    Wheat & 5990  & 114079& 0.0064  & 38.09 & 1313                \\
    Maize & 5990  & 69708 & 0.0039  & 23.27 & 486              \\
    Soyabeans& 5990  & 70674 & 0.0039  & 23.60 & 381              \\
    Rice  & 5990  & 224657& 0.0125  & 75.01 & 2746                \\
    Barley& 5990  & 141793& 0.0079  & 47.34 & 810             \\ \bottomrule
    \end{tabular}
    \label{Table:GOIVG:MacroProperties}
\end{table}

Fig.~\ref{Fig:GOIVG:VisibilityGraph} is the VGs of the six GOI indices. We can observe the community structures in six VGs. It may indicate that time series have different behavior patterns on different time scales. The community structures in five VGs (except Rice VG) are like a sparse star network. It implies that those time series may have significant peaks and low short-term correlation. The connection of the community structures in Rice VG are more dense. The dense connection shows a strong correlation between the data points in the time series. This correlation may be reflected in the long-range dependence, that is, the values between any two points in the sequence may affect each other, rather than being limited to adjacent or similar points.

\subsection{Degree distribution}

To find the properties of node degree distribution for six VGs, we calculated the empirical distribution density function $f(k)$ of node degree $k_i$. To avoid the problem of fitting parameter deviation caused by using linear binning for fat-tailed distributions \cite{Milojevic-2010-JAmSocInfSciTechnol}, we choose logarithmic binning to obtain the equivalent empirical distribution functions $g(\ln k)$ of 
$\ln k$, since \cite{Press-Teukolsky-Vetterling-Flannery-1996}
\begin{equation}
    f(k)~\mathrm{d} k = p(\ln k)~\mathrm{d}\ln k,
\end{equation}
then we have
\begin{equation}
    f(k) = \frac{1}{k}p(\ln k).
\end{equation}

\begin{figure}[!ht]
    \centering
    \includegraphics[width=0.325\linewidth]{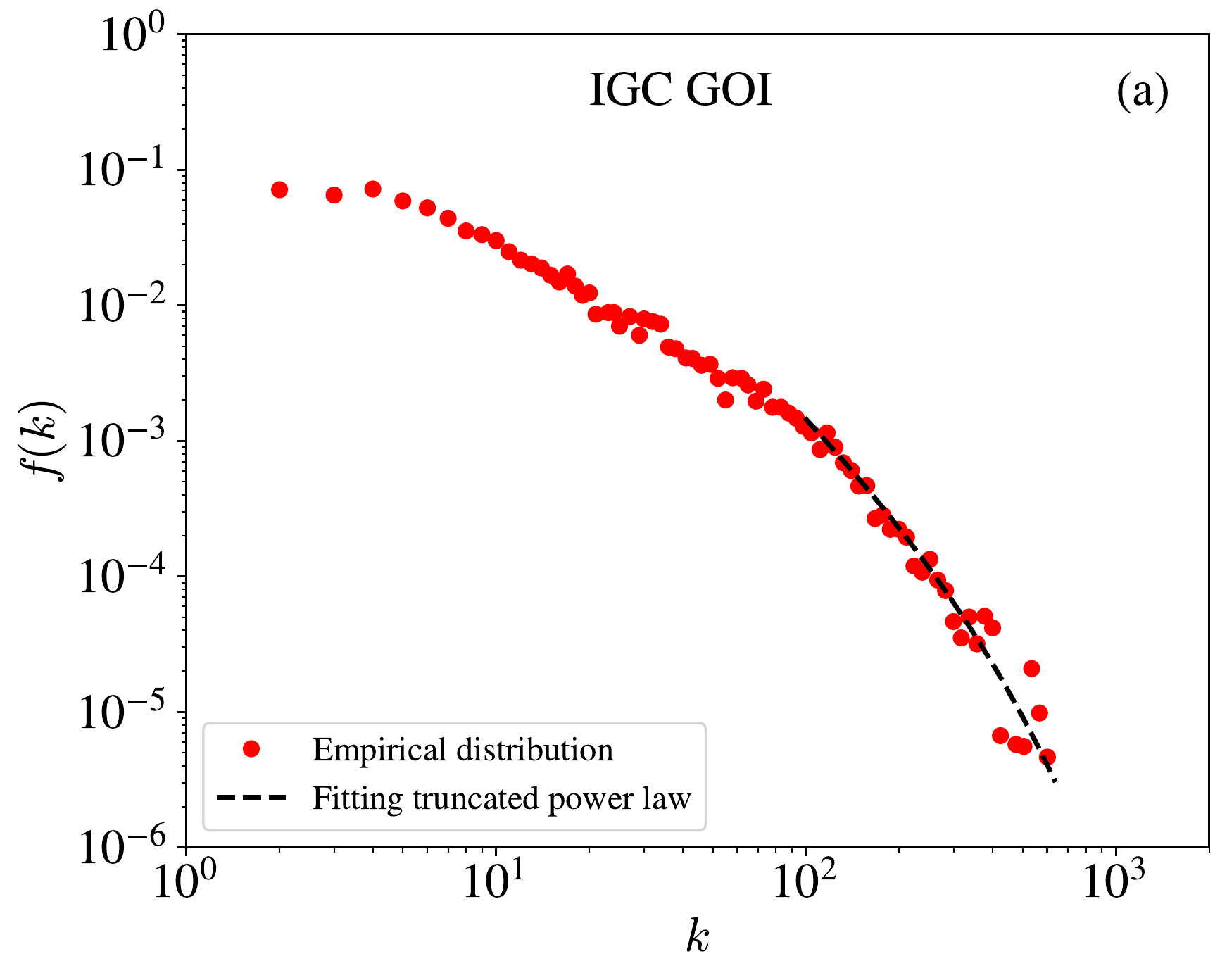}
    \includegraphics[width=0.325\linewidth]{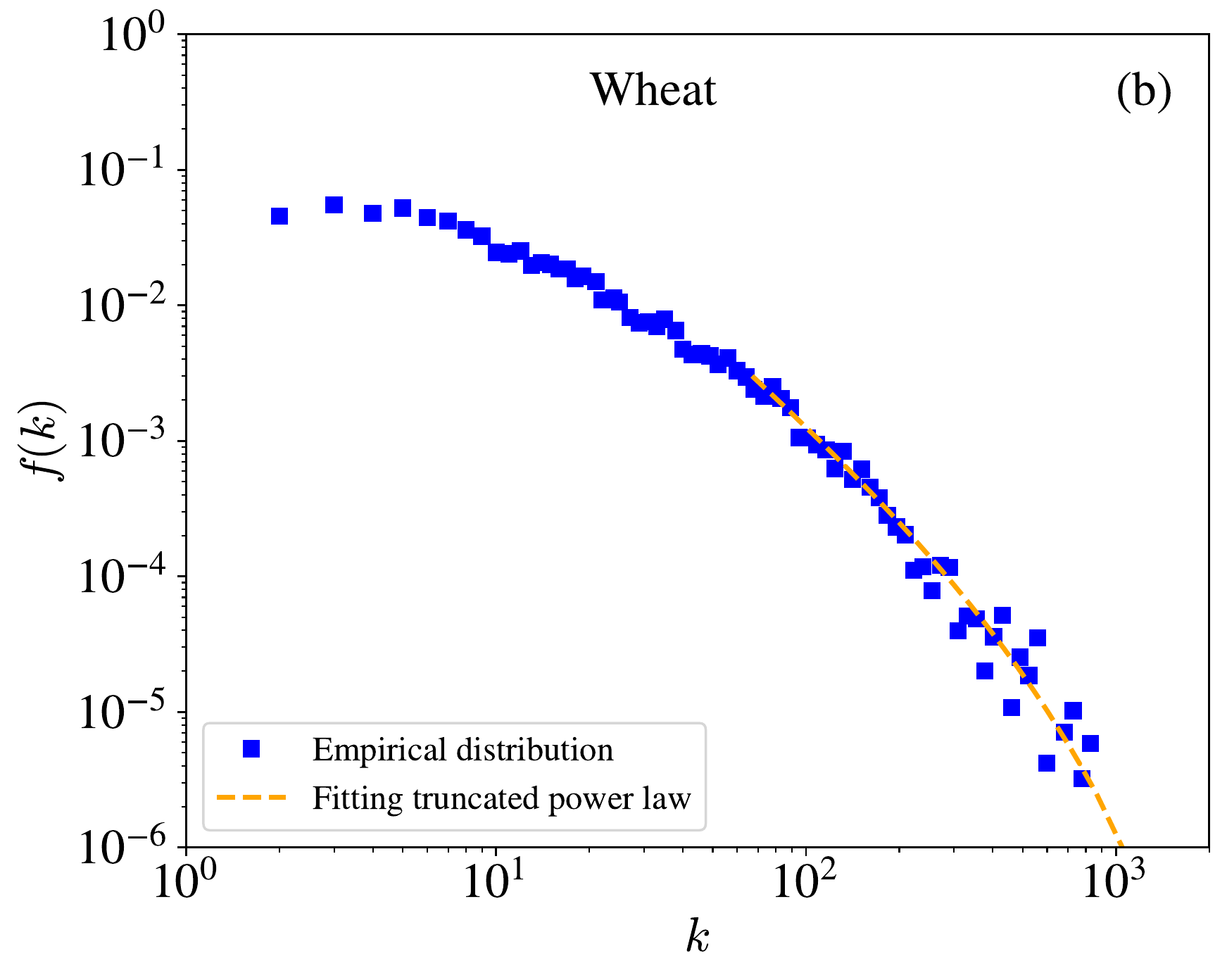}
    \includegraphics[width=0.325\linewidth]{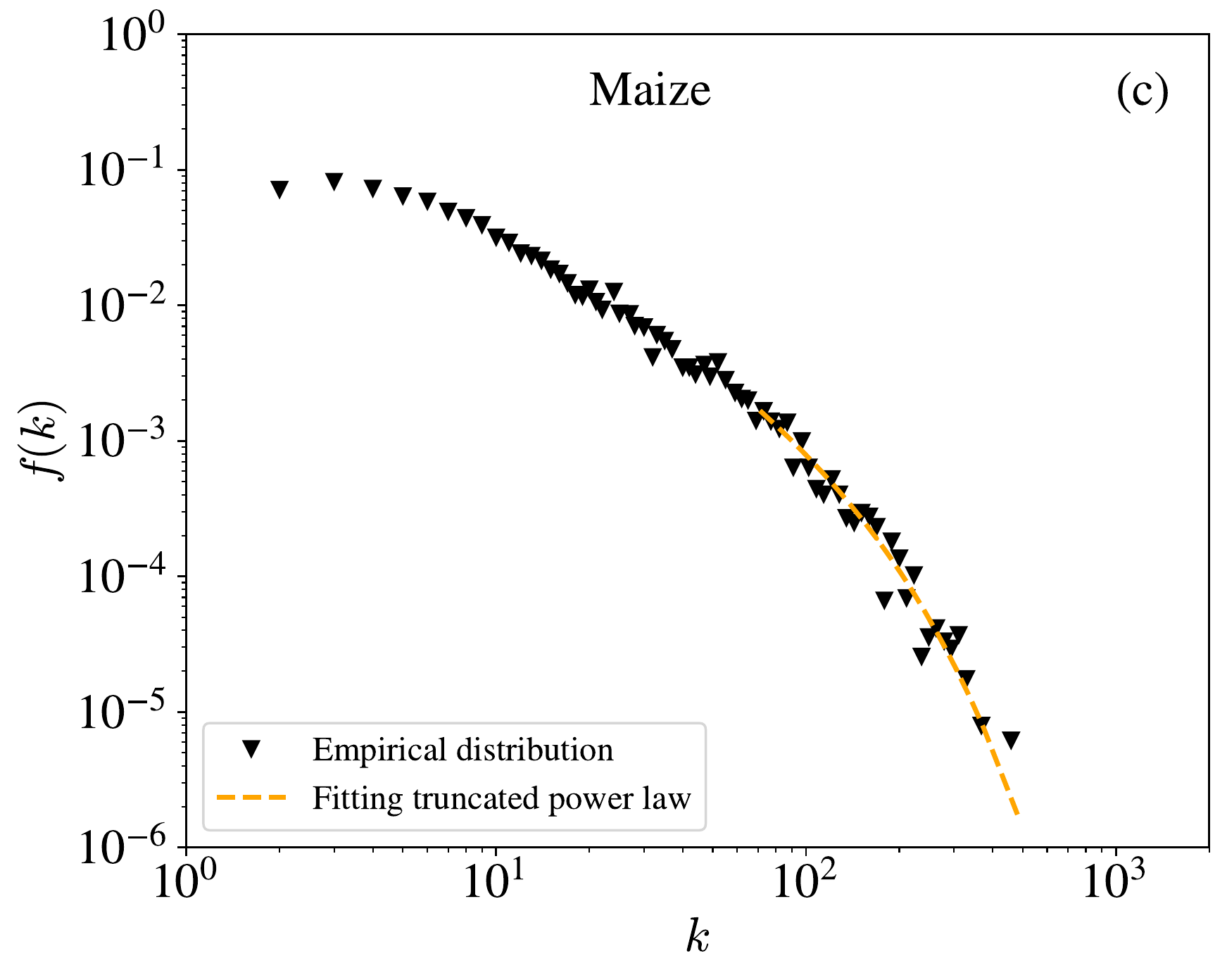}\\
    \includegraphics[width=0.325\linewidth]{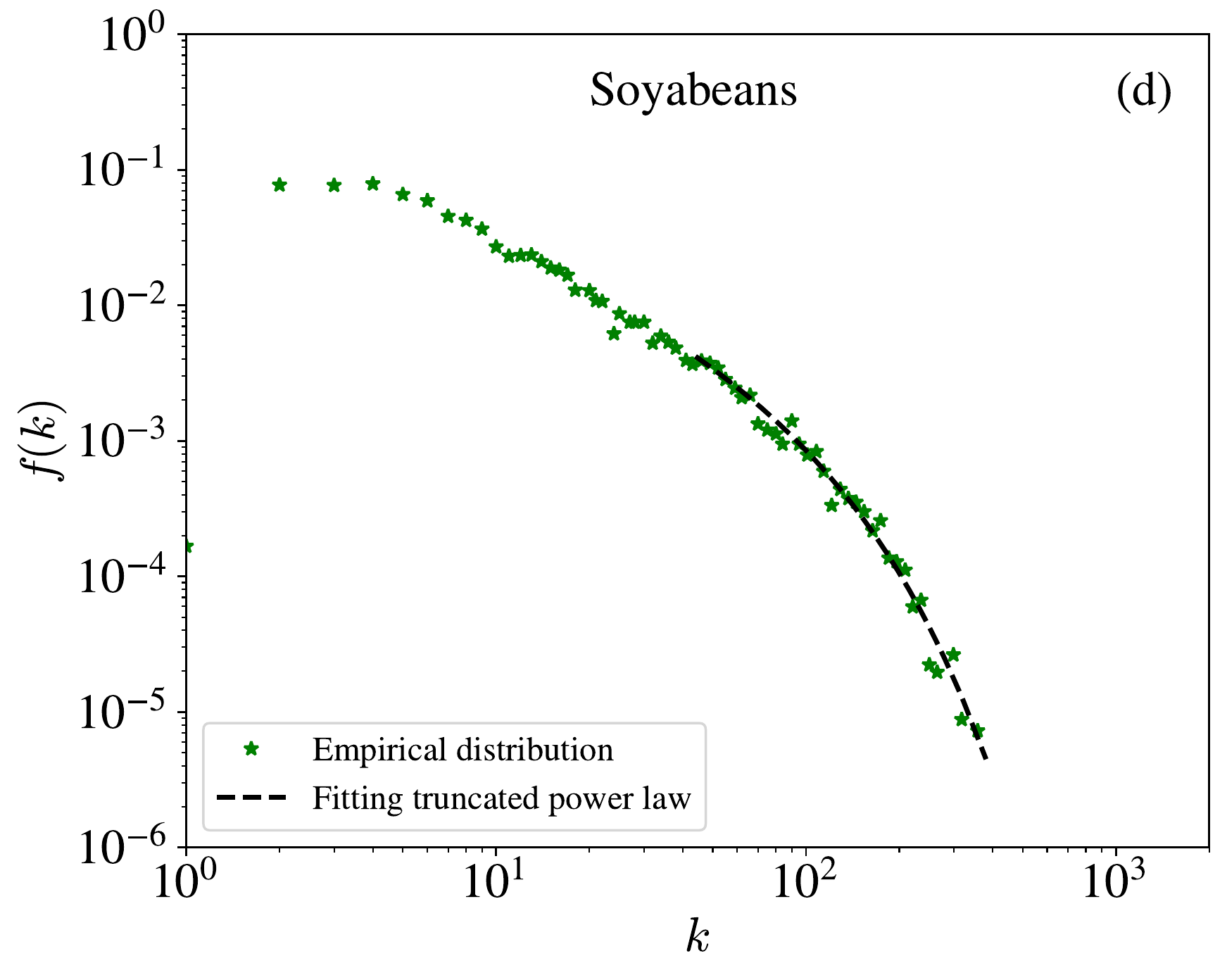}
    \includegraphics[width=0.325\linewidth]{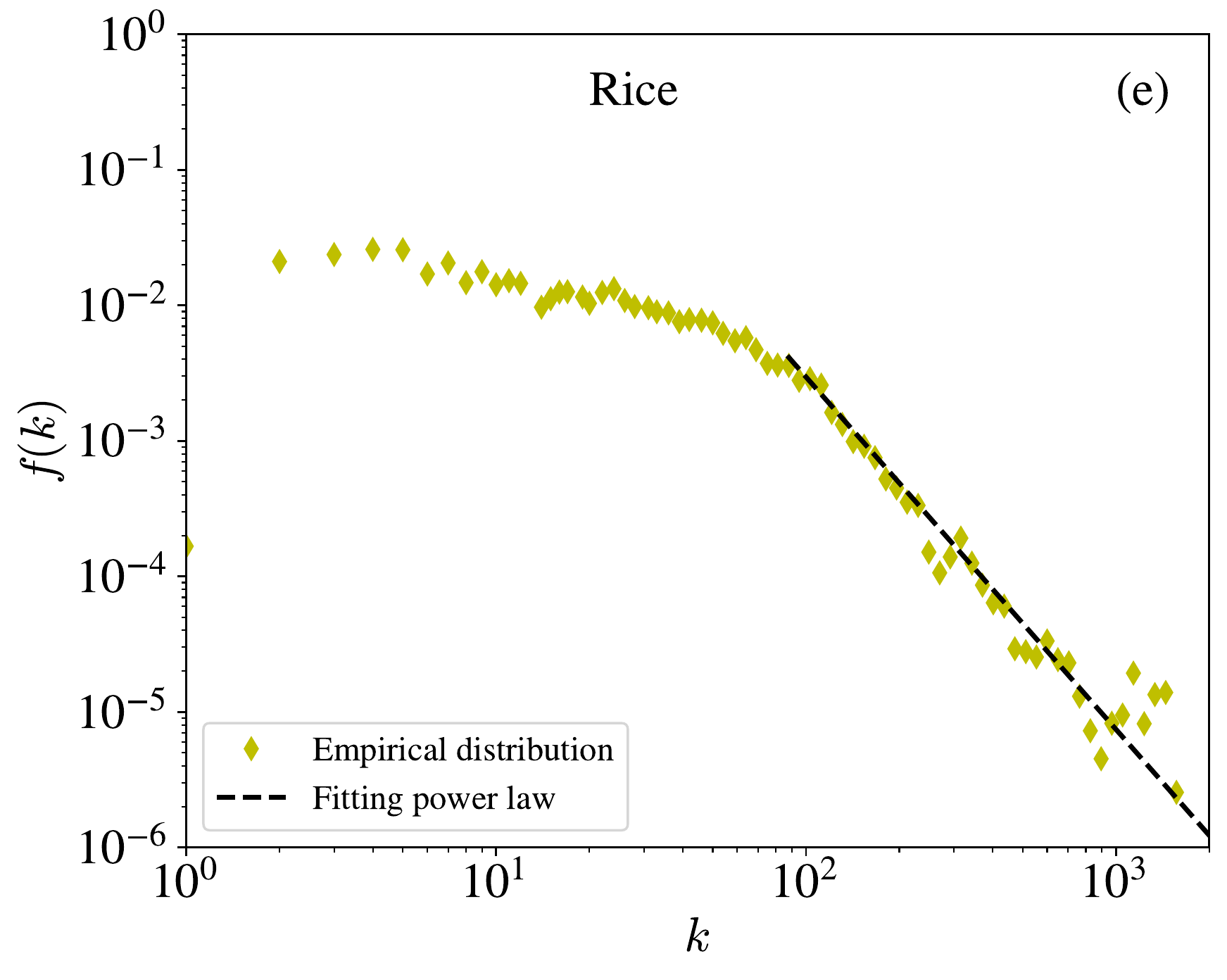}
    \includegraphics[width=0.325\linewidth]{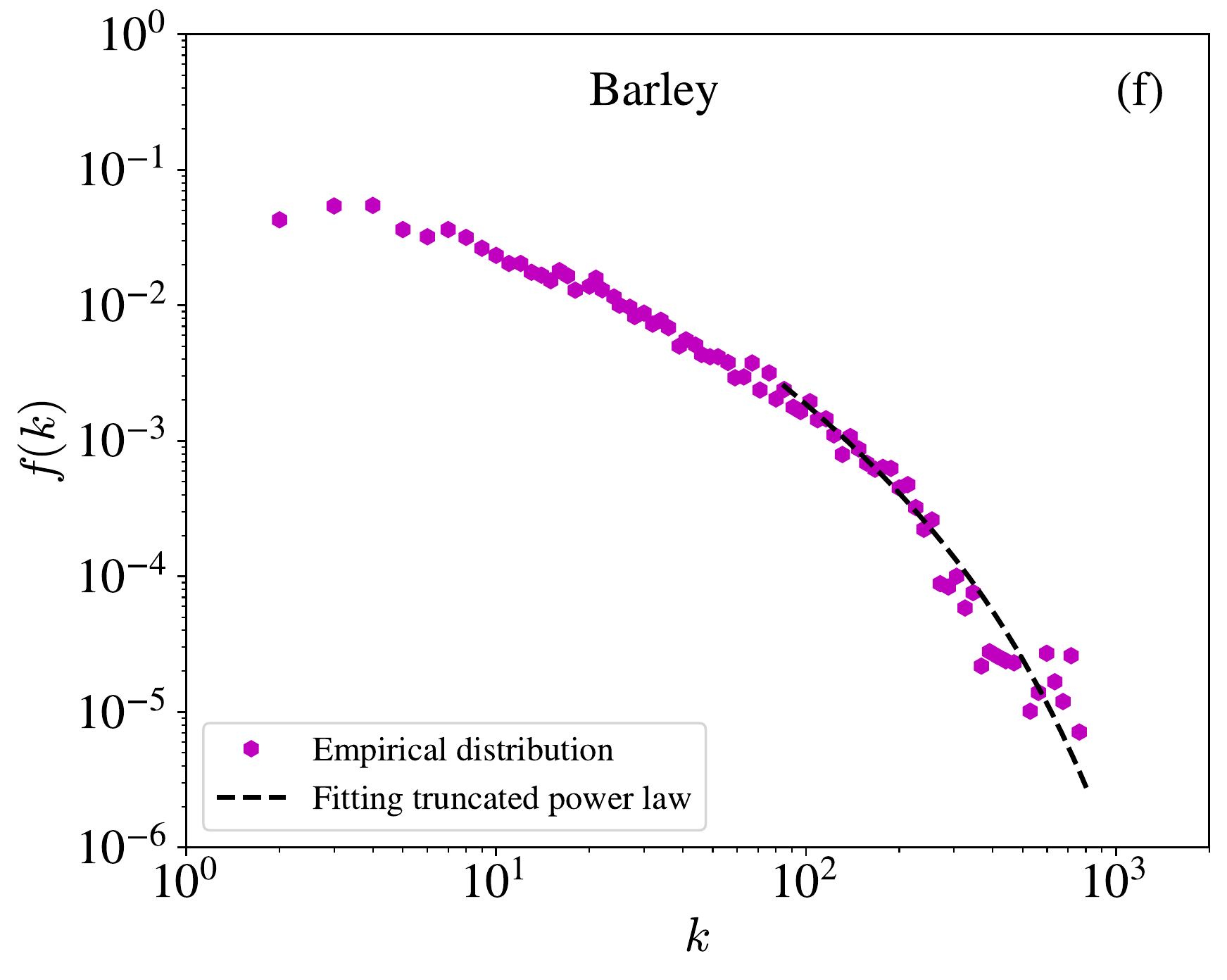}
    \caption{Empirical distributions $f(k)$ of node degree $k$ for six VGs and the fitting distributon.
     (a) IGC GOI VG. (b) Wheat VG. (c) Maize VG. (d) Soyabeans VG. (e) Rice VG. (f) Barley VG.}
    \label{Fig:GOIVG:DegreeDistributions:fk:k}
\end{figure}

The empirical degree distribution of six VGs obtained using logarithmic binning method is illustrated in log-log scales in Fig.~\ref{Fig:GOIVG:DegreeDistributions:fk:k}. Intuitively, we find that the degree distribution of rice VG is close to a power-law distribution \cite{Clauset-Shalizi-Newman-2009-SIAMRev}
\begin{equation}
    f(k) \propto k^{-\alpha},
\end{equation}
where $\alpha$  a constant parameter of the distribution known as the tail exponent.
However, the other five VGs is close to an exponentially truncated power-law distribution
\begin{equation}
    f(k) \propto k^{-\alpha}e^{-\lambda k},
    \label{Eq:GOIVG:Powerlaw:fk:k}
\end{equation}
where $\alpha$ here is the shape parameter controlling the slope of the power-law part, determining the fatness of the tail. $\lambda$ is the scale parameter from the exponential part, influencing how quickly the distribution decays. We use $k_{\min}$ as the lower bound to the distribution. Then, we use the method of maximum likelihood to estimate the parameter $\alpha$ and $\lambda)$ \cite{Clauset-Shalizi-Newman-2009-SIAMRev,Alstott-Bullmore-Plenz-2014-PLoSOne}. 

\begin{table}[!ht]
    \centering
    \caption{The fitting results of power-law (Rice VG) and exponentially truncated power-law (other 5 VGs). The shape parameter $\alpha$, scale parameter $\lambda$, lower bound $k_{\min}$ and $p$-value for KS test of six VGs.}
    \smallskip
    \setlength{\tabcolsep}{11.3mm}
    \begin{tabular}{cccccc}
    \toprule
    Series & $\alpha$ & $\lambda$ & $k_{\min}$& $p$\\ \midrule
    IGC GOI & 2.0310 & 0.0045 & 99 & 0.2373  \\
    Wheat & 1.9638 & 0.0026 & 67 & 0.1165\\
    Maize & 1.3027 & 0.0106 & 71 & 0.7489  \\
    Soyabeans & 1.0001 & 0.0139 & 44 & 0.1134  \\
    Rice & 2.5994 & - & 87 & 0.0761  \\
    Barley & 1.4707 & 0.0049 & 84 & 0.1523  \\ \bottomrule
    \end{tabular}
    \label{Table:GOIVG:PowerLaw:FittingResult}
\end{table}

In this way, we can calculate power-law and exponentially truncated power-law distribution functions obtained from the fit and compare it with the empirical distribution function, which is illustrated in log-log scale Fig.~\ref{Fig:GOIVG:DegreeDistributions:fk:k}. We also use the Kolmogorov-Smirnov (KS) test to check the significance of all the fits. The $p$-values for each fit are shown in Table~\ref{Table:GOIVG:PowerLaw:FittingResult}. It can be seen that the $p$-values for the six VGs are all greater than 0.05. It means that the empirical distribution function fits well. 

The degree distribution of rise VG follows a power-law distribution with $\alpha \approx 2.6$. We find that $\alpha = 4 - 2H$, which means that the rise VGs is  between a fractal Brwonian series ($\alpha = 3-2H$) and a Gaussian series ($\alpha = 5-2H)$. The other five VGs have an exponentially truncated power-law fit corresponding to the small Hurst exponent.

\subsection{Clustering coefficient}
\label{S2:Clusteringcoefficient}

The clustering coefficient $c_i$ for node $i$ of a VG is defined as
\begin{equation}
    c_i = \frac{2T_i}{k_i(k_i-1)},
\end{equation}
where $k_i$ is the degree of node $i$ and $T_i$ is the number of triangles through node $i$:
\begin{equation}
    T_i = \sharp \left( \{(i,j,u)~|~ \{(i,j), (j,u),(i,u)\} \subset \mathscr{E}  \}\right).
\end{equation}
For a VG, the average clustering coefficient $C$ is defined as the average of all node clustering coefficients
\begin{equation}
    C = \frac{1}{N_{\mathscr{V}}} \sum_{i \in \mathscr{V}} c_i.
\end{equation}
For the classical random graph model $G(N_\mathscr{V},p)$, each edge is independently in place with probability $p$ \cite{Fagiolo-2007-PhysRevE}
\begin{equation}
    p = \frac{N_{\mathscr{E}}}{N_{\mathscr{V}}(N_{\mathscr{V}}-1)}=\rho,
\end{equation}
which indicates that the expectation of the average clustering coefficient $C$ in random networks $G(N_{\mathscr{V}},\rho)$ is its density $\rho$ \cite{Fagiolo-2007-PhysRevE,Fagiolo-Reyes-Schiavo-2010-JEvolEcon}. 

For comparison, the average clustering coefficient $C$ and density $\rho$ for six VGs are illustrated in Table~\ref{Table:GOIVG:ClusteringCoefficient}. As shown in Table~\ref{Table:GOIVG:ClusteringCoefficient}, the average clustering coefficient of all the six VGs is very large
($>0.5$), and it is significantly greater than that of random graphs. The results indicate that the six VGs are statistically clustered and tend to form ternary cliques.

\begin{table}[!ht]
    \centering
    \caption{The average clustering coefficient $C$ and density $\rho$ of six VGs.}
    \smallskip
    \setlength{\tabcolsep}{6.7mm}
    \begin{tabular}{ccccccc}
    \toprule
      & IGC GOI & Wheat  & Maize  & Soyabeans & Rice& Barley \\ \midrule
    $C$ & 0.6293  & 0.6037 & 0.6657 & 0.6595 & 0.5230 & 0.5738  \\
    $\rho$ & 0.0054  & 0.0064 & 0.0039 & 0.0039 & 0.0125 & 0.0079 \\\bottomrule
    \end{tabular}
    \label{Table:GOIVG:ClusteringCoefficient}
\end{table}

\begin{figure}[!ht]
    \centering
    \includegraphics[width=0.325\linewidth]{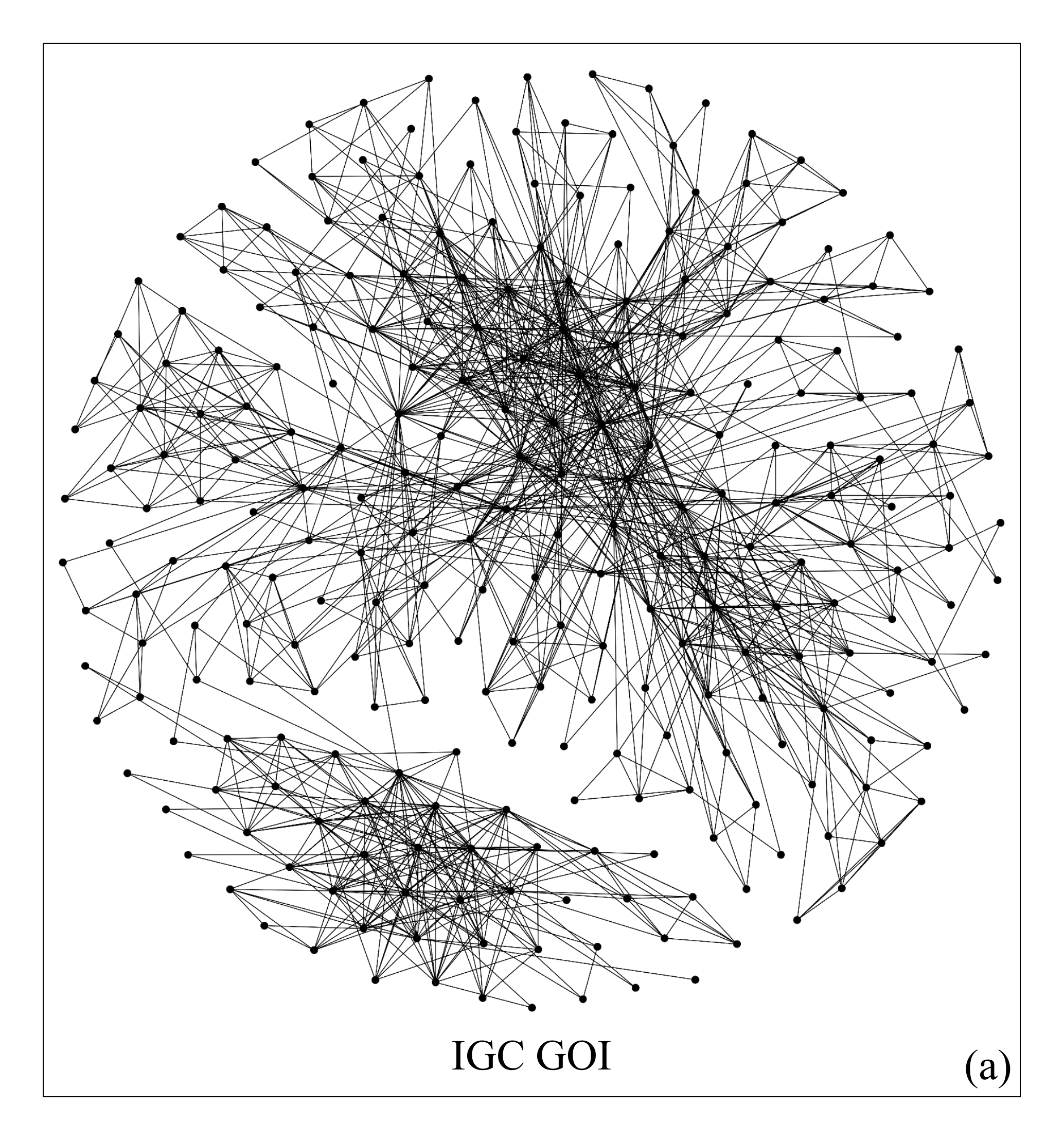}
    \includegraphics[width=0.325\linewidth]{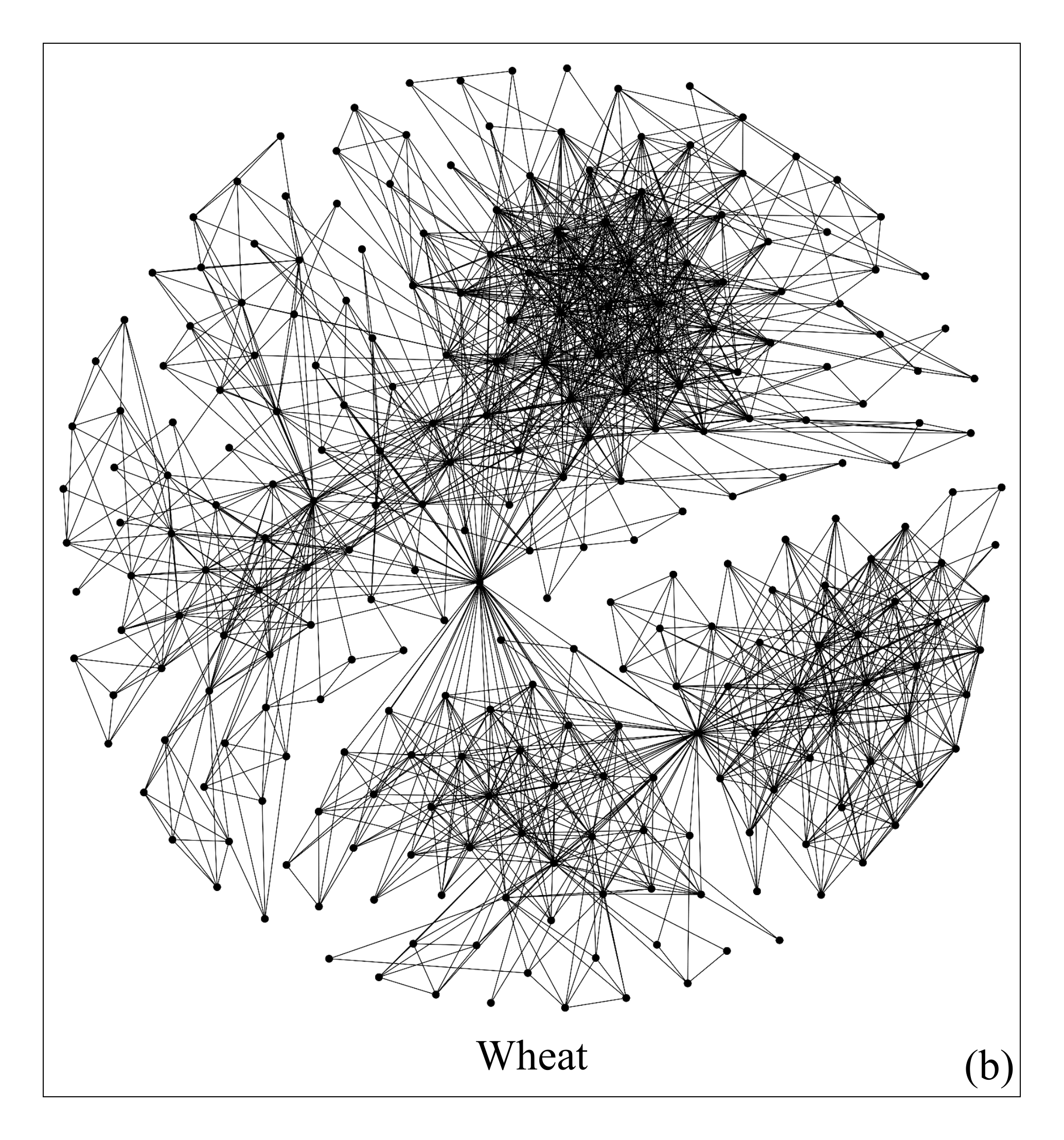}
    \includegraphics[width=0.325\linewidth]{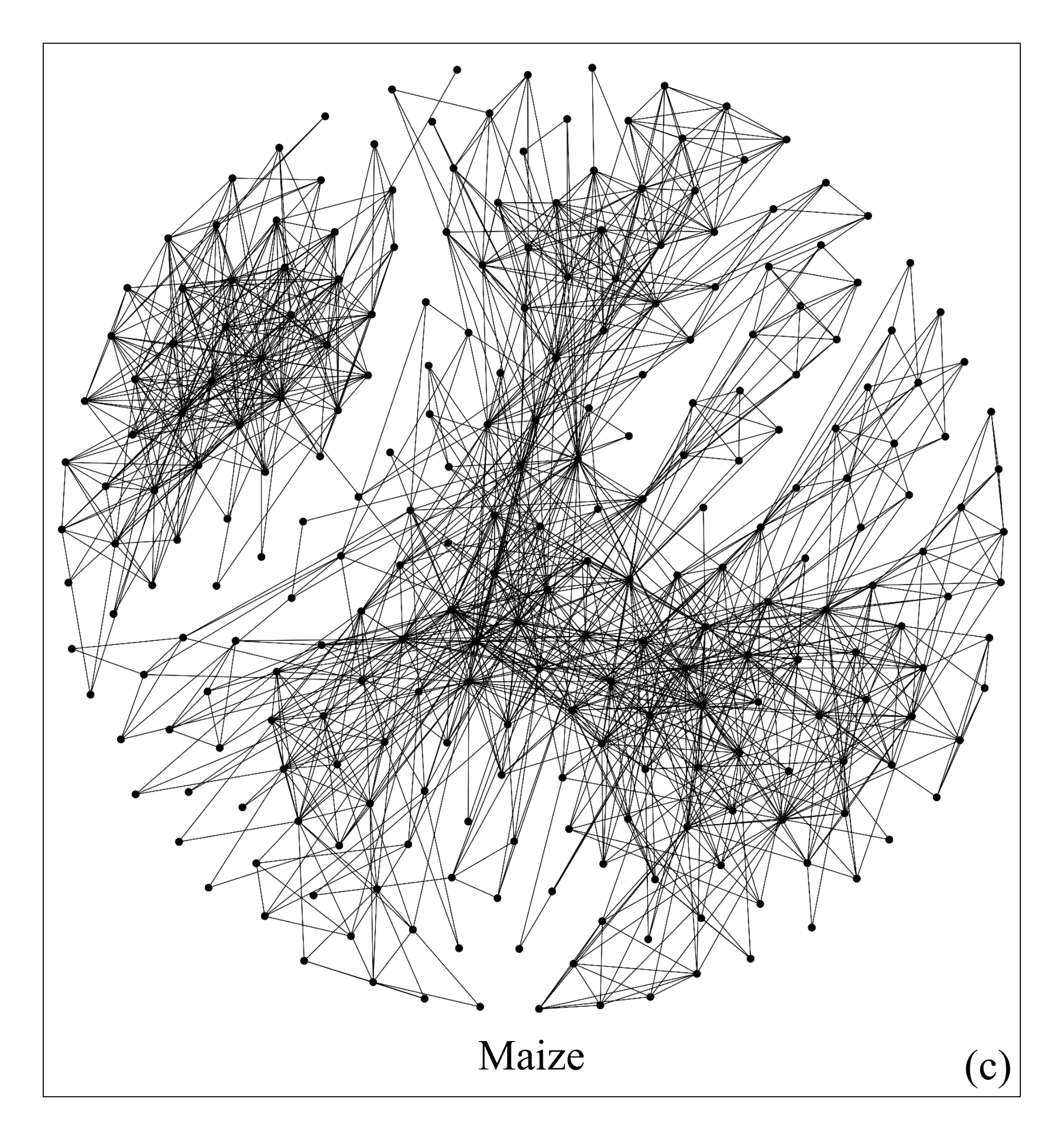}\\
    \includegraphics[width=0.325\linewidth]{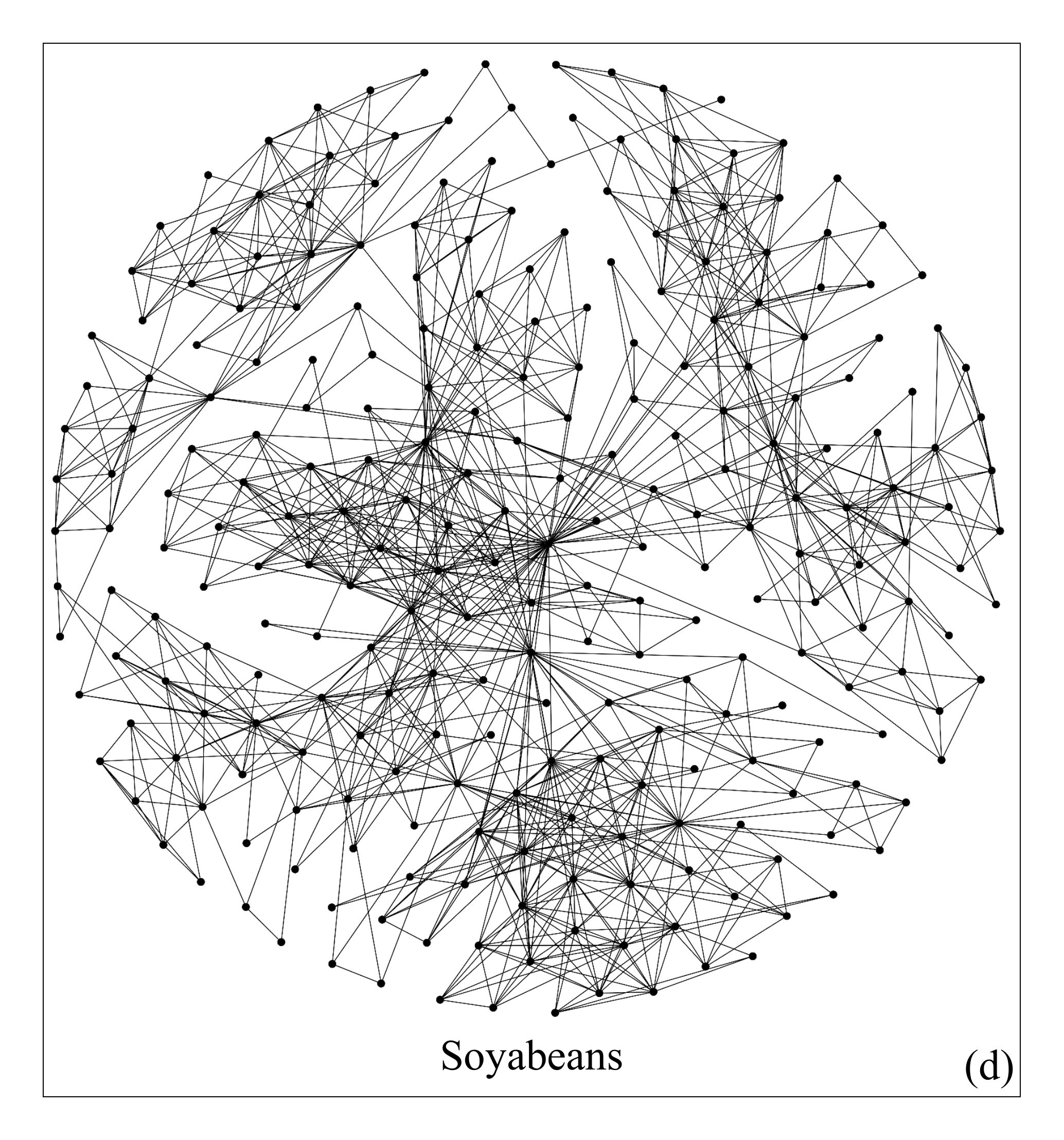}
    \includegraphics[width=0.325\linewidth]{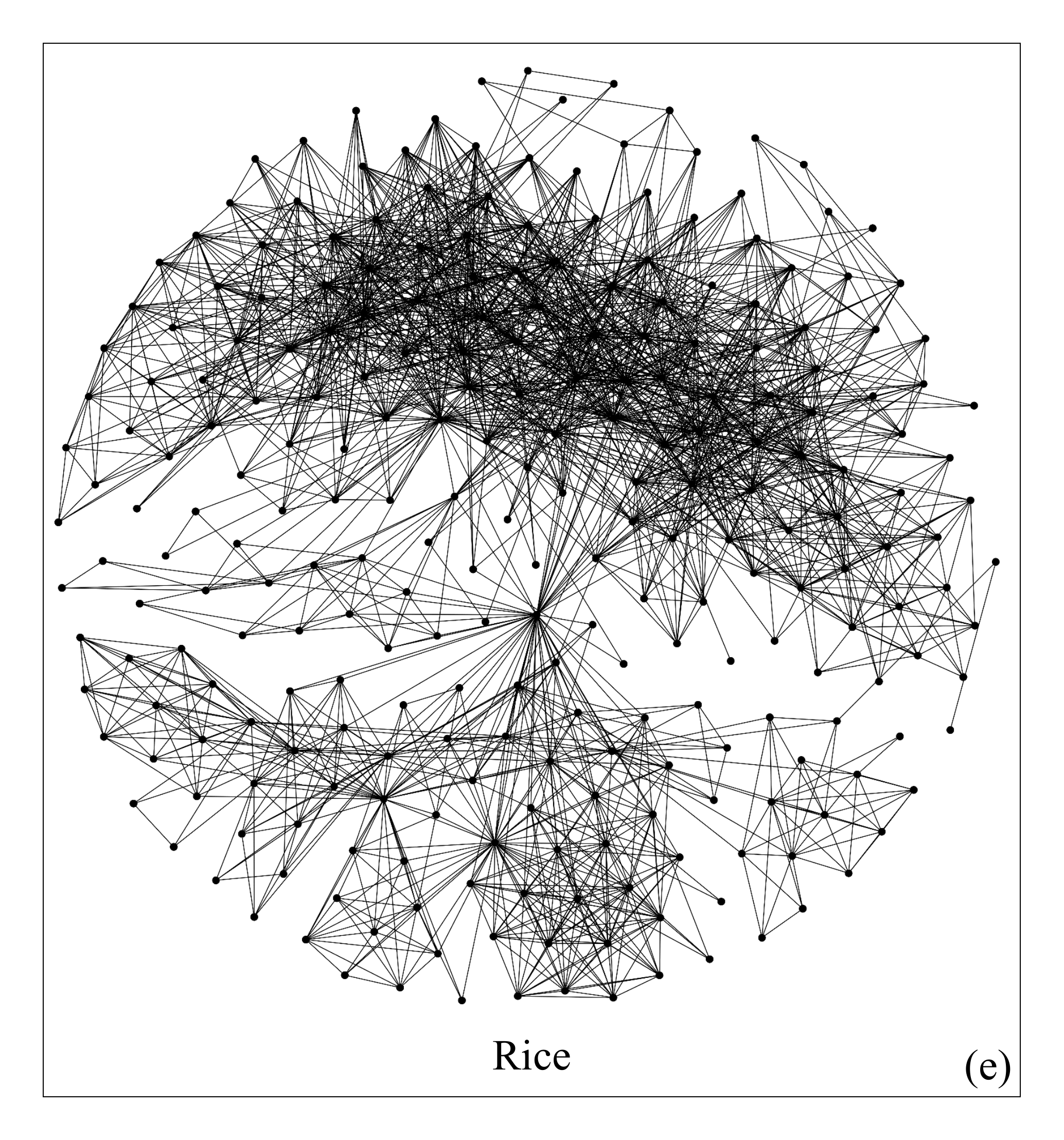}
    \includegraphics[width=0.325\linewidth]{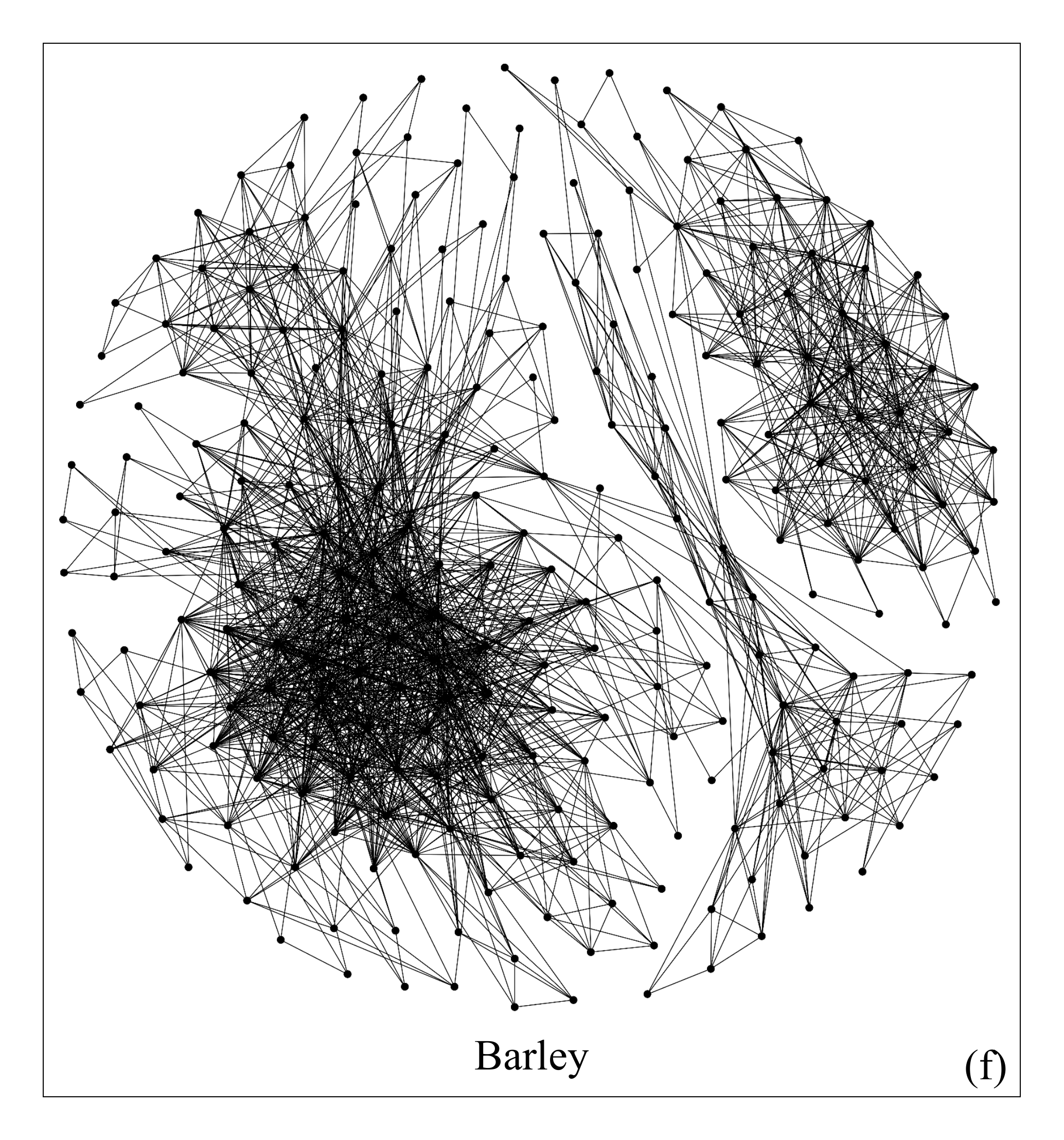}
    \caption{The sub-graphs of six VGs constructed by the GOI Index time series of 2022. (a) IGC GOI VG. (b) Wheat VG. (c) Maize VG. (d) Soyabeans VG. (e) Rice VG. (f) Barley VG.}
    \label{Fig:GOIVG:VisibilityGraph2022}
\end{figure}

Fig.~\ref{Fig:GOIVG:VisibilityGraph2022} visually explains this property. In Fig.~\ref{Fig:GOIVG:VisibilityGraph2022}, we illustrate the sub-graphs of six VGs constructed by the GOI Index time series of 2022. We observed the presence of a large number of triangular structures in the sub-VG, which means the value of the GOI Index between two points is often less than the value of the linear function through the GOI Index of those two points.

\begin{figure}[!ht]
    \centering
    \includegraphics[width=0.5\linewidth]{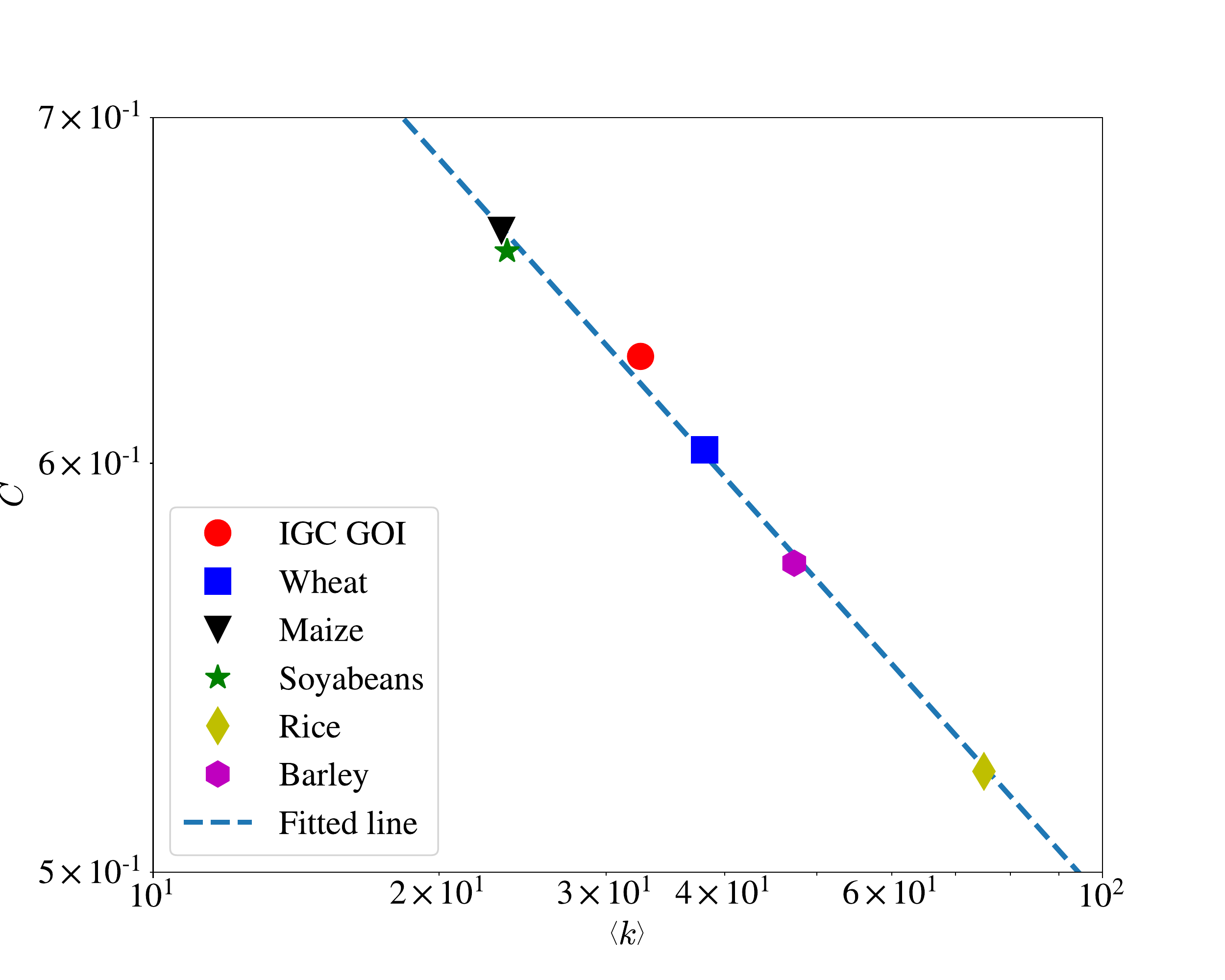}
    \caption{The relationship between average clustering coefficient $C$ and average degree $\left\langle k \right\rangle$ of six VGs in log-log scale.}
    \label{Fig:GOIVG:ClusteringCoefficient:C:avek}
\end{figure}

To illustrate the relationship between the average clustering coefficient $C$ and average degree $\left\langle k\right\rangle$, we plot $C$ and $\left\langle k\right\rangle$ in log-log scale Fig.~\ref{Fig:GOIVG:ClusteringCoefficient:C:avek} and observe a linear correlation in log-log scale, which means
\begin{equation}
    \ln C \sim \ln \left \langle k \right\rangle.
\end{equation}
We calculated the Pearson correlation coefficient between $C$ and $k$, which is -0.9973. Then, we fitted $\ln C$ and $\ln \left \langle k \right\rangle$ using the OLS method and obtained the result as
\begin{equation}
    \ln C = 0.2400 - 0.2052 \ln \left \langle k \right\rangle,
    \label{Eq:GOIVG:lnC:lnavek}
\end{equation}
where the $p$-value of the fitting is smaller than 0.05.

\begin{figure}[!ht]
    \centering
    \includegraphics[width=0.325\linewidth]{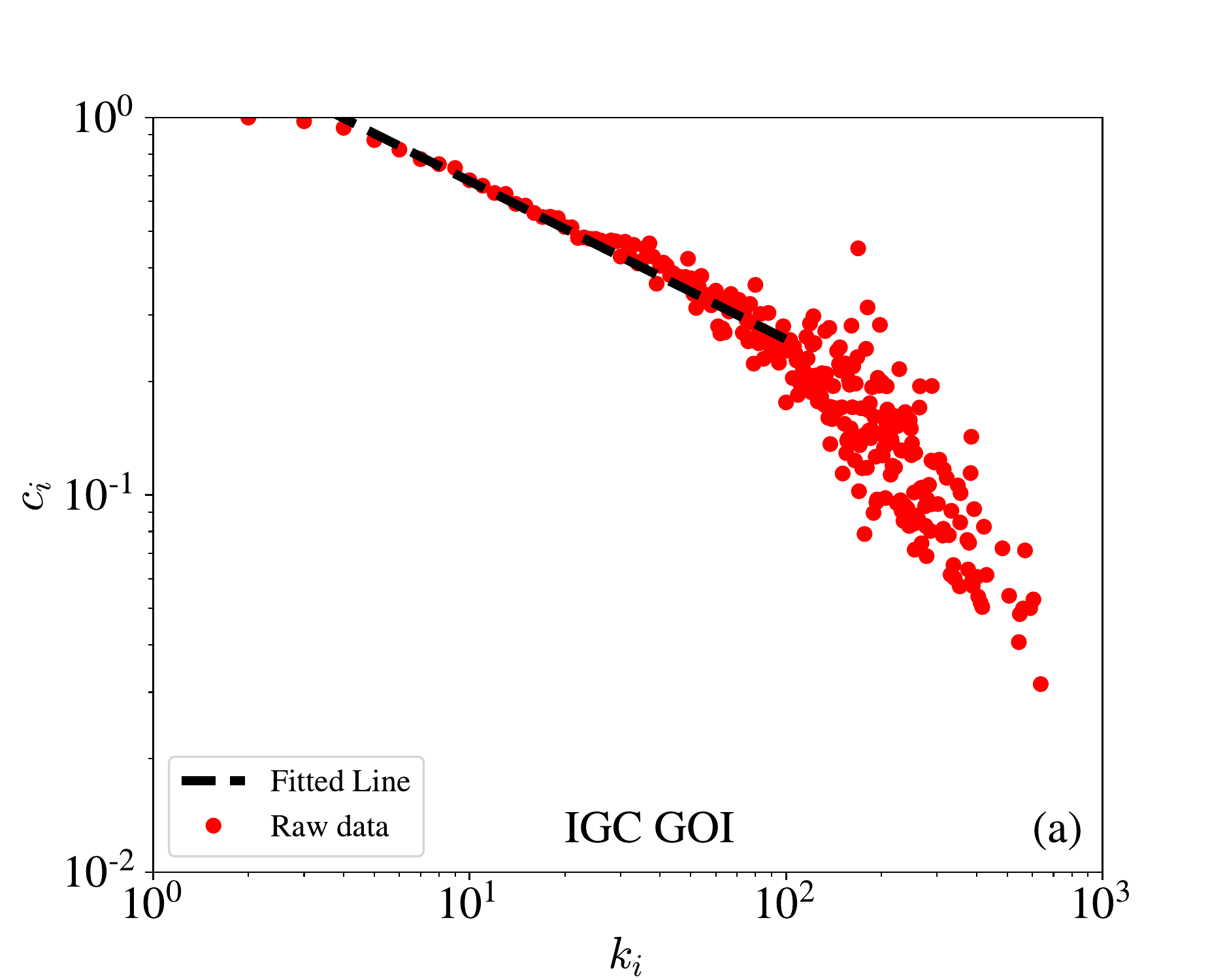}
    \includegraphics[width=0.325\linewidth]{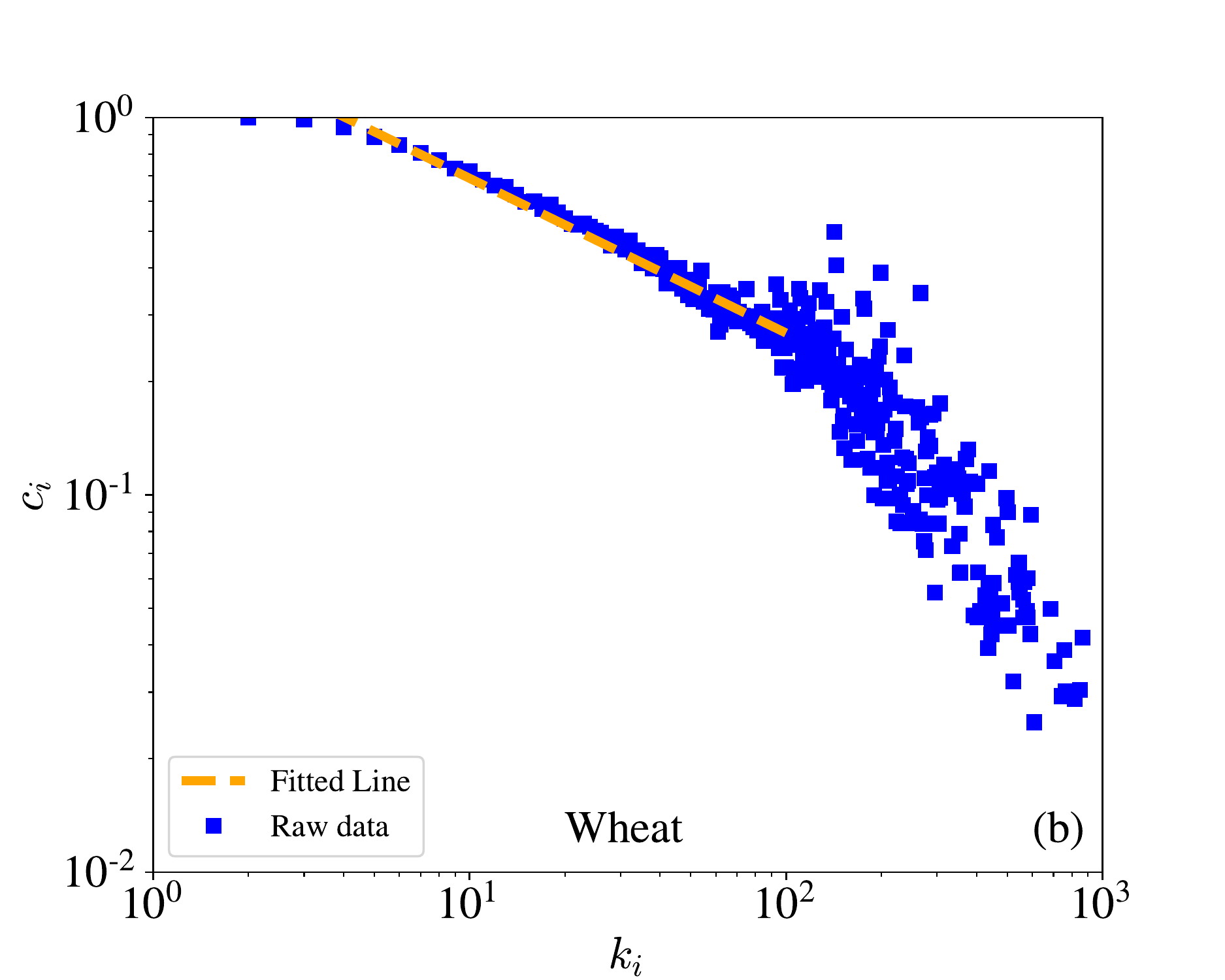}
    \includegraphics[width=0.325\linewidth]{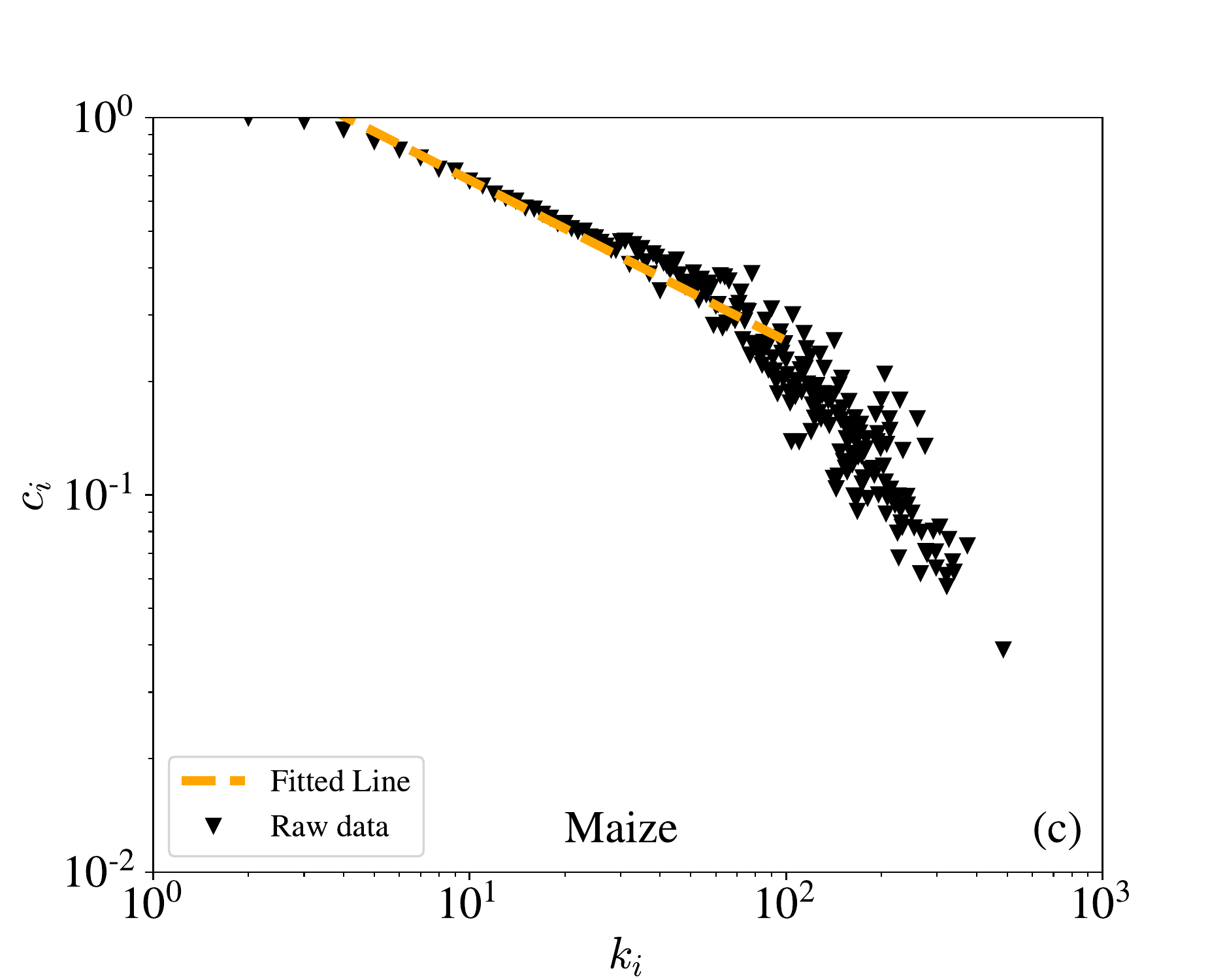}\\
    \includegraphics[width=0.325\linewidth]{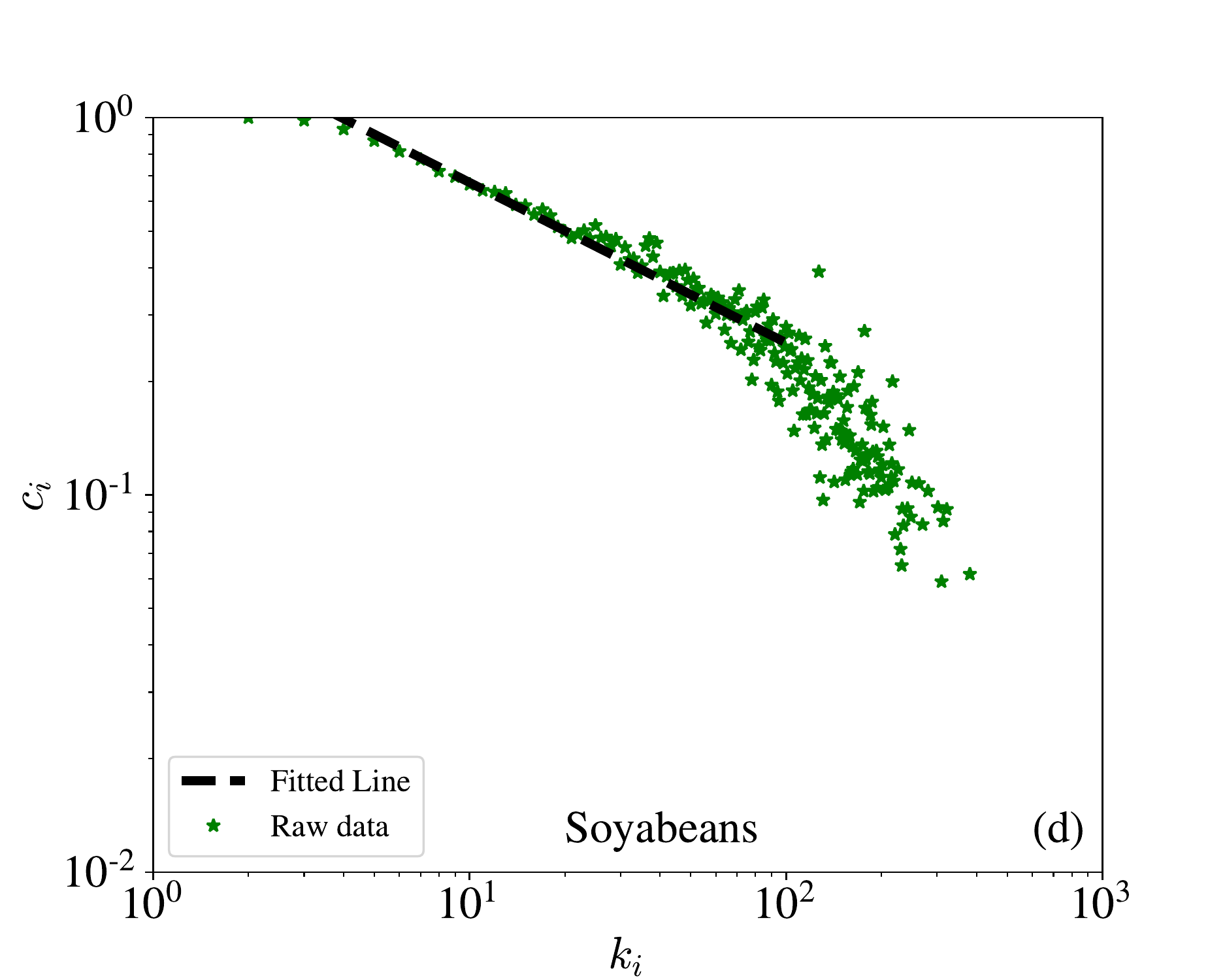}
    \includegraphics[width=0.325\linewidth]{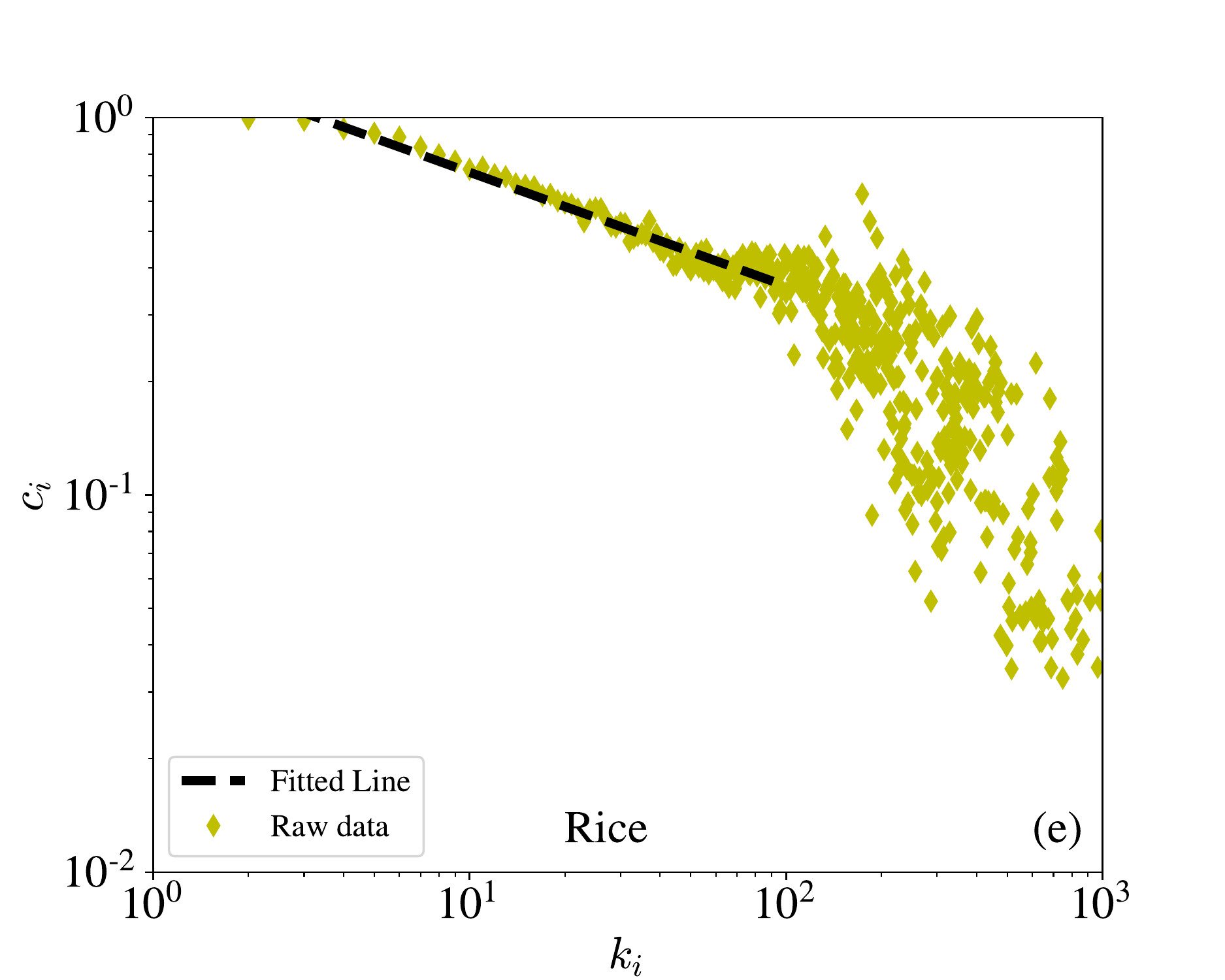}
    \includegraphics[width=0.325\linewidth]{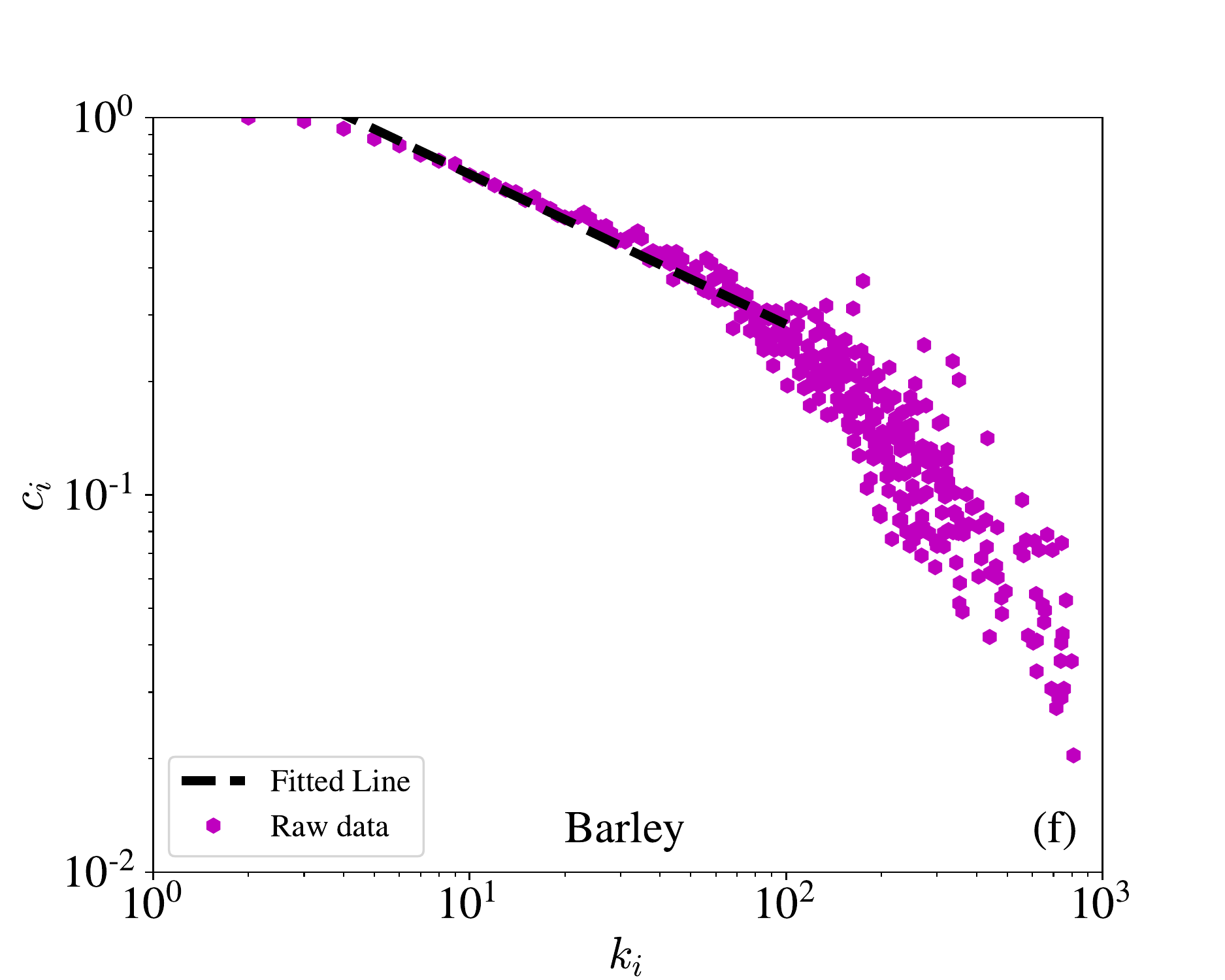}
    \caption{The relationship between node clustering coefficient $c_i$ and node degree $k_i$ of six VGs in log-log scale. (a) IGC GOI VG. (b) Wheat VG. (c) Maize VG. (d) Soyabeans VG. (e) Rice VG. (f) Barley VG.}
    \label{Fig:GOIVG:ClusteringCoefficient:ci:ki}
\end{figure}

In order to explore the relationship between local clustering coefficients $c_i$ and node degree $k_i$, we plot them on the log-log scale Fig.~\ref{Fig:GOIVG:ClusteringCoefficient:ci:ki}. As shown in Fig.~\ref{Fig:GOIVG:ClusteringCoefficient:ci:ki}, the correlation between the clustering coefficient $c_i$ and node degree $k_i$ is negative, which indicates that the nodes with lower degree are more likely to form local denser clusters.

Moreover, we observed a linear correlation in Fig.~\ref{Fig:GOIVG:ClusteringCoefficient:ci:ki} when node degree $k_i$ is not very large ($k_i \leqslant 100$), which means
\begin{equation}
    \ln c_i =\beta_0 + \beta_1 \ln k_i.
    \label{Eq:GOIVG:lnci:lnki}
\end{equation}
We calculated the Pearson correlation coefficient $r$ between $\ln c_i$ and $\ln k_i$ for six VGs and fitted Eq.~(\ref{Eq:GOIVG:lnci:lnki}) using the OLS method. The fitting results are illustrated in Table~\ref{Table:GOIVG:ClusteringCoefficient:FittingResult}.

\begin{table}[!ht]
    \centering
    \caption{The fitting results of coefficient $\beta_0,\beta_1$, adjust R-square $\overline{R^2}$, Pearson correlation coefficient $r$ of $\ln c_i$ and $\ln k_i$ and the significance of the fitting $p$ of six VGs.}
    \smallskip
    \setlength{\tabcolsep}{8.2mm}
    \begin{tabular}{cccccc}
    \toprule
    Series & $\beta_0$ & $\beta_1$ & $\overline{R^2}$ & $r$ & $p$   \\ \midrule
    IGC GOI& 0.2508 & -0.4192& 0.9305 & -0.9650 & 0.0000 \\
    Wheat  & 0.2492 & -0.4101& 0.9488 & -0.9743 & 0.0000 \\
    Maize  & 0.2606 & -0.4262& 0.9085 & -0.9536 & 0.0000\\
    Soyabeans & 0.2529 & -0.4249& 0.9123 & -0.9556 & 0.0000\\
    Rice& 0.1566 & -0.3015& 0.9285 & -0.9640 & 0.0000\\
    Barley & 0.2476 & -0.3978& 0.9348 & -0.9672 & 0.0000\\ \bottomrule
    \end{tabular}
    \label{Table:GOIVG:ClusteringCoefficient:FittingResult}
\end{table}

\begin{figure}[!ht]
    \centering
    \includegraphics[width=0.325\linewidth]{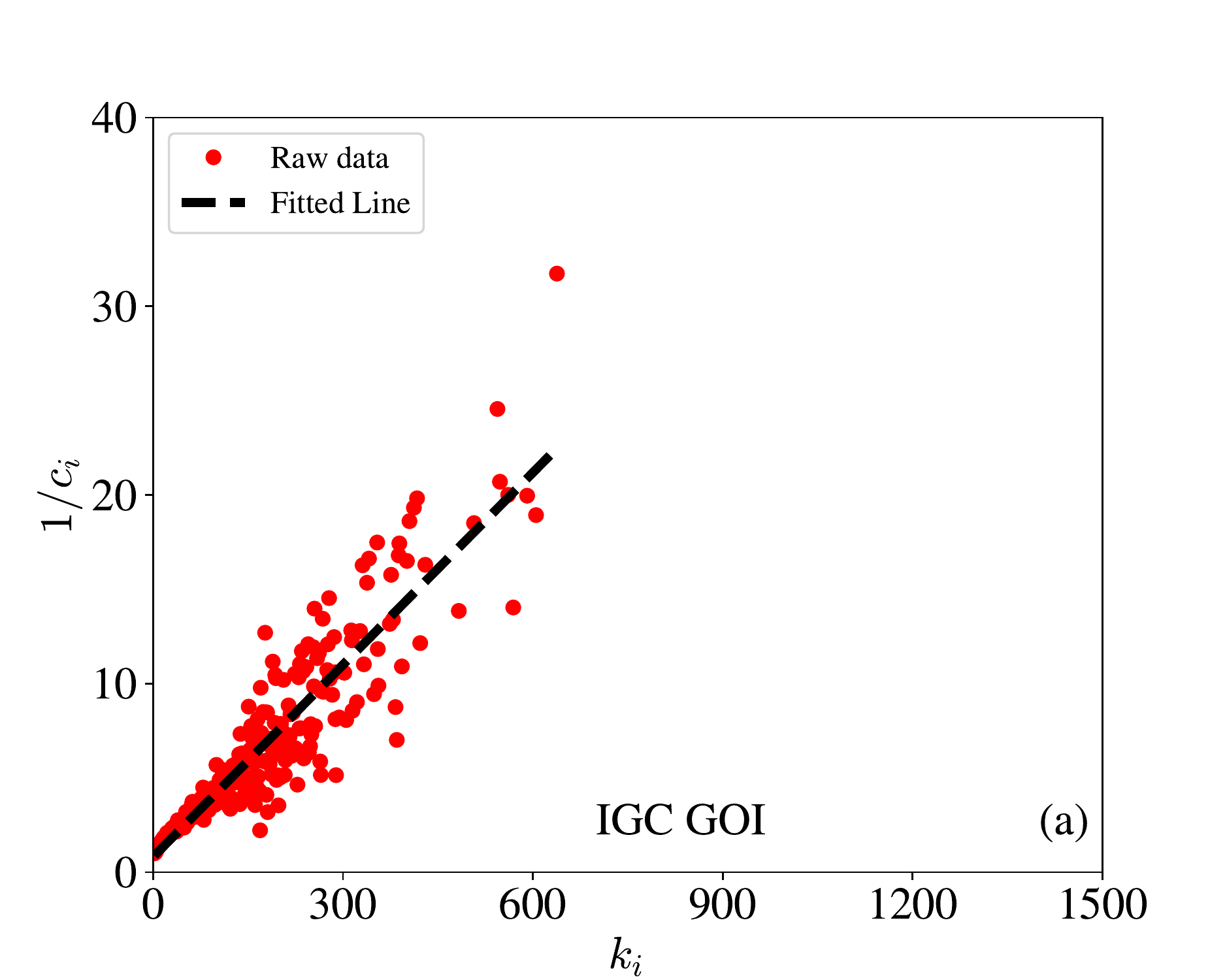}
    \includegraphics[width=0.325\linewidth]{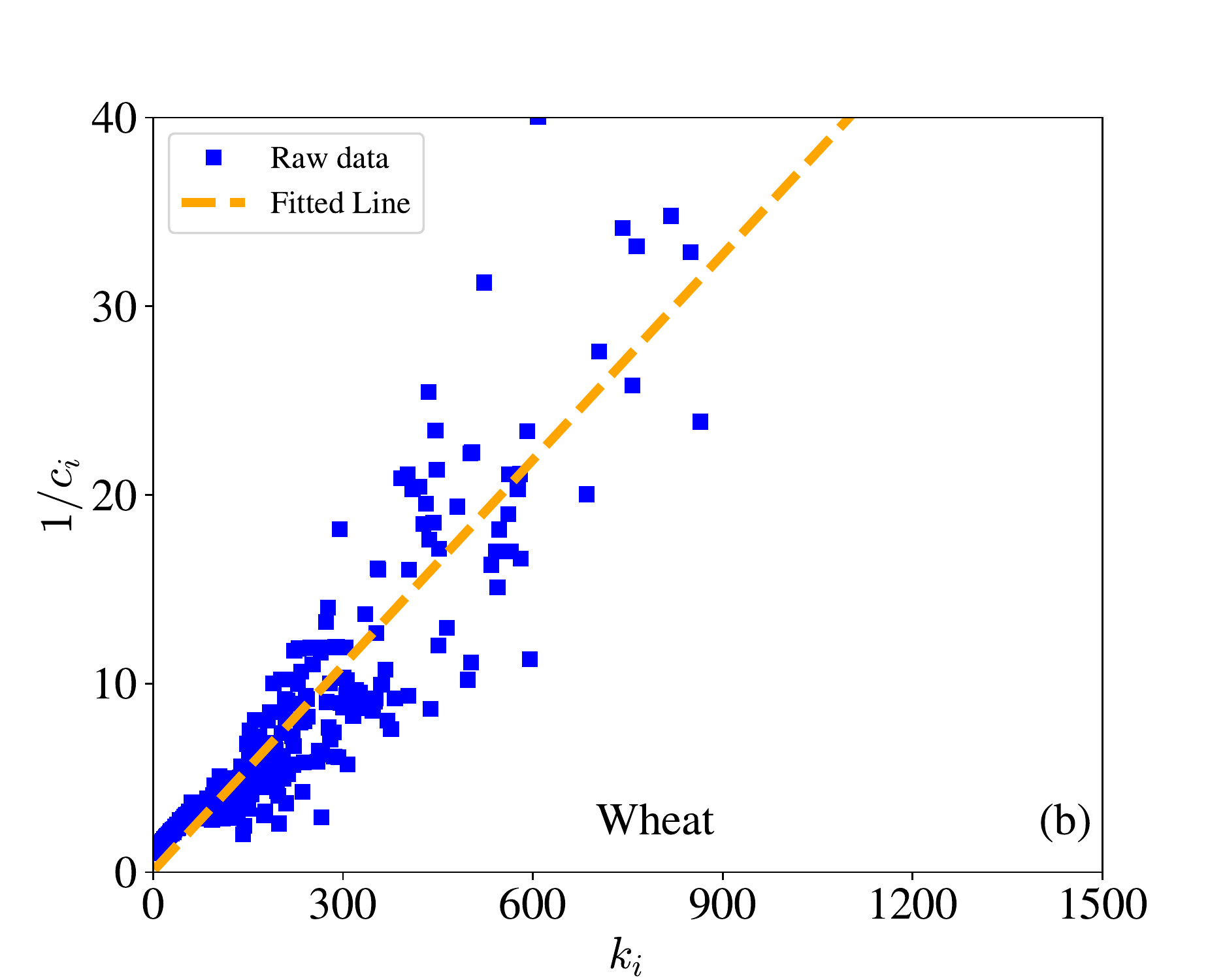}
    \includegraphics[width=0.325\linewidth]{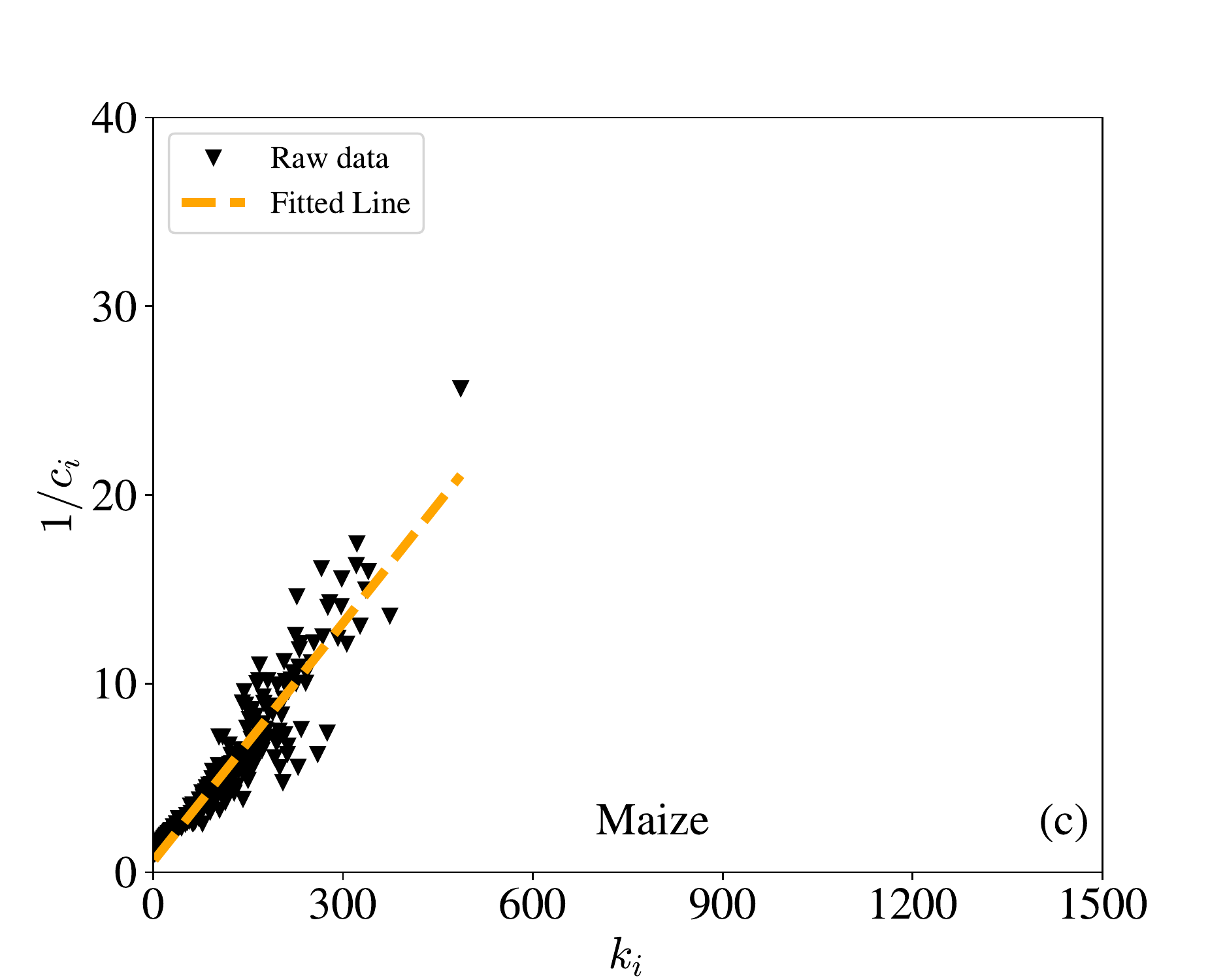}\\
    \includegraphics[width=0.325\linewidth]{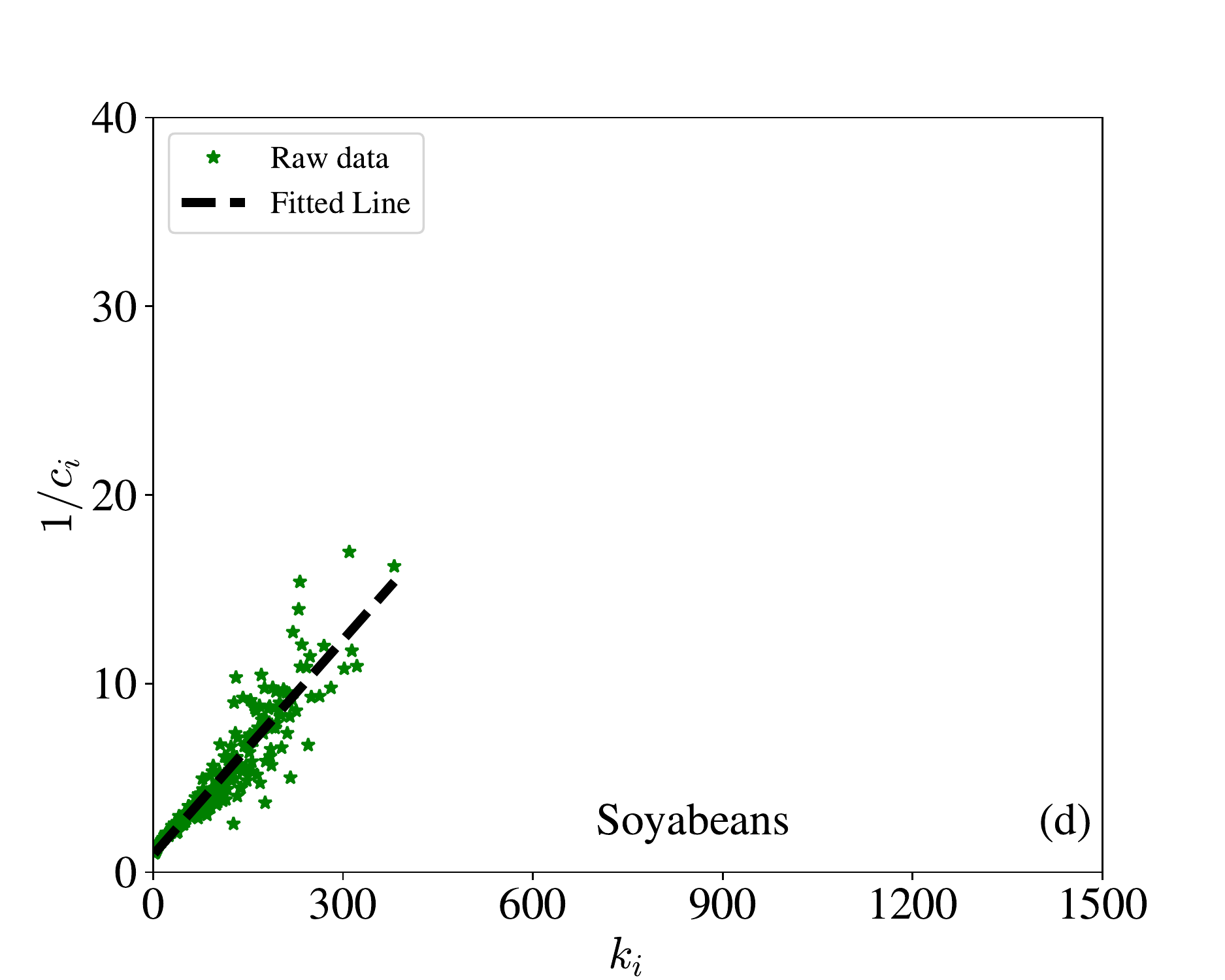}
    \includegraphics[width=0.325\linewidth]{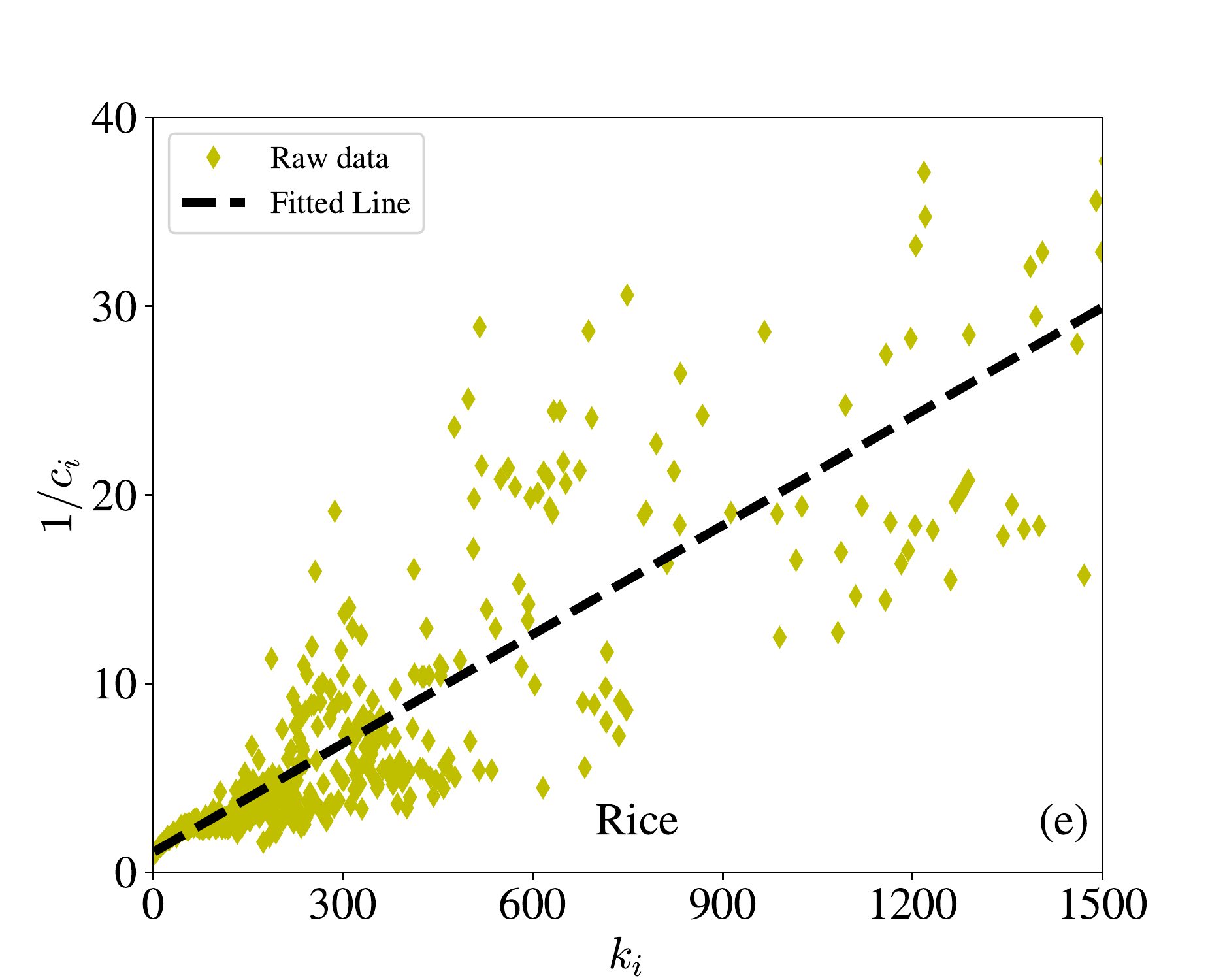}
    \includegraphics[width=0.325\linewidth]{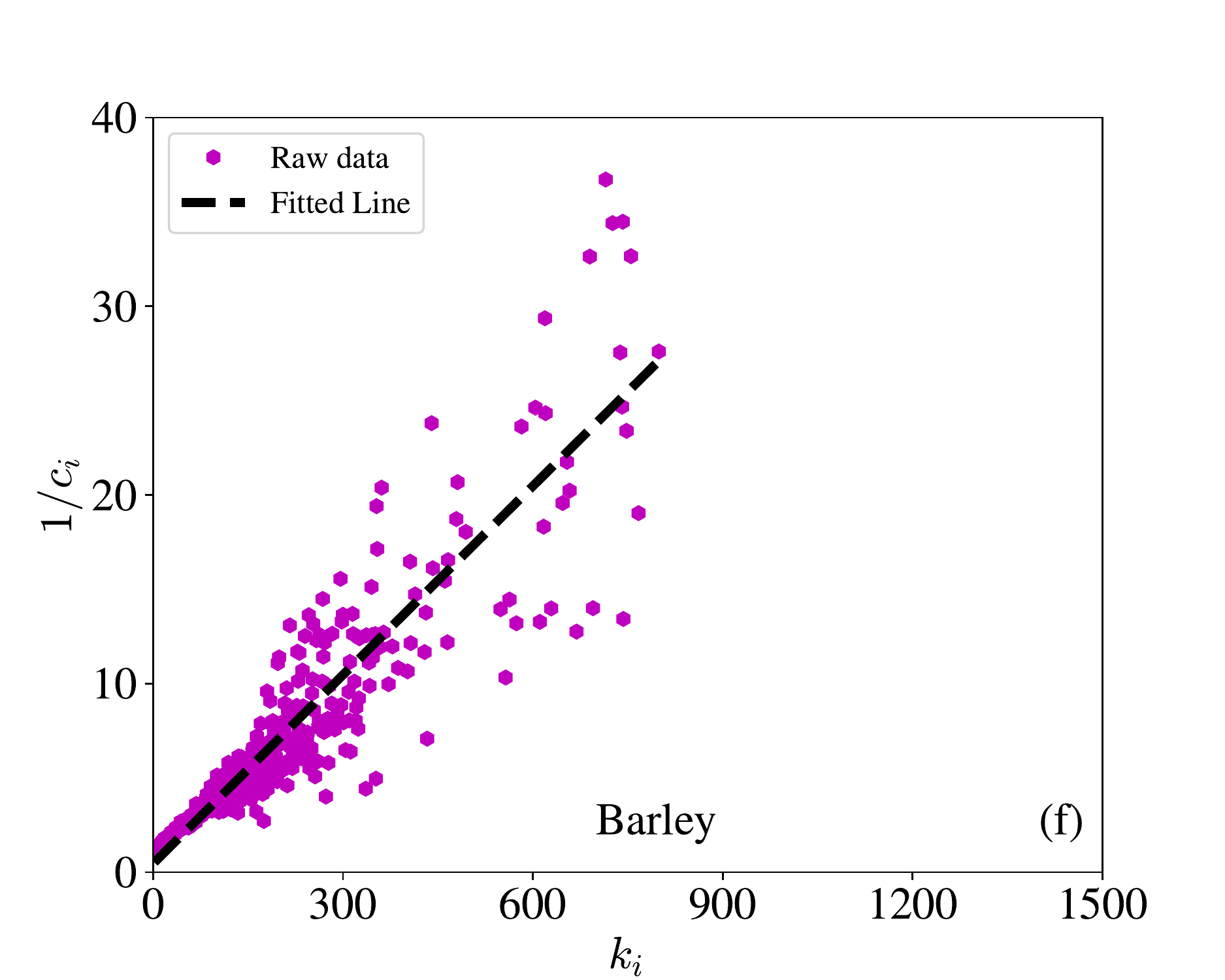}
    \caption{The relationship between the reciprocal of clustering coefficient $\frac{1}{c_i}$ and node degree $k_i$ of six VGs. (a) IGC GOI VG. (b) Wheat VG. (c) Maize VG. (d) Soyabeans VG. (e) Rice VG. (f) Barley VG.}
    \label{Fig:GOIVG:ClusteringCoefficient:1/ci:ki}
\end{figure}

In Fig.~\ref{Fig:GOIVG:ClusteringCoefficient:1/ci:ki}, we plot the reciprocal of clustering coefficient $\frac{1}{c_i}$ and the node degree $k_i$ for all range of node degree $k_i$. We observed a linear correlation between the reciprocal of clustering coefficient $\frac{1}{c_i}$ and the node degree $k_i$, where the $p$-values for six VGs are smaller than 0.05. It means
\begin{equation}
    \frac{1}{c_i} = \frac{k_i(k_i-1)}{2T_i} \sim k_i,
\end{equation}
and then we have
\begin{equation}
    T_i \sim k_i.
\end{equation}
It indicates that nodes with small degree in the network have high clustering coefficient and belong to small modules with high connection. The hub nodes with high degree have low clustering coefficient, and they mainly connect different modules. The six VGs exhibit characteristics of hierarchical modularity.

\subsection{Small-world property}
A complex network that conforms to small-world property usually conforms to two main characteristics \cite{Watts-Strogatz-1998-Nature}. One is that the clustering coefficients of the network are large, which is demonstrated in Section~\ref{S2:Clusteringcoefficient}, and the other is that the average shortest path length $L(N)$ of the network satisfies
\begin{equation}
    L(N) \sim \log N,
    \label{Eq:GOIVG:SmallWorldNetwork:LN:N}
\end{equation}
and $L(N)$ is defined as
\begin{equation}
    L(N) = \frac{2}{N(N-1)}\sum_{i,j \in \mathscr{V}} d(i,j),
    \label{Eq:GOIVG:SmallWorldNetwork:L(N)}
\end{equation}
where $N$ is the number of nodes in the network and $d(i,j)$ is the shortest path length between two nodes $i$ and $j$.

To determine if six VGs satisfy the small-world property, we picked 50 lengths $N$, and we estimated $L(N)$ by averaging the $L(N)$ of the VGs constructed by all $\lfloor {T}/{N} \rfloor$ sub-time series.

\begin{figure}[!ht]
    \centering
    \includegraphics[width=0.325\linewidth]{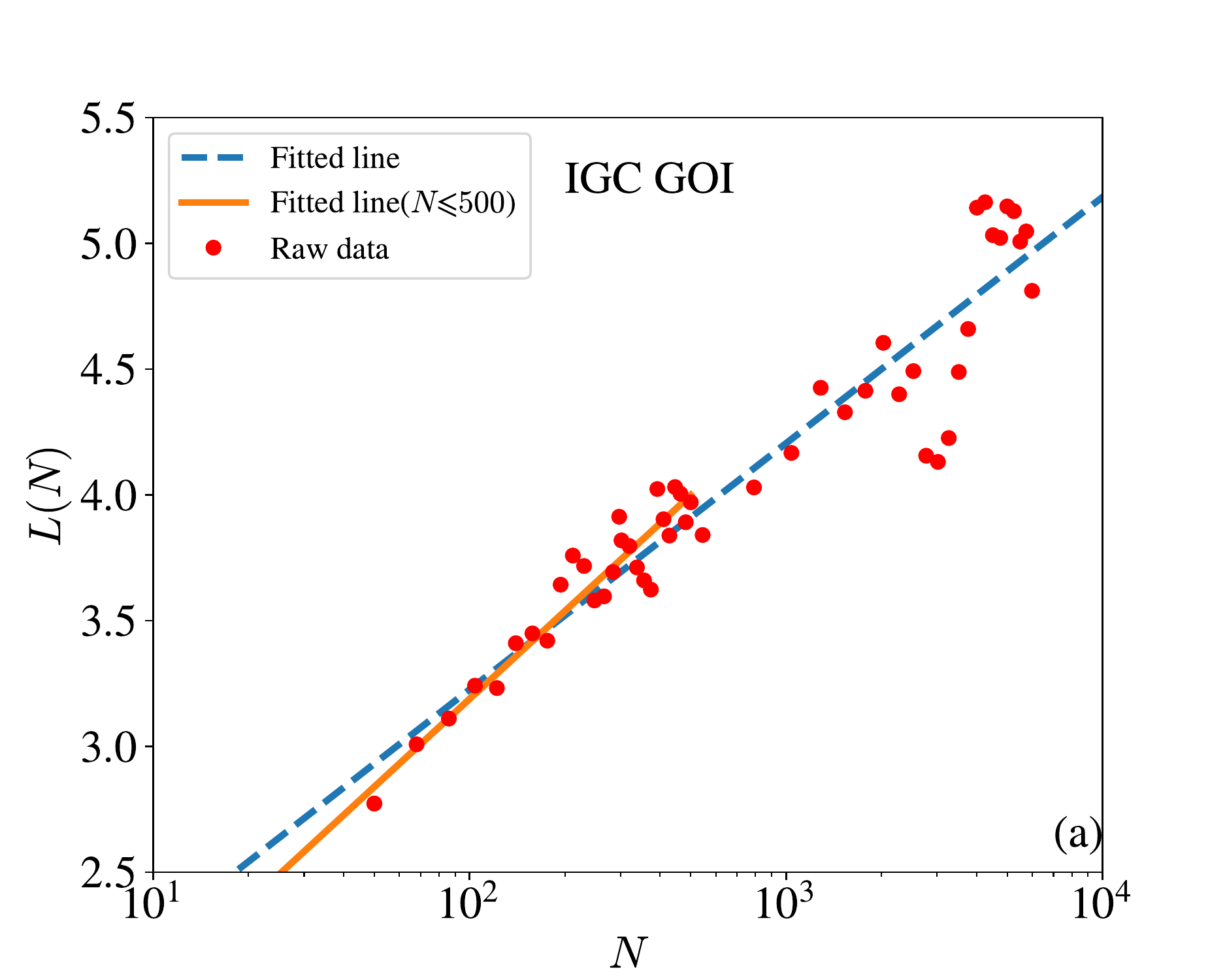}
    \includegraphics[width=0.325\linewidth]{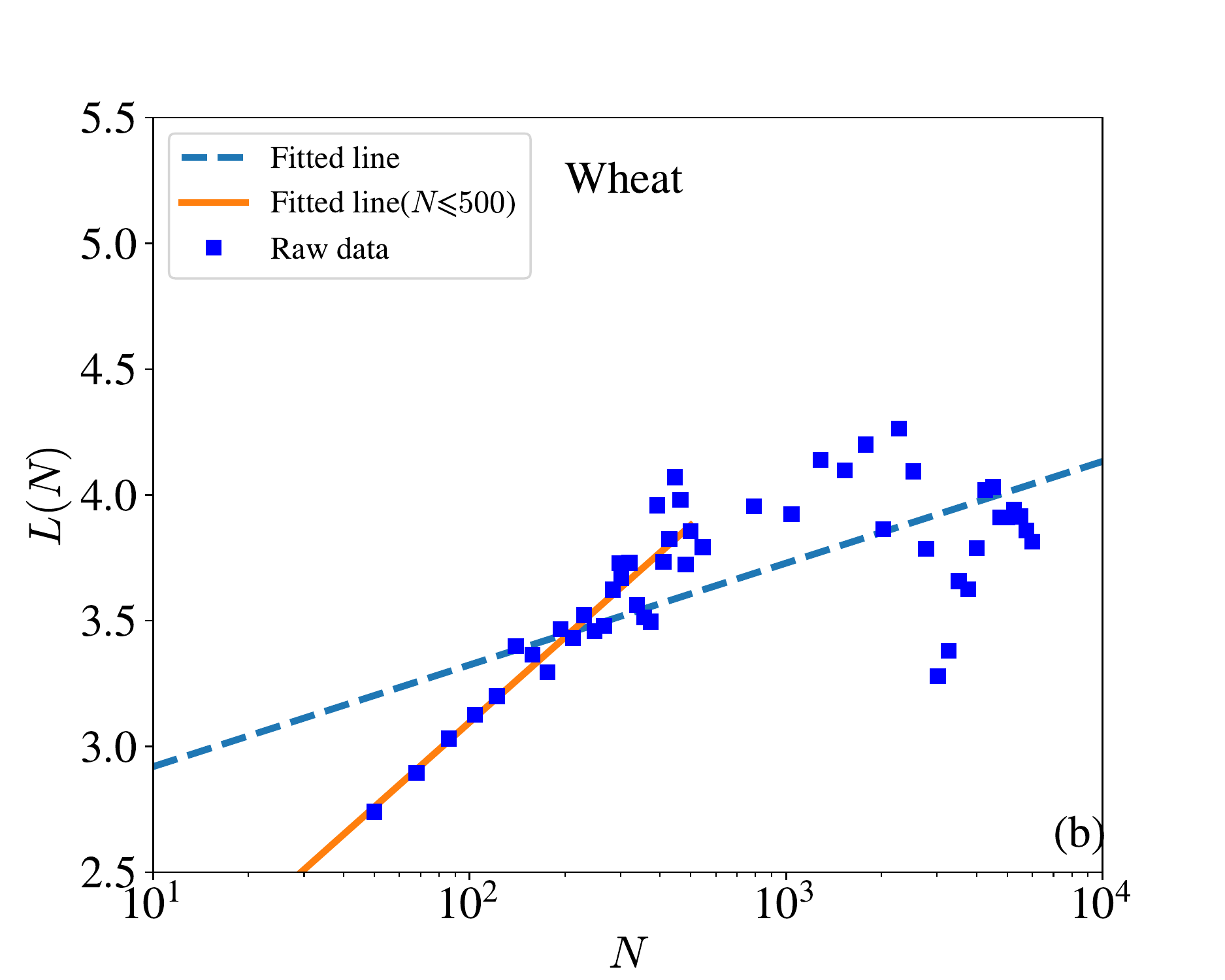}
    \includegraphics[width=0.325\linewidth]{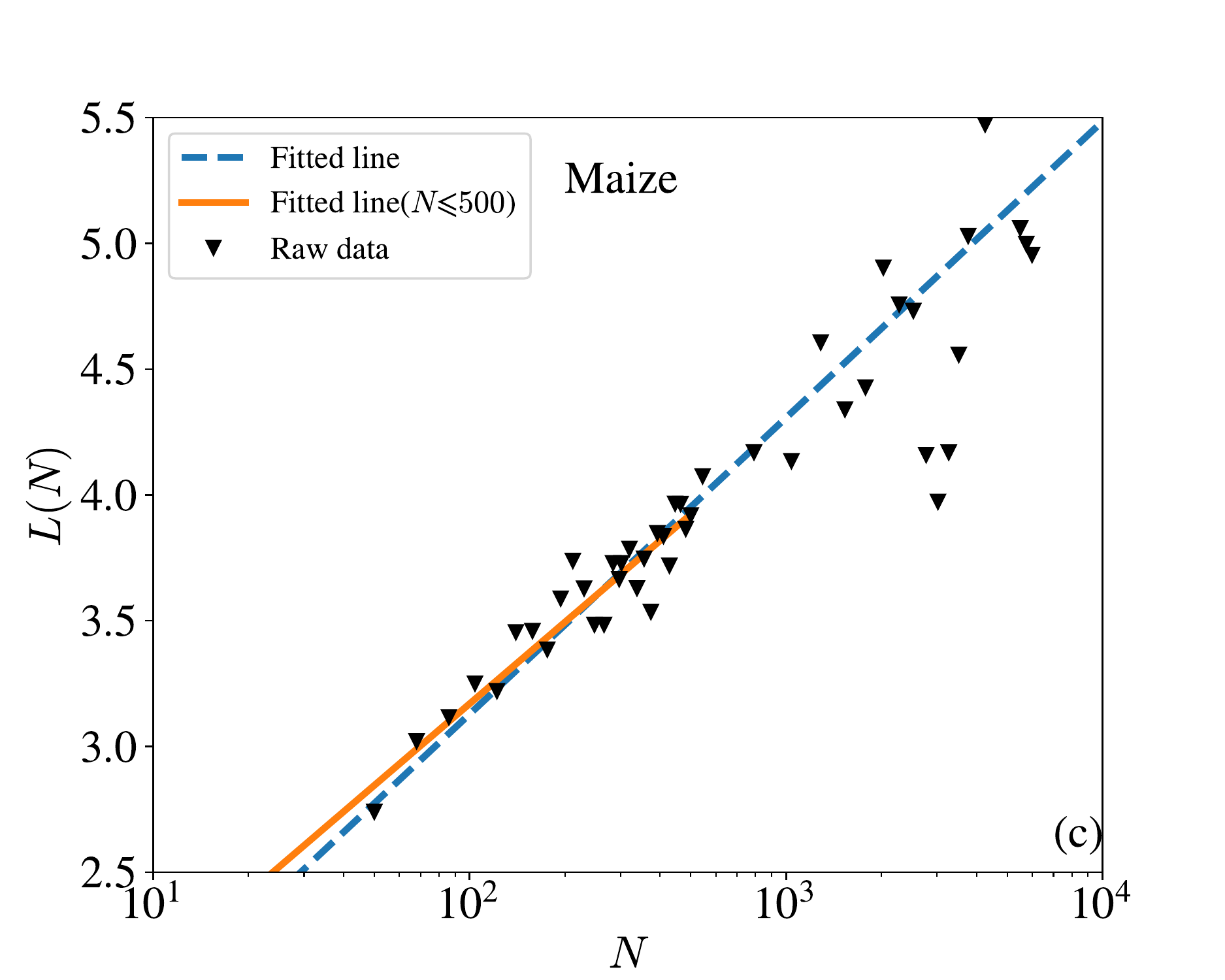}\\
    \includegraphics[width=0.325\linewidth]{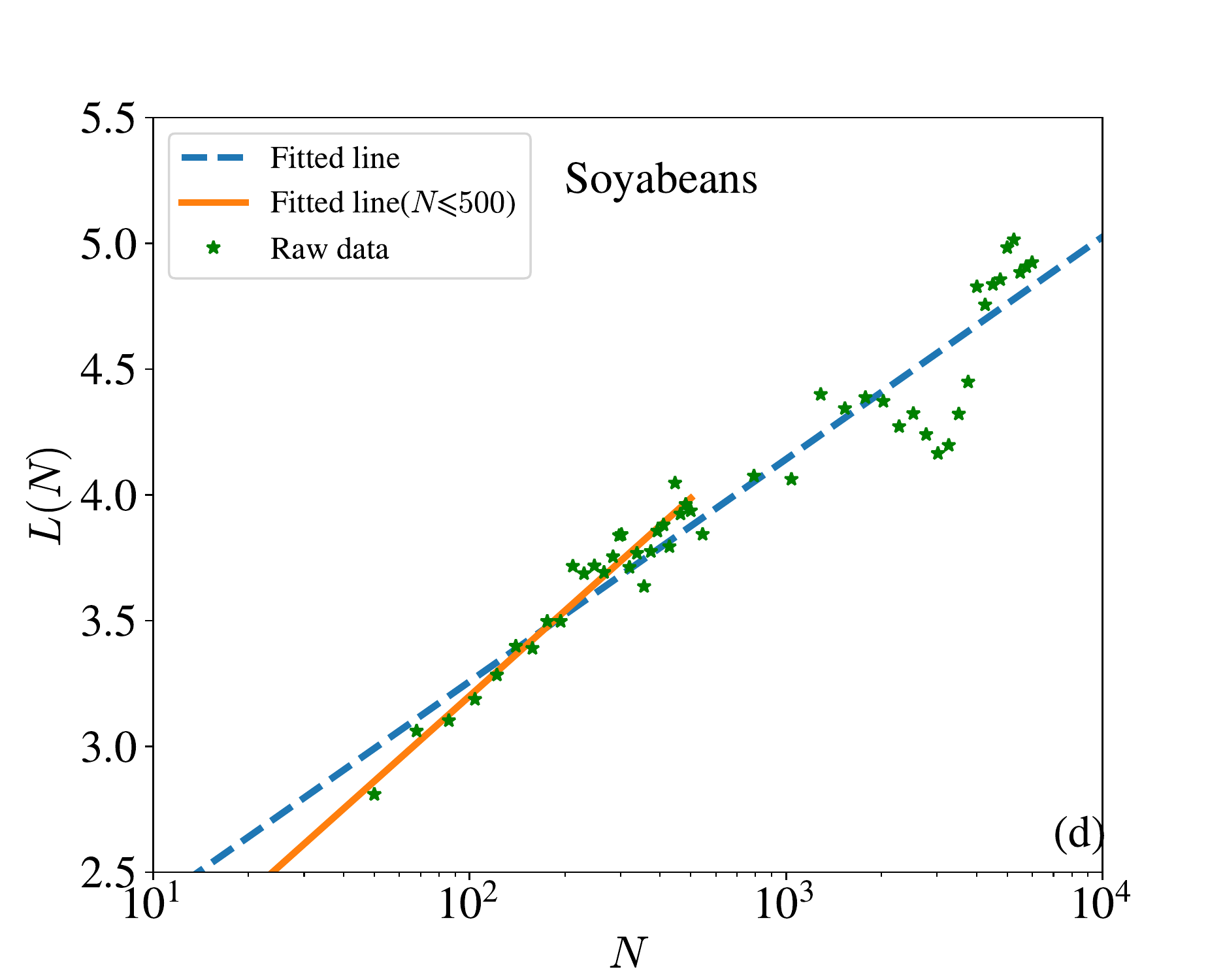}
    \includegraphics[width=0.325\linewidth]{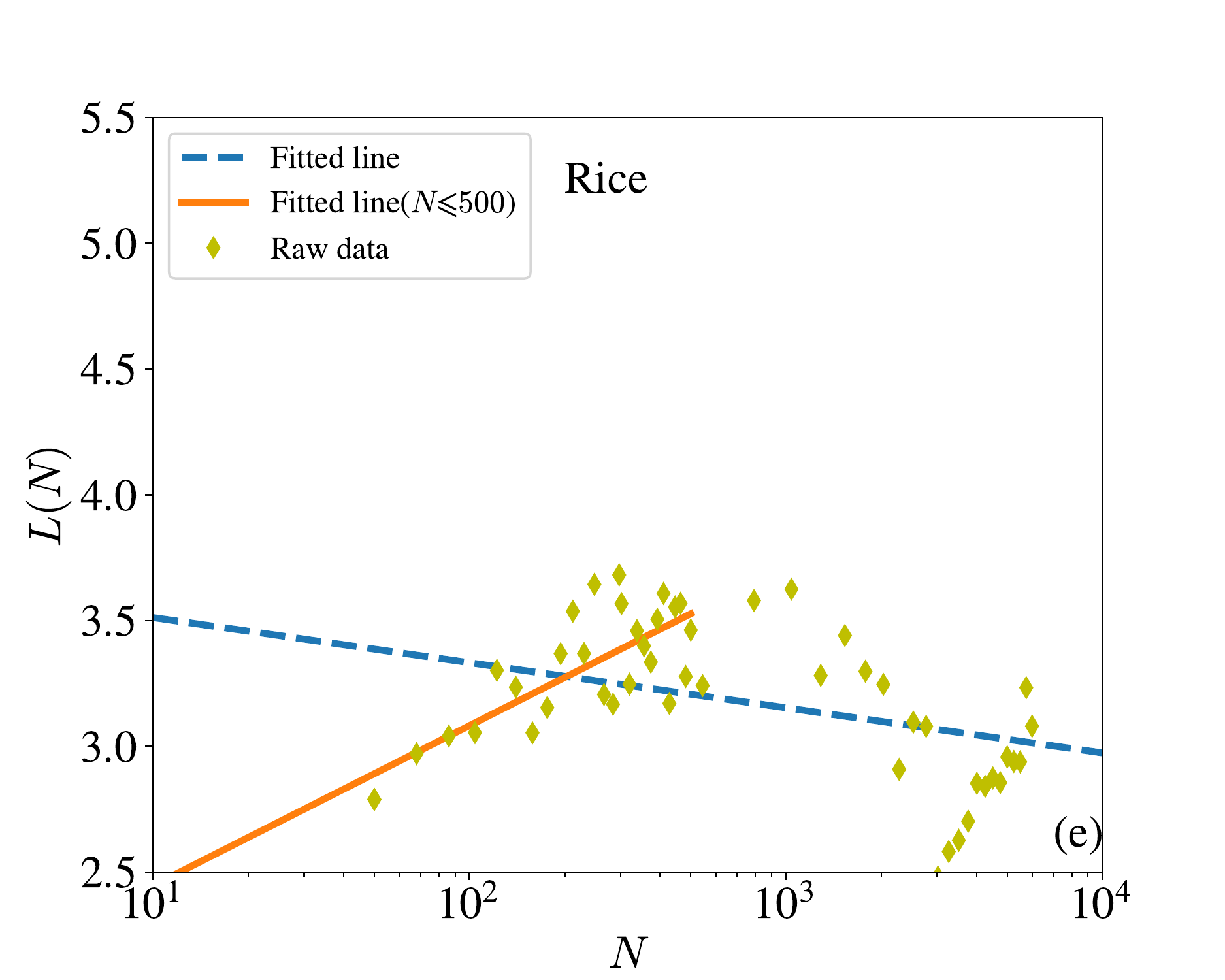}
    \includegraphics[width=0.325\linewidth]{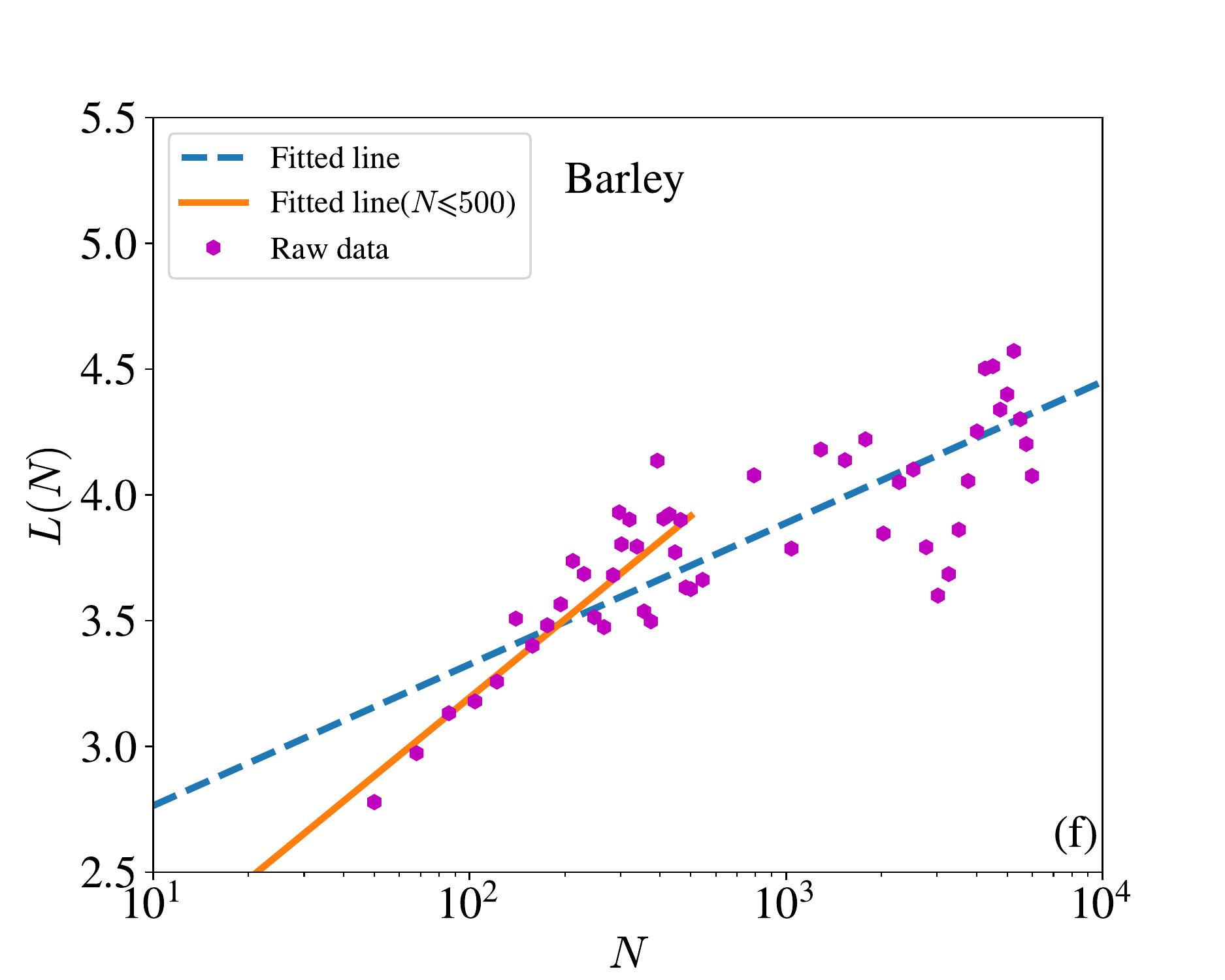}
    \caption{The relationship between average shortest path length $L(N)$ and logarithm of node number $N$ of six VGs. (a) IGC GOI VG. (b) Wheat VG. (c) Maize VG. (d) Soyabeans VG. (e) Rice VG. (f) Barley VG.}
    \label{Fig:GOIVG:SmallWorld:LN:N}
\end{figure}

In Fig.~\ref{Fig:GOIVG:SmallWorld:LN:N}, we plot $L(N)$ and $N$ and observe a linear correlation, especially when $N$ is not large ($N \leqslant 500$). We assume that
\begin{equation}
    L(N) = \beta_0 + \beta_1 \lg N,
    \label{Eq:GOIVG:Smallworld:FittedEquation}
\end{equation}
and fit Eq.~(\ref{Eq:GOIVG:Smallworld:FittedEquation}) using the OLS method. The results are illustrated in Table~\ref{Table:GOIVG:SmallWorld:FittingResult}. Note that the parameters $\beta_0$, $\beta_1$ and $\overline{R^2}$ in Table~\ref{Table:GOIVG:SmallWorld:FittingResult} are the fitting results using all data, while the parameters $\beta_0^*$, $\beta_1^*$ and $\overline{R^2}^*$ are the fitting results using the data with $N \leqslant 500$.

\begin{table}[!ht]
    \centering
    \caption{The fitting result of intercept coefficient $\beta_0,\beta_0^*$, slope coefficient $\beta_1,\beta_1^*$, adjust R-square $\overline{R^2},\overline{R^2}^*$ and $p$-value $p,p*$ of $L(N)$ and $\ln N$ of six VGs.}
    \medskip
    \setlength{\tabcolsep}{3.8mm}
    \begin{tabular}{ccccccccc}
    \toprule
    \multirow{2}{*}{Series} & \multicolumn{2}{c}{Intercept coefficient} & \multicolumn{2}{c}{Slope coefficient}  & \multicolumn{2}{c}{Adjusted R-square} & \multicolumn{2}{c}{$p$-value}                 \\ \cline{2-9} 
& $\beta_0$& $\beta_0^*$& $\beta_1$ & $\beta_1^*$ & $\overline{R^2}$ & $\overline{R^2}^*$ & $p$ & $p^*$ \\ \midrule
    IGC GOI & 1.2682& 0.8694 & ~~0.9789 & 1.1597& 0.9123  & 0.9195  & 0.0000 & 0.0000 \\
    Wheat   & 2.5153& 0.8558 & ~~0.4045 & 1.1202& 0.4751  & 0.8995  & 0.0000 & 0.0000 \\
    Maize   & 0.7719& 1.0148 & ~~1.1782 & 1.0774& 0.8587  & 0.9065  & 0.0000 & 0.0000     \\
    Soyabeans  & 1.4902& 0.9506 & ~~0.8839 & 1.1248& 0.9285  & 0.9524 & 0.0000 & 0.0000  \\
    Rice    & 3.6915& 1.8110 & $-0.1793$& 0.6360& 0.1171  & 0.5744 & 0.0080 & 0.0000  \\
    Barley  & 2.2024& 1.1276 & ~~0.5620 & 1.0331& 0.6951  & 0.7801 & 0.0000 & 0.0000  \\ 
    \bottomrule
    \end{tabular}
    \label{Table:GOIVG:SmallWorld:FittingResult}
\end{table}

Comparing the parameters in Table~\ref{Table:GOIVG:SmallWorld:FittingResult}, the adjust R-square is larger when $N \leqslant 500$, which indicates that the characteristic of Eq.~(\ref{Eq:GOIVG:SmallWorldNetwork:LN:N}) is better maintained when the series length $N$ is not large ($N \leqslant 500$). As shown in Table~\ref{Table:GOIVG:SmallWorld:FittingResult}, the $L(N)$ and $\lg N$ in the other five VGs are significantly and positively correlated, with the exception of the Rice VG. We plot the estimated values with the fitted lines in Fig.~\ref{Fig:GOIVG:SmallWorld:LN:N} and find that the fitted lines for these five VGs fit well. Moreover, we observe that a significant and positive linear correlation between $L(N)$ and $\lg N$ is maintained for all the VGs when $N$ is not large.

\begin{table}[!ht]
    \centering
    \caption{The clustering coefficients $C$ and average shortest path lengths $L$ of six VGs, compared to random graphs with the same number of nodes and number of edges.}
    \medskip
    \setlength{\tabcolsep}{7.5mm}
    \begin{tabular}{ccccc}
    \toprule
    Series & $L_\mathrm{actual}$ & $L_\mathrm{random}$ & $C_\mathrm{actual}$ & $C_\mathrm{random}$ \\ \midrule
    IGC GOI       & 4.8113     & 2.8334           & 0.6293     & 0.0053           \\
    Wheat         & 3.8134     & 2.7742           & 0.6033     & 0.0065           \\
    Maize         & 4.9548     & 3.0407           & 0.6639     & 0.0038           \\
    Soyabeans     & 4.9241     & 3.0277           & 0.6584     & 0.0039           \\
    Rice          & 3.0806     & 2.3737           & 0.5224     & 0.0126           \\
    Barley        & 4.0756     & 2.6748           & 0.5725     & 0.0079          \\ \bottomrule
    \end{tabular}
    \label{Table:GOIVG:SmallWorld:RandomCompare}
\end{table}

To further elucidate the structural nuances of the VGs, we embarked on a comparative analysis against null models. Specifically, we juxtaposed the clustering coefficients ($C$) and average shortest path lengths ($L$) of the six VGs against null models. According to the results in Table~\ref{Table:GOIVG:SmallWorld:RandomCompare}, a compelling manifestation of the small-world phenomenon can be observed in all VGs. Notably, the clustering coefficients of our VGs far exceeded those characteristic of random graphs, underscoring the pronounced tendency of nodes to form tightly-knit clusters. Concurrently, the average shortest path lengths within our VGs, though longer than in random graphs, still maintained relatively short distances between nodes, facilitating efficient information dissemination and connectivity. This dual characteristic of high clustering and short path lengths epitomizes the quintessence of small-world networks.


\subsection{Mixing pattern}

Assortativity reflects the possibility that nodes in a graph are connected to nodes with comparable degrees \cite{Newman-2002-PhysRevLett}. The assortativity is quantified by the Pearson correlation coefficient between node degrees
\begin{equation}
    r=\frac{1}{N_{\mathscr{E}}}\sum_{(i,j) \in \mathscr{E}} \frac{(k_i-\left\langle k_i \right\rangle)(k_j-\left\langle k_j \right\rangle)}{\sigma_{i,\mathscr{E}}\sigma_{j,\mathscr{E}}},
\end{equation}
where $k_i$ is the degree of node $i$ and $\sigma_{i,\mathscr{E}}$ is defined by
\begin{equation}
    \sigma_{i,\mathscr{E}} = \sqrt{\frac{1}{N_{\mathscr{E}}} \sum_{(i,j) \in \mathscr{E}}k_i^2 - {\left\langle k \right\rangle}^2}.
\end{equation}
The range of values of assortativity $r$ is $-1 \leqslant r \leqslant 1$. If $r>0$, the VG is assortative mixing. If $r=0$, the VG has no assortative mixing. If $r<0$, the GOI visiblity graph is disassortative mixing.

\begin{table}[!ht]
    \centering
    \caption{The degree assortativity coefficients $r$ of the six VGs.}
    \smallskip
    \setlength{\tabcolsep}{5.1mm}
    \begin{tabular}{ccccccc}
    \toprule
      & IGC GOI & Wheat  & Maize  & Soyabeans & Rice& Barley \\ \midrule
    Assortativity & $-0.0413$  & $-0.1394$ & 0.0617 & 0.1506 & $-0.2606$ & $-0.1315$  \\ \bottomrule
    \end{tabular}
    \label{Table:GOIVG:Assortativity}
\end{table}

The assortativity coefficients of the six VGs are presented in Table~\ref{Table:GOIVG:Assortativity}. As shown in the table, the assortativity coefficients of the maize VG and the soyabeans VG are positive, which indicates that those two graph exhibit an assortative mixing pattern. Other VGs is negative, which indicates that the other four graphs exhibit a disassortative mixing pattern. In addition, all the VGs obtain small values of assortativity, which means that all the VGs only have weak assortative or disassortative mixing patterns.

The mixing pattern of a VG can also be measured by the average nearest neighbor degree $\left\langle k_{nn}|k\right\rangle$ as a function of the degree $k$ \cite{Li-Jiang-Xie-Micciche-Tumminello-Zhou-Mantegna-2014-SciRep}. The average nearest neighbor degree $k_{nn,i}$ of node $i$ in a VG is defined as
\begin{equation}
    k_{nn,i} = \frac{1}{k_i} \sum_{(i,j) \in \mathscr{E}} k_j,
\end{equation}
where $k_i,k_j$ is the node degree of node $i,j$ and the average nearest neighbor degree $\left\langle k_{nn}|k\right\rangle$ as a function of the degree $k$ is defined as
\begin{equation}
    \left\langle k_{nn}|k\right\rangle = \frac{1}{N_{k_i}} \sum_{k_i = k} k_{nn,i},
\end{equation}
where $N_{k_i}$ is the number of nodes whose node degree is $k$:
\begin{equation}
    N_{k_i} = \sharp(\{k_i~|~ k_i=k\}).
\end{equation}
If $\left\langle k_{nn}|k\right\rangle$ is an increasing function of $k$ in a VG, this VG is said to be assortatively mixed. If $\left\langle k_{nn}|k\right\rangle$ is a decreasing function of $k$ in a VG, this VG is said to be disassortatively mixed.

\begin{figure}[!ht]
    \centering
    \includegraphics[width=0.325\linewidth]{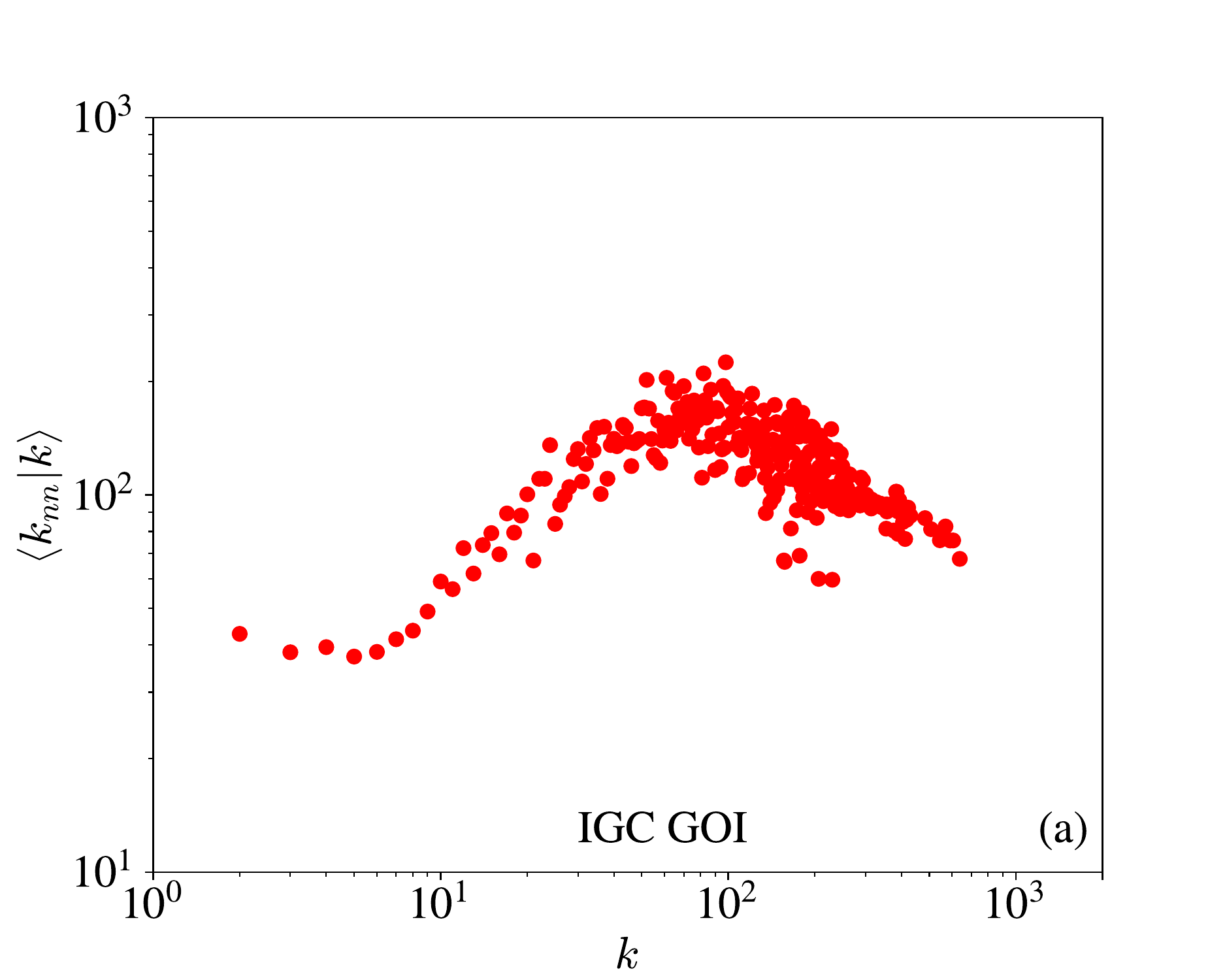}
    \includegraphics[width=0.325\linewidth]{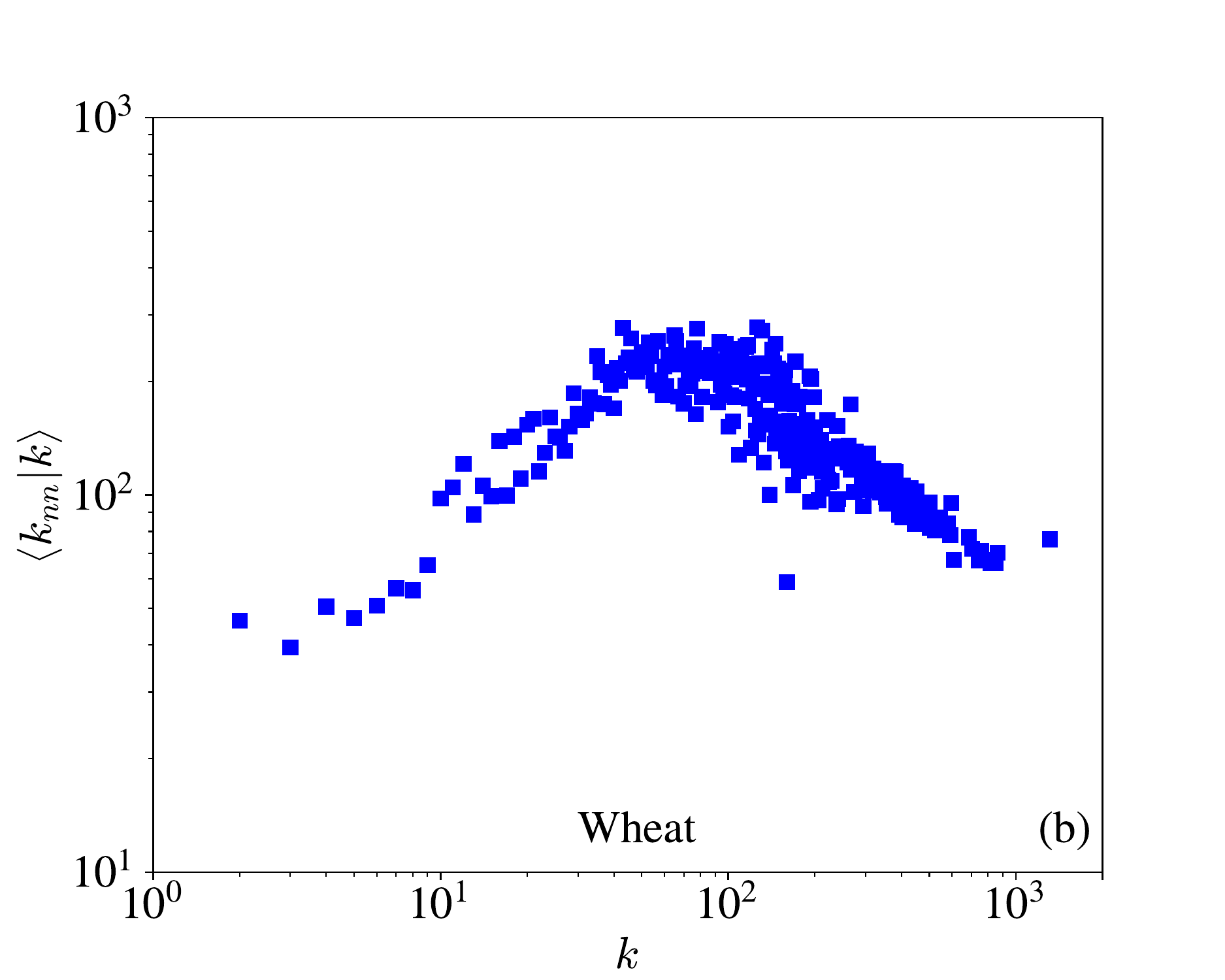}
    \includegraphics[width=0.325\linewidth]{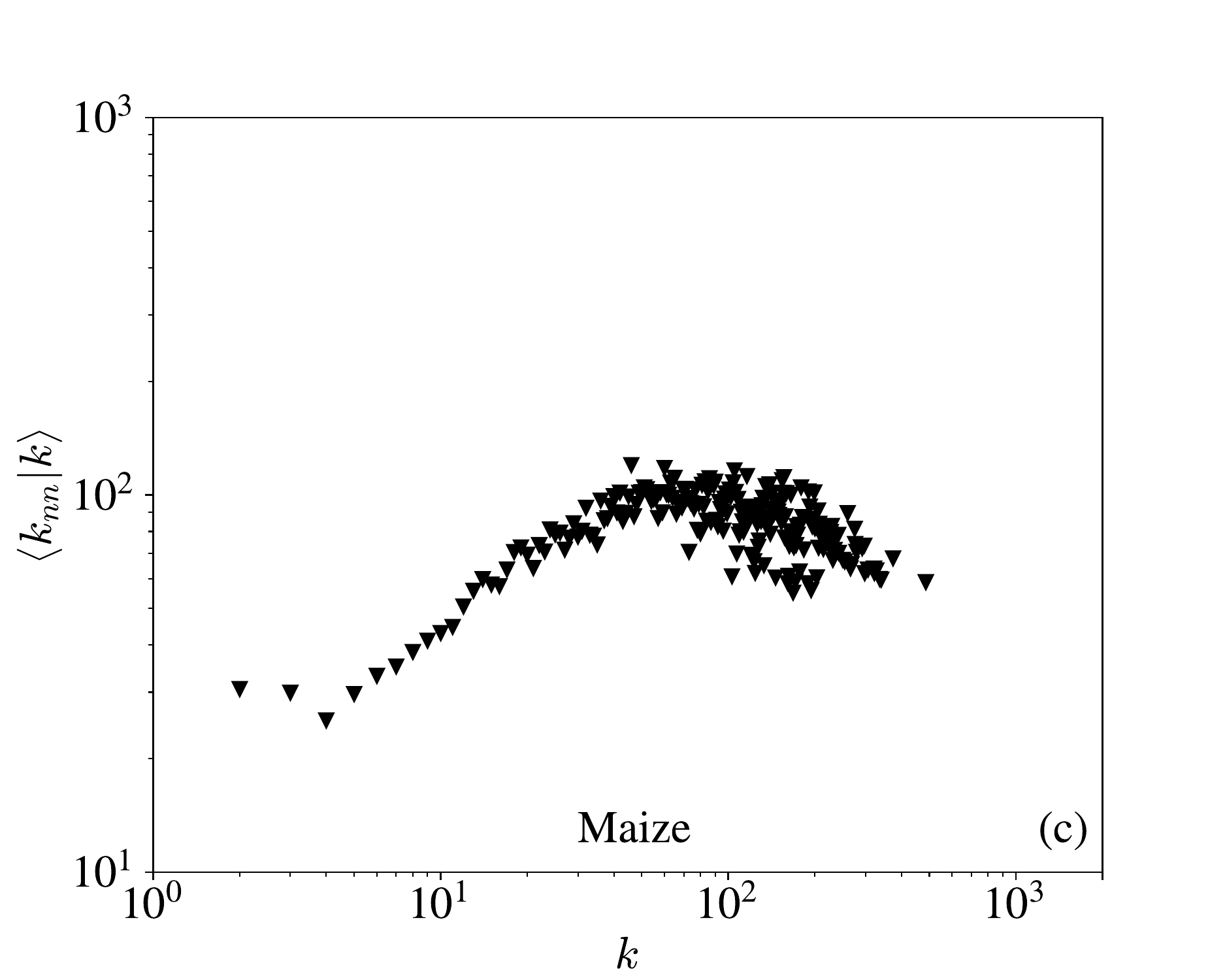}\\
    \includegraphics[width=0.325\linewidth]{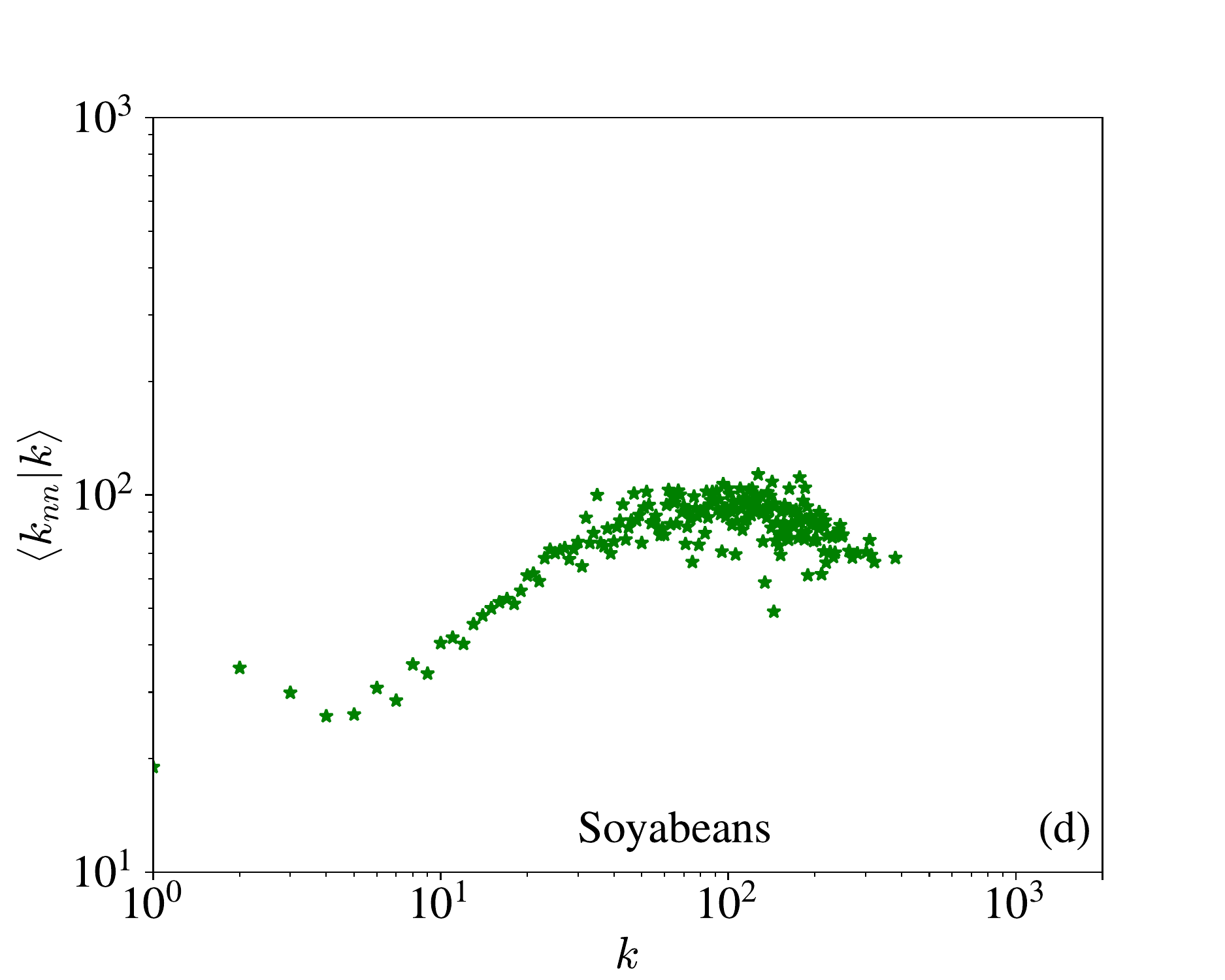}
    \includegraphics[width=0.325\linewidth]{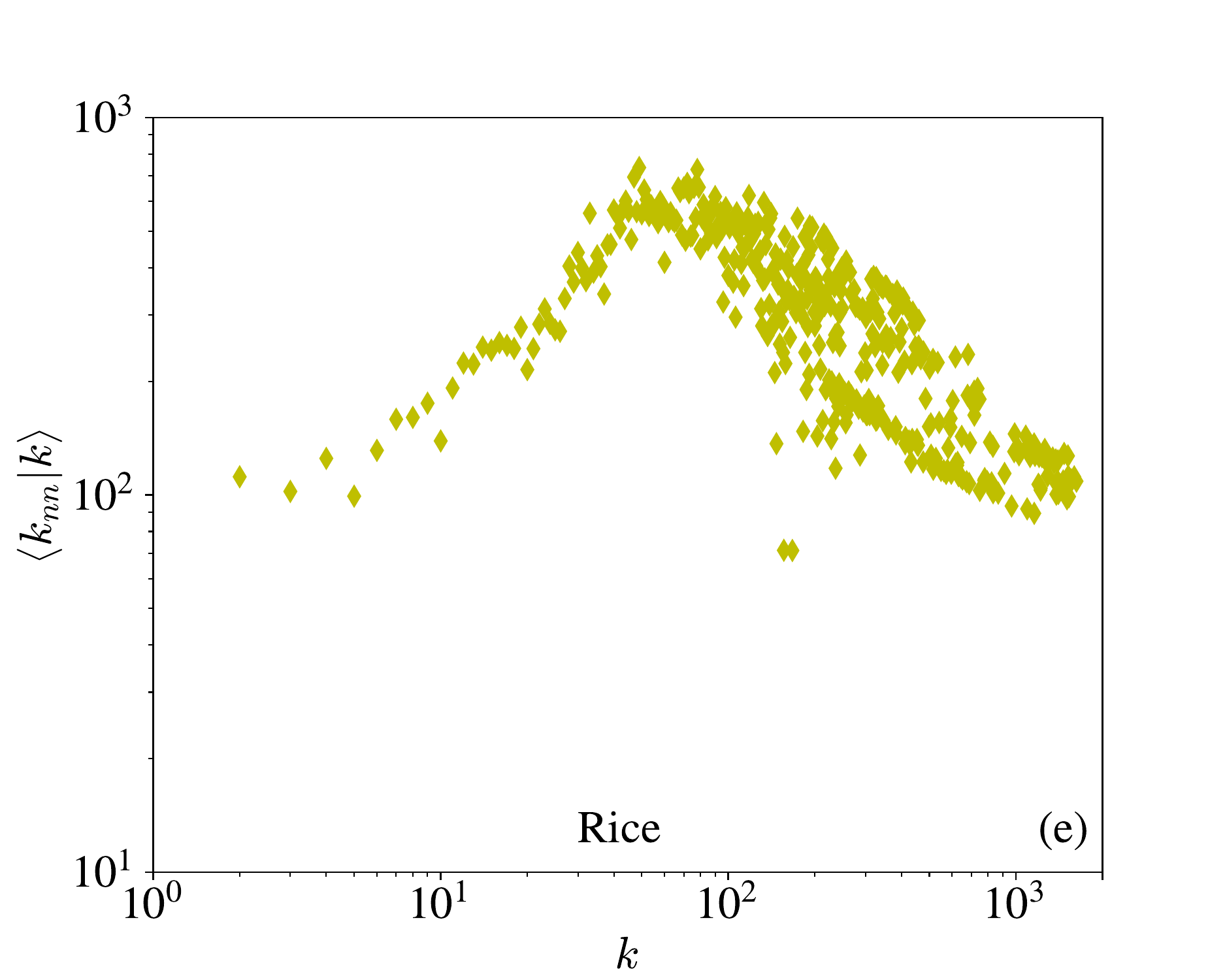}
    \includegraphics[width=0.325\linewidth]{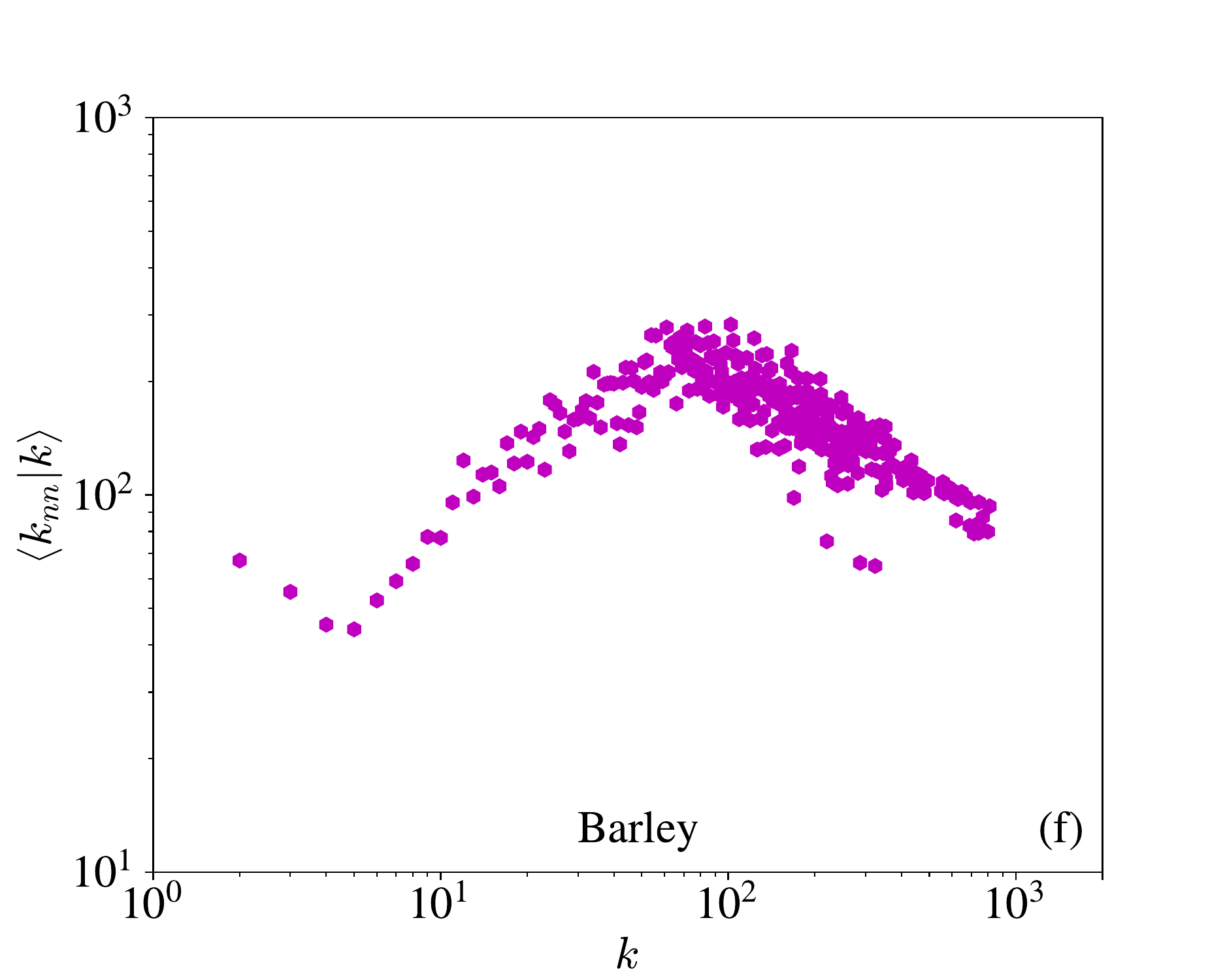}
    \caption{The average nearest neighbor degree $\left\langle k_{nn}|k\right\rangle$ as a function of degree $k$ for six VGs in log-log scale. (a) IGC GOI VG. (b) Wheat VG. (c) Maize VG. (d) Soyabeans VG. (e) Rice VG. (f) Barley VG.}
    \label{Fig:GOIVG:MixingPattern:knn:k}
\end{figure}

In Fig.~\ref{Fig:GOIVG:MixingPattern:knn:k}, we plot the average nearest neighbor degree $\left\langle k_{nn}|k\right\rangle$ and degree $k$ of six VGs. The average nearest neighbor degree functions $\left\langle k_{nn}|k\right\rangle$ of all the six VGs have similar shapes. With the increasing of node degree $k$, each function firstly decreases, then increases, and finally decreases again, showing a more complex mixing pattern.

\section{Summary}
\label{S1:Summary}

In this paper, we converted the index time series into six VGs using the visibility graph algorithm. We then investigated the structural properties of the six VGs, including the Hurst exponents, the global network quantities, the degree distributions, the clustering coefficients, the shortest path lengths, and the mixing pattern.

We found the Rice index has the largest Hurst exponent of 0.7540, which indicates that the Rice index time series is strongly persistent. The GOI indices of maize and soyabeans are 0.5405 and 0.5200, which indicates that those two indices are close to Brownian motions. We observed that the degree distribution of the Rice VG show a power-law tail using the maximum likelihood estimation method. The results indicate that the Rice VG is a scale-free network. The degree distributions of the remaining five VGs are found to conform admirably to an exponentially truncated power-law model. This implies that while these networks do exhibit scale-free characteristics in the lower to moderate degree range, their topologies are subject to limitations that prevent the unbounded growth of node degrees observed in scale-free networks.

Under the condition that the densities of all the six VGs are very small, their average clustering coefficients are quite large, which indicates that the VGs tend to form ternary cliques locally. We also observed a power-law correlation between the average clustering coefficient and the average degree, as shown in Eq.~(\ref{Eq:GOIVG:lnC:lnavek}). In addition, the node clustering coefficient of six VGs is inversely proportional to the node degree. When the node degree is not large, the node clustering coefficient of six VGs power-law correlated to the node degree.

We used a null model and compared path length and clustering to verify the small-world property of VG. The average shortest path length of the six VGs was observed to be slightly greater than that of the null model. However, the clustering coefficient for each VG was markedly higher. It implies that the small-world property was indeed confirmed across our VG ensemble. Yet, the differences in clustering coefficients and path lengths highlighted the intricate balance between local connectivity and global efficiency within these networks. Particularly intriguing was the observation that the average shortest path length of all VGs increased with the number of nodes when the series length was modest ($\leqslant 500$). This trend underscores the preservation of small-world characteristics even as network size expands, albeit within certain limits. However, a notable exception emerged in the case of the Rice VG. As the series length grew, the increase in average shortest path length was not as pronounced as in the other VGs. This observation dovetails with the smaller disparity in clustering coefficients between the Rice VG and the null model, suggesting a more uniform distribution of connectivity and potentially different mechanisms governing network evolution and topology.

We used two metrics to measure the mixing pattern of six VGs: the degree-degree correlation coefficient and the average nearest neighbor degree. For the degree-degree correlation coefficient, the maize VG and soyabeans VG exhibit weak assortative mixing patterns, while the other VGs exhibit weak disassortative mixing patterns. For comparison, the average nearest neighbor degree functions have similar patterns, and each function shows a more complex mixing pattern which decreases for small degrees, increases for mediate degrees, and decreases again for large degrees.

Our findings underscore the multifaceted nature of commodity price dynamics when viewed through the lens of network science. The identification of persistent trends, scale-free structures, and small-world properties within these VGs opens up new avenues for understanding and predicting market behaviors. However, to solidify these insights, we will do deeper research that includes extended time series analysis, comparative studies with other financial instruments, and detailed investigations into the mechanisms driving these network properties. Additionally, incorporating external factors such as geopolitical events, climate change impacts, and economic indicators could enrich our understanding of how these variables interact with the internal network structures, providing a more holistic view of commodity markets.

\section*{Acknowledgements}
This work was supported by the National Natural Science Foundation of China (72171083), the Shanghai Outstanding Academic Leaders Plan, and the Fundamental Research Funds for the Central Universities.
   







\begin{thebibliography}{10}
\expandafter\ifx\csname url\endcsname\relax
  \def\url#1{\texttt{#1}}\fi
\expandafter\ifx\csname urlprefix\endcsname\relax\def\urlprefix{URL }\fi
\expandafter\ifx\csname href\endcsname\relax
  \def\href#1#2{#2} \def\path#1{#1}\fi

\bibitem{Zou-Donner-Marwan-Donges-Kurths-2019-PhysRep}
Y.~Zou, R.~V. Donner, N.~Marwan, J.~F. Donges, J.~Kurths, {Complex network
  approaches to nonlinear time series analysis}, Phys. Rep. 787 (2019) 1--97.
\newblock \href {https://doi.org/10.1016/j.physrep.2018.10.005}
  {\path{doi:10.1016/j.physrep.2018.10.005}}.

\bibitem{Li-Wang-2006-CSB}
P.~Li, B.-H. Wang, {An approach to Hang Seng Index in Hong Kong stock market
  based on network topological statistics}, Chin. Sci. Bull. 51~(5) (2006)
  624--629.
\newblock \href {https://doi.org/10.1007/s11434-006-0624-4}
  {\path{doi:10.1007/s11434-006-0624-4}}.

\bibitem{Li-Wang-2007-PhysicaA}
P.~Li, B.-H. Wang, {Extracting hidden fluctuation patterns of Hang Seng stock
  index from network topologies}, Physica A 378~(2) (2007) 519--526.
\newblock \href {https://doi.org/10.1016/j.physa.2006.10.089}
  {\path{doi:10.1016/j.physa.2006.10.089}}.

\bibitem{Zhang-Small-2006-PhysRevLett}
J.~Zhang, M.~Small, {Complex network from pseudoperiodic time series: Topology
  versus dynamics}, Phys. Rev. Lett. 96~(23) (2006) 238701.
\newblock \href {https://doi.org/10.1103/PhysRevLett.96.238701}
  {\path{doi:10.1103/PhysRevLett.96.238701}}.

\bibitem{Xu-Zhang-Small-2008-ProcNatlAcadSciUSA}
X.-K. Xu, J.~Zhang, M.~Small, {Superfamily phenomena and motifs of networks
  induced from time series}, Proc. Natl. Acad. Sci. U.S.A. 105~(50) (2008)
  19601--19605.
\newblock \href {https://doi.org/10.1073/pnas.0806082105}
  {\path{doi:10.1073/pnas.0806082105}}.

\bibitem{Yang-Yang-2008-PhysicaA}
Y.~Yang, H.-J. Yang, {Complex network-based time series analysis}, Physica A
  387 (2008) 1381--1386.
\newblock \href {https://doi.org/10.1016/j.physa.2007.10.055}
  {\path{doi:10.1016/j.physa.2007.10.055}}.

\bibitem{Lacasa-Luque-Ballesteros-Luque-Nuno-2008-ProcNatlAcadSciUSA}
L.~Lacasa, B.~Luque, F.~Ballesteros, J.~Luque, J.~C. Nu{\~n}o, {From time
  series to complex networks: The visibility graph}, Proc. Natl. Acad. Sci.
  U.S.A. 105~(13) (2008) 4972--4975.
\newblock \href {https://doi.org/10.1073/pnas.0709247105}
  {\path{doi:10.1073/pnas.0709247105}}.

\bibitem{Luque-Lacasa-Ballesteros-Luque-2009-PhysRevE}
B.~Luque, L.~Lacasa, F.~Ballesteros, J.~Luque, {Horizontal visibility graphs:
  Exact results for random time series}, Phys. Rev. E 80 (2009) 046103.
\newblock \href {https://doi.org/10.1103/PhysRevE.80.046103}
  {\path{doi:10.1103/PhysRevE.80.046103}}.

\bibitem{Lacasa-Luque-Luque-Nuno-2009-EPL}
L.~Lacasa, B.~Luque, J.~Luque, J.~C. Nu{\~n}o, {The visibility graph: A new
  method for estimating the Hurst exponent of fractional Brownian motion}, EPL
  (Europhys. Lett.) 86 (2009) 30001.
\newblock \href {https://doi.org/10.1209/0295-5075/86/30001}
  {\path{doi:10.1209/0295-5075/86/30001}}.

\bibitem{Xu-Zhang-Deng-2018-ChaosSolitonsFractals}
P.~Xu, R.~Zhang, Y.~Deng, {A novel visibility graph transformation of time
  series into weighted networks}, Chaos Solitons Fractals 117 (2018) 201--208.
\newblock \href {https://doi.org/10.1016/j.chaos.2018.07.039}
  {\path{doi:10.1016/j.chaos.2018.07.039}}.

\bibitem{Ahadpour-Sadra-2012-IS}
S.~Ahadpour, Y.~Sadra, {Randomness criteria in binary visibility graph and
  complex network perspective}, Inf. Sci. 197 (2012) 161--176.
\newblock \href {https://doi.org/10.1016/j.ins.2012.02.022}
  {\path{doi:10.1016/j.ins.2012.02.022}}.

\bibitem{Ahadpour-Sadra-ArastehFard-2014-IS}
S.~Ahadpour, Y.~Sadra, Z.~Arasteh~Fard, {Markov-binary visibility graph: A new
  method for analyzing complex systems}, Inf. Sci. 274 (2014) 286--302.
\newblock \href {https://doi.org/10.1016/j.ins.2014.03.007}
  {\path{doi:10.1016/j.ins.2014.03.007}}.

\bibitem{Bezsudnov-Snarskii-2014-PhysicaA}
I.~V. Bezsudnov, A.~A. Snarskii, {From the time series to the complex networks:
  The parametric natural visibility graph}, Physica A 414 (2014) 53--60.
\newblock \href {https://doi.org/10.1016/j.physa.2014.07.002}
  {\path{doi:10.1016/j.physa.2014.07.002}}.

\bibitem{Bianchi-Livi-Alippi-Jenssen-2017-SR}
F.~M. Bianchi, L.~Livi, C.~Alippi, R.~Jenssen, {Multiplex visibility graphs to
  investigate recurrent neural network dynamics}, Sci. Rep. 7 (2017) 44037.
\newblock \href {https://doi.org/10.1038/srep44037}
  {\path{doi:10.1038/srep44037}}.

\bibitem{Zhou-Jin-Gao-Luo-2012-APS}
T.-T. Zhou, N.-D. Jin, Z.-K. Gao, Y.-B. Luo, {Limited penetrable visibility
  graph for establishing complex network from time series}, Acta Phys. Sin.
  61~(3) (2012) 030506.
\newblock \href {https://doi.org/10.7498/aps.61.030506}
  {\path{doi:10.7498/aps.61.030506}}.

\bibitem{Gao-Cai-Yang-Dang-2017-PhysicaA}
Z.-K. Gao, Q.~Cai, Y.-X. Yang, W.-D. Dang, {Time-dependent limited penetrable
  visibility graph analysis of nonstationary time series}, Physica A 476 (2017)
  43--48.
\newblock \href {https://doi.org/10.1016/j.physa.2017.02.038}
  {\path{doi:10.1016/j.physa.2017.02.038}}.

\bibitem{Zou-Donner-Marwan-Small-Kurths-2014-NPG}
Y.~Zou, R.~V. Donner, N.~Marwan, M.~Small, J.~Kurths, Long-term changes in the
  north-south asymmetry of solar activity: a nonlinear dynamics
  characterization using visibility graphs, Nonlin. Process. Geophys. 21 (2014)
  1113--1126.
\newblock \href {https://doi.org/10.5194/npg-21-1113-2014}
  {\path{doi:10.5194/npg-21-1113-2014}}.

\bibitem{Gao-Cai-Yang-Dang-Zhang-2016-SR}
Z.-K. Gao, Q.~Cai, Y.-X. Yang, W.-D. Dang, S.-S. Zhang, {Multiscale limited
  penetrable horizontal visibility graph for analyzing nonlinear time series},
  Sci. Rep. 6 (2016) 35622.
\newblock \href {https://doi.org/10.1038/srep35622}
  {\path{doi:10.1038/srep35622}}.

\bibitem{Chen-Hu-Mahadevan-Deng-2014-PhysicaA}
S.~Chen, Y.~Hu, S.~Mahadevan, Y.~Deng, {A visibility graph averaging
  aggregation operator}, Physica A 403 (2014) 1--12.
\newblock \href {https://doi.org/10.1016/j.physa.2014.02.015}
  {\path{doi:10.1016/j.physa.2014.02.015}}.

\bibitem{Snarskii-Bezsudnov-2016-PhysRevE}
A.~A. Snarskii, I.~V. Bezsudnov, {Phase transition in the parametric natural
  visibility graph}, Phys. Rev. E 94~(4) (2016) 042137.
\newblock \href {https://doi.org/10.1103/PhysRevE.94.042137}
  {\path{doi:10.1103/PhysRevE.94.042137}}.

\bibitem{Qian-Jiang-Zhou-2010-JPhysA}
M.-C. Qian, Z.-Q. Jiang, W.-X. Zhou, Universal and nonuniversal allometric
  scaling behaviors in the visibility graphs of world stock market indices, J.
  Phys. A 43~(33) (2010) 335002.
\newblock \href {https://doi.org/10.1088/1751-8113/43/33/335002}
  {\path{doi:10.1088/1751-8113/43/33/335002}}.

\bibitem{Ni-Jiang-Zhou-2009-PhysLettA}
X.-H. Ni, Z.-Q. Jiang, W.-X. Zhou, Degree distributions of the visibility
  graphs mapped from fractional {B}rownian motions and multifractal random
  walks, Phys. Lett. A 373~(42) (2009) 3822--3826.
\newblock \href {https://doi.org/10.1016/j.physleta.2009.08.041}
  {\path{doi:10.1016/j.physleta.2009.08.041}}.

\bibitem{Zhang-Shang-Xiong-Xia-2018-FNL}
Y.-P. Zhang, P.-J. Shang, H.~Xiong, J.-A. Xia, {Multiscale analysis of time
  irreversibility based on phase-space reconstruction and horizontal visibility
  graph approach}, Fluct. Noise Lett. 17~(1) (2018) 1850006.
\newblock \href {https://doi.org/10.1142/S0219477518500062}
  {\path{doi:10.1142/S0219477518500062}}.

\bibitem{Xie-Han-Zhou-2019-EPL}
W.-J. Xie, R.-Q. Han, W.-X. Zhou, Triadic time series motifs, EPL 125~(1)
  (2019) 18002.
\newblock \href {https://doi.org/10.1209/0295-5075/125/18002}
  {\path{doi:10.1209/0295-5075/125/18002}}.

\bibitem{Nguyen-Nguyen-Nguyen-2019-PhysicaA}
Q.~Nguyen, N.~K.~K. Nguyen, L.~H.~N. Nguyen, {Dynamic topology and allometric
  scaling behavior on the Vietnamese stock market}, Physica A 514 (2019)
  235--243.
\newblock \href {https://doi.org/10.1016/j.physa.2018.09.061}
  {\path{doi:10.1016/j.physa.2018.09.061}}.

\bibitem{Vamvakaris-Pantelous-Zuev-2018-PhysicaA}
M.~D. Vamvakaris, A.~A. Pantelous, K.~M. Zuev, {Time series analysis of S\&P
  500 index: A horizontal visibility graph approach}, Physica A 497 (2018)
  41--51.
\newblock \href {https://doi.org/10.1016/j.physa.2018.01.010}
  {\path{doi:10.1016/j.physa.2018.01.010}}.

\bibitem{Yang-Wang-Yang-Mang-2009-PhysicaA}
Y.~Yang, J.-B. Wang, H.-J. Yang, J.-S. Mang, {Visibility graph approach to
  exchange rate series}, Physica A 388 (2009) 4431--4437.
\newblock \href {https://doi.org/10.1016/j.physa.2009.07.016}
  {\path{doi:10.1016/j.physa.2009.07.016}}.

\bibitem{Wang-Zheng-Wang-2019-PhysicaA}
G.-C. Wang, S.-Z. Zheng, J.~Wang, {Complex and composite entropy fluctuation
  behaviors of statistical physics interacting financial model}, Physica A 517
  (2019) 97--113.
\newblock \href {https://doi.org/10.1016/j.physa.2018.11.014}
  {\path{doi:10.1016/j.physa.2018.11.014}}.

\bibitem{Elsner-Jagger-Fogarty-2009-GRL}
J.~B. Elsner, T.~H. Jagger, E.~A. Fogarty, {Visibility network of United States
  hurricanes}, Geophys. Res. Lett. 36 (2009) L16702.
\newblock \href {https://doi.org/10.1029/2009GL039129}
  {\path{doi:10.1029/2009GL039129}}.

\bibitem{Liu-Zhou-Yuan-2010-PhysicaA}
C.~Liu, W.-X. Zhou, W.-K. Yuan, Statistical properties of visibility graph of
  energy dissipation rates in three-dimensional fully developed turbulence,
  Physica A 389~(13) (2010) 2675--2681.
\newblock \href {https://doi.org/10.1016/j.physa.2010.02.043}
  {\path{doi:10.1016/j.physa.2010.02.043}}.

\bibitem{Zhang-Zou-Zhou-Gao-Guan-2017-CNSNS}
R.~Zhang, Y.~Zou, J.~Zhou, Z.-K. Gao, S.-G. Guan, {Visibility graph analysis
  for re-sampled time series from auto-regressive stochastic processes},
  Commun. Nonlinear Sci. Numer. Simul. 42 (2017) 396--403.
\newblock \href {https://doi.org/10.1016/j.cnsns.2016.04.031}
  {\path{doi:10.1016/j.cnsns.2016.04.031}}.

\bibitem{Xie-Han-Jiang-Wei-Zhou-2017-EPL}
W.-J. Xie, R.-Q. Han, Z.-Q. Jiang, L.~Wei, W.-X. Zhou, Analytic degree
  distributions of horizontal visibility graphs mapped from unrelated random
  series and multifractal binomial measures, EPL 119~(4) (2017) 48008.
\newblock \href {https://doi.org/10.1209/0295-5075/119/48008}
  {\path{doi:10.1209/0295-5075/119/48008}}.

\bibitem{Wang-Li-Wang-2012-PhysicaA}
N.~Wang, D.~Li, Q.~Wang, {Visibility graph analysis on quarterly macroeconomic
  series of China based on complex network theory}, Physica A 391~(24) (2012)
  6543--6555.
\newblock \href {https://doi.org/10.1016/j.physa.2012.07.054}
  {\path{doi:10.1016/j.physa.2012.07.054}}.

\bibitem{Xie-Zhou-2011-PhysicaA}
W.-J. Xie, W.-X. Zhou, Horizontal visibility graphs transformed from fractional
  {B}rownian motions: topological properties versus the {H}urst index, Physica
  A 390~(20) (2011) 3592--3601.
\newblock \href {https://doi.org/10.1016/j.physa.2011.04.020}
  {\path{doi:10.1016/j.physa.2011.04.020}}.

\bibitem{Ravetti-Carpi-Goncalves-Frery-Rosso-2014-PLoS1}
M.~G. Ravetti, L.~C. Carpi, B.~A. Goncalves, A.~C. Frery, O.~A. Rosso,
  Distinguishing noise from chaos: objective versus subjective criteria using
  horizontal visibility graph, PLoS ONE 9 (2014) e108004.
\newblock \href {https://doi.org/10.1371/journal.pone.0108004}
  {\path{doi:10.1371/journal.pone.0108004}}.

\bibitem{Dai-Xiong-Zhou-2019-PhysicaA}
P.-F. Dai, X.~Xiong, W.-X. Zhou, Visibility graph analysis of economy policy
  uncertainty indices, Physica A 531 (2019) 121748.
\newblock \href {https://doi.org/10.1016/j.physa.2019.121748}
  {\path{doi:10.1016/j.physa.2019.121748}}.

\bibitem{Yu-2013-PhysicaA}
L.~Yu, {Visibility graph network analysis of gold price time series}, Physica A
  392~(16) (2013) 3374--3384.
\newblock \href {https://doi.org/10.1016/j.physa.2013.03.063}
  {\path{doi:10.1016/j.physa.2013.03.063}}.

\bibitem{Partida-Gerassis-Criado-Romance-Giraldez-Taboada-2022-ChaosSolitonsFractals}
A.~Partida, S.~Gerassis, R.~Criado, M.~Romance, E.~Giraldez, J.~Taboada, The
  chaotic, self-similar and hierarchical patterns in bitcoin and ethereum price
  series, Chaos Solitons Fractals 165 (2022) 112806.
\newblock \href {https://doi.org/10.1016/j.chaos.2022.112806}
  {\path{doi:10.1016/j.chaos.2022.112806}}.

\bibitem{Xie-Han-Zhou-2019-CommunNonlinearSciNumerSimul}
W.-J. Xie, R.-Q. Han, W.-X. Zhou, Tetradic motif profiles of horizontal
  visibility graphs, Commun. Nonlinear Sci. Numer. Simul. 72 (2019) 544--551.
\newblock \href {https://doi.org/10.1016/j.cnsns.2019.01.012}
  {\path{doi:10.1016/j.cnsns.2019.01.012}}.

\bibitem{Fan-Li-Yin-Tian-Liang-2019-AEn}
X.~Fan, X.~Li, J.~Yin, L.~Tian, J.~Liang, {Similarity and heterogeneity of
  price dynamics across China's regional carbon markets: A visibility graph
  network approach}, Appl. Energy 235 (2019) 739--746.
\newblock \href {https://doi.org/10.1016/j.apenergy.2018.11.007}
  {\path{doi:10.1016/j.apenergy.2018.11.007}}.

\bibitem{Sun-Wang-Gao-2016-PhysicaA}
M.~Sun, Y.~Wang, C.~Gao, {Visibility graph network analysis of natural gas
  price: The case of North American market}, Physica A 462 (2016) 1--11.
\newblock \href {https://doi.org/10.1016/j.physa.2016.06.051}
  {\path{doi:10.1016/j.physa.2016.06.051}}.

\bibitem{Kantelhardt-Zschiegner-KoscielnyBunde-Havlin-Bunde-Stanley-2002-PhysicaA}
J.~W. Kantelhardt, S.~A. Zschiegner, E.~Koscielny-Bunde, S.~Havlin, A.~Bunde,
  H.~E. Stanley, {Multifractal detrended fluctuation analysis of nonstationary
  time series}, Physica A 316~(1-4) (2002) 87--114.
\newblock \href {https://doi.org/10.1016/S0378-4371(02)01383-3}
  {\path{doi:10.1016/S0378-4371(02)01383-3}}.

\bibitem{Gao-Shao-Yang-Zhou-2022-ChaosSolitonsFractals}
X.-L. Gao, Y.-H. Shao, Y.-H. Yang, W.-X. Zhou, Do the global grain spot markets
  exhibit multifractal nature?, Chaos Solitons Fractals 164 (2022) 112663.
\newblock \href {https://doi.org/10.1016/j.chaos.2022.112663}
  {\path{doi:10.1016/j.chaos.2022.112663}}.

\bibitem{Milojevic-2010-JAmSocInfSciTechnol}
S.~Milojevic, Power law distributions in information science: making the case
  for logarithmic binning, J. Am. Soc. Inf. Sci. Technol. 61~(12) (2010)
  2417--2425.
\newblock \href {https://doi.org/10.1002/asi.21426}
  {\path{doi:10.1002/asi.21426}}.

\bibitem{Press-Teukolsky-Vetterling-Flannery-1996}
W.~Press, S.~Teukolsky, W.~Vetterling, B.~Flannery, {Numerical Recipes in
  FORTRAN: The Art of Scientific Computing}, Cambridge University Press,
  Cambridge, 1996.

\bibitem{Clauset-Shalizi-Newman-2009-SIAMRev}
A.~Clauset, C.~R. Shalizi, M.~E.~J. Newman, Power-law distributions in
  empirical data, SIAM Rev. 51~(4) (2009) 661--703.
\newblock \href {https://doi.org/10.1137/070710111}
  {\path{doi:10.1137/070710111}}.

\bibitem{Alstott-Bullmore-Plenz-2014-PLoSOne}
J.~Alstott, E.~T. Bullmore, D.~Plenz, Powerlaw: a python package for analysis
  of heavy-tailed distributions, PLoS One 9~(1) (2014) e85777.
\newblock \href {https://doi.org/10.1371/journal.pone.0085777}
  {\path{doi:10.1371/journal.pone.0085777}}.

\bibitem{Fagiolo-2007-PhysRevE}
G.~Fagiolo, Clustering in complex directed networks, Phys. Rev. E 76~(2) (2007)
  026107.
\newblock \href {https://doi.org/10.1103/PhysRevE.76.026107}
  {\path{doi:10.1103/PhysRevE.76.026107}}.

\bibitem{Fagiolo-Reyes-Schiavo-2010-JEvolEcon}
G.~Fagiolo, J.~Reyes, S.~Schiavo, The evolution of the world trade web: A
  weighted-network analysis, J. Evol. Econ. 20 (2010) 479--514.
\newblock \href {https://doi.org/10.1007/s00191-009-0160-x}
  {\path{doi:10.1007/s00191-009-0160-x}}.

\bibitem{Watts-Strogatz-1998-Nature}
D.~J. Watts, S.~H. Strogatz, Collective dynamics of `small-world' networks,
  Nature 393 (1998) 440--442.
\newblock \href {https://doi.org/10.1038/30918} {\path{doi:10.1038/30918}}.

\bibitem{Newman-2002-PhysRevLett}
M.~Newman, Assortative mixing in networks, Phys. Rev. Lett. 89~(20) (2002)
  208701.
\newblock \href {https://doi.org/10.1103/PhysRevLett.89.208701}
  {\path{doi:10.1103/PhysRevLett.89.208701}}.

\bibitem{Li-Jiang-Xie-Micciche-Tumminello-Zhou-Mantegna-2014-SciRep}
M.-X. Li, Z.-Q. Jiang, W.-J. Xie, S.~Micciche, M.~Tumminello, W.-X. Zhou, R.~N.
  Mantegna, A comparative analysis of the statistical properties of large
  mobile phone calling networks, Sci. Rep. 4 (2014) 5132.
\newblock \href {https://doi.org/10.1038/srep05132}
  {\path{doi:10.1038/srep05132}}.

\end{thebibliography}
\end{document}